\documentclass[12pt,preprint]{aastex}
\slugcomment{Accepted and scheduled for publication  
in  A\&A}
\def\lax {\ifmmode{_<\atop^{\sim}}\else{${_<\atop^{\sim}}$}\fi}  
\def\gax {\ifmmode{_>\atop^{\sim}}\else{${_>\atop^{\sim}}$}\fi}  
\def\gtorder{\mathrel{\raise.3ex\hbox{$>$}\mkern-14mu
             \lower0.6ex\hbox{$\sim$}}}

\def\cm2{cm$^{-2}$}
\def\s1{s$^{-1}$}

\usepackage{graphicx}
\usepackage{colortbl}
\usepackage{txfonts}

\begin{document}

\title{Changing  Look AGN: An X-ray Look}

\author{Lev Titarchuk\altaffilmark{1}, Elena Seifina\altaffilmark{2}  and Yegor Mishin\altaffilmark{2}}

\altaffiltext{1}{Dipartimento di Fisica, Universit\`a di Ferrara, Via Saragat 1, I-44100 Ferrara, Italy, email:titarchuk@fe.infn.it; George Mason University Fairfax, VA 22030}
\altaffiltext{2}{Moscow M.V.~Lomonosov State University/Sternberg Astronomical Institute, Universitetsky Prospect 13, Moscow, 119992, Russia; seif@sai.msu.ru}

\begin{abstract}
{To date, a number of changing-look (CL) active galactic nuclei (AGNs) are known. We studied, in detail what happens to the X-ray spectrum during the CL events using the example of the nearby CL Seyfert  NGC~1566, which was observed by {\it Swift, NuSTAR, XMM-Newton}, and {\it Suzaku}. We applied the Comptonization model to describe an evolution of NGC~1566 X-ray spectra during outbursts and compared these results with a typical behavior for other AGNs to identify some differences and common properties that can ultimately help us  to  better understand the physics of the CL phenomenon. 
We found that changes in the X-ray properties of NGC~1566 are characterized by a different combination of Sy1 (using 1H~0707--495 as a representative)  and Sy2 properties (using NGC~7679 and Mrk~3 as  their representatives). At high X-ray luminosities NGC~1566 exhibits the behavior typical for  Sy1, and at low luminosities we see a transition of NGC~1566 from the Sy1 behavior to the Sy2 pattern. 
We revealed the  saturation of the spectral indices, $\alpha$  for these four AGNs during outbursts ($\alpha_{1566}\sim 1.1$, $\alpha_{0707}\sim 2$, $\alpha_{7679}\sim 0.9$ and $\alpha_{mrk3}\sim0.9$) and determined the masses of the 
black holes (BHs) in the centers of these AGNs namely, M$_{0707}\sim 6.8\times 10^7$ M$_ {\odot}$, M$_{7679}\sim8.4\times 10^6$ M$_{\odot}$, M$_{mrk3}\sim2.2\times 10^8$ M$_{\odot}$ and M$_{1566} \sim2\times 10^5 $ M$_{\odot}$, applying the scaling method. 
Our spectral analysis shows that the changing-look of NGC1566 from Sy1.2 to Sy1.9 in 2019 was accompanied by the transition of NGC~1566 to an accretion regime which is typical for the intermediate and highly soft spectral states of other BHs. 
We also find that when going from Sy2 to Sy1, the spectrum of NGC~1566 shows an increase in the soft excess accompanied by a decrease in the Comptonized fraction ($0.1<f<0.5$), which is consistent with the typical behavior of BH sources during X-ray outburst decay. Our results strongly suggest that the large diversity in its behavior observed among CL, Sy1, and Sy2 AGNs with different X-ray luminosities can be explained by changes in a single variable parameter, such as the ratio of the AGN's X-ray luminosity to its Eddington luminosity,  without any need for additional differences in Sy AGN parameters, such as its inclination, thereby blurring the distinction between the Sy1, Sy2 and CL-AGN subclasses.
}
\end{abstract}


\section{Introduction}
Recent detections of numerous changing-look  active galactic nucleus (CL-AGN) events have led to a surge of interest in the study of this phenomenon. The origin of the CL-AGN events is still unclear. NGC~1566  galaxy is a prominent representative of the CL-AGN subclass of AGNs. Its supermassive black hole (SMBH) is estimated to be (1.3$\pm$0.6)$\times 10^7$ M$_{\odot}$. As the closest  galaxy ($z=0.005$, see also Table~\ref{tab:parameters_binaries}) among the CL-AGNs, NGC~1566 has been intensively studied  over the past 70 years. It was first classified as a Seyfert 1 (Sy1) with the broad H$_{\alpha}$ and H$_{\beta}$ lines \citep{deVaucouleurs61}. The H$_{\beta}$ line was later found to be weak, leading to the source being classified as Seyfert 2 (Sy2) \citep {Pastoriza+Gerola70}. This is consistent with the generally accepted classification of AGN as Sy1 or Sy2, depending on the presence or absence of broad optical emission lines. 

The existence of different classes of AGN can be explained using a unified model \citep{Antonucci93}, based on the orientation of an optically thick torus relative to the line of our vision. However, NGC~1566 exhibits the appearance or disappearance of broad optical emission lines, transitioning from Sy 2 (or Sy 1.8--1.9) to Sy 1 (or Sy 1.2--1.5) and vice versa within a few months \citep{da Silva17}, and does not fit into the generally accepted classification and constitutes a significant  problem in our understanding of AGN. 

Since 2018, observations of NGC~1566 have been carried out in many wavelength ranges: from  hard X-rays (see X-ray image of NGC~1566 in Fig.~\ref{image_1566}) to infrared rays \citep{Ducci18}. It became clear that the NGC~1566 flux varies in all wavelengths. In particular, in July 2018, its flux increased strongly and reached its peak \citep{Parker19,Oknyansky19,Oknyansky20}. Long-term light curves of ASAS-SN and NEOWISE showed that the IR and optical flux began to increase as early as September 2017 \citep{Cutri18,Dai18}. The Swift/XRT flux increased by about 30 times (Fig.~\ref{ev_1566}) when the source changed from Sy1.8--1.9 to Sy1.2  \citep{Oknyansky20,Oknyansky19}. The source became Sy1 with an appearance of strong broad emission lines   \citep{Oknyansky19,Ochmann20}. Having reached their peak, the source fluxes decreased in all wavebands. Several smaller flares were observed after the main outburst \citep{Grupe18,Grupe19}.

This 2018--2019 outburst of NGC~1566 with a CL-effect  was observed using {\it NuSTAR} and {\it XMM-Newton}. \cite{Jana21} investigated these observations applying the {\tt power-law, NTHCOMP, OPTXAGNF} and {\tt RELXILL} models with an addition of {\tt Gaussian} lines and taking into account X-ray absorption effects. Each of these models shed light on the CL-peculiarities and the change in the object itself. In the {\tt power-law} model, \cite{Jana21} found the presence of a soft excess ($<$ 2 keV), which they approximated by an additional {\tt bbody} component. This model fits well the {\it XMM-Newton} data for about 2.5 years before the 2018 outburst 
with parameters $N_H = (3.53 \pm 0.06) \times 10^{21}$ cm$^{-2}$ and a 
 photon index $\Gamma\sim 1.7$ ($\Gamma=\alpha-1$). 
Moreover, an iron K$_{\alpha}$ emission line was detected at 6.4 keV with an equivalent width (EW) of 200 eV. The rise phase of the 2018 outburst was analyzed using simultaneous {\it XMM-Newton} and {\it NuSTAR} observations of NGC~1566, and in this model the fit yielded $\Gamma=1.8$.  

The Fe K$_{\alpha}$ line was detected at 6.38 keV with an $EW > 110$ eV, as well as the Fe~XXVI emission feature at 6.87 keV with an $EW <$~37 eV. It is interesting to note that two ionized absorbers were required to fit the source spectra in the rise phase, one low-ionization absorber  ($\xi \sim 101.7\pm 0.1$) with $N_{H, 1} = (8.1 \pm 2.2) \times 10^{20}$ cm$^{-2}$, and one high-ionization absorber ($\xi\sim 104.7\pm 0.4$) with $N_{H, 2} = (4.3\pm 0.4)\times 10^{21}$ cm$^{-2}$. 

In the outburst decay phase the photon index was almost constant ($\Gamma\sim 1.7$),  the blackbody temperature was constant $kT_{bb}\sim 110$ eV and the Fe K$_{\alpha}$ line was detected with EW $>$~100 eV. Although  column density of the weakly ionizing absorber varied in the range of $N_{H, 1}\sim(1.2-1.3)\times 10^{21}$ cm$^{-2}$, no highly ionizing absorber was required to fit the spectra during the outburst decay. In addition, a weak reflection hump was detected as an excess of emission at energies 15--40 keV. In the {\tt NTHCOMP} model, \cite{Jana21}  fixed the seed photon temperature  at  $kT_s = 30$ eV. They again needed to account for two absorption components during the outburst rise and initial decay phases. They found that as the corona size $R_{cor}$ decreased, the plasma electron temperature $kT_e$ increased from 60 to 100 keV with a nearly constant photon index ($\Gamma = 1.7-1.9$).

To model this NGC~1566 outburst with {\tt OPTXAGNF} \cite{Jana21} again accounted for two absorption components during the outburst rise and initial decay phases. The photon index $\Gamma$ and optical depth $\tau$ were held almost constant at $\Gamma = 1.7-1.8$ and $\tau\sim 4-5$  throughout the 2018 outburst. Before this outburst, the Eddington ratio and the size of the X-ray corona were found to be small  ($L/L_{Edd}\sim 0.04$ and $R_{cor} = 12 R_g$) compared to the rise phase outburst 
($L/L_{Edd}\sim 0.23$ and $R_{cor} = 43 R_g$). In later observations, they decreased  $L/L_{Edd}\sim 0.06$ and $R_{cor} \sim 20  R_g$.

This model fit all observations well and indicated an increase in the accretion rate and corona size at the outburst peak, as well as a high degree of the plasma Comptonization during the outburst. The {\tt RELXILL} model again required taking into account two absorption components during the rising and initial decay phases of the outburst. In addition, the model included the reprocessed X-ray emission from the disk as a reflectivity parameter $R_{refl}$, which turned out to be relatively weak ($R_{refl}\sim  0.1-0.2$) throughout the outburst. The inner disk edge $R_{in}$ varied from  $4R_g$ to $7R_g$. 

\cite{Tripathi+Dewangan22a}, hereafter TD22a  also analyzed NGC~1566 using {\it XMM-Newton, Swift} and {\it  NuSTAR} data at different epochs during the decay phase of the 2018 outburst using a  broadband continuum model {\tt OPTXAGNF} \citep{Done12} taking into account the thermal Comptonization ({\tt THCOMP}) and the reflection ({\tt RELXILL}) model, as well as investigated the correlations between the accretion disk X-ray emission, soft X-ray excess and the power-law continuum. They argued that at low X-ray flux levels, the source soft X-ray excess was absent and only the disk emission provided seed photons for thermal Comptonization in the corona, while at high flux levels, both the soft X-ray excess and the disk emission were present, providing seed photons for the thermal Comptonization in the corona. 

TD22a  found that the X-ray photon index remained constant ($\Gamma\sim 1.66-1.72$), although the electron temperature of the corona increased from 22 to 200 keV from June 2018 to August 2019. At the same time, the optical depth of the corona $\tau$ decreased from 4 to 0.7, and the scattering fraction increased from 1\% to 10\%. TD22a  interpreted this as an increase in the size of the corona and its heating with a decrease in the mass accretion rate during the decay phase.

Different models are used to study AGN, including those specifically designed for AGN (assuming a specific structure of active galactic nuclei) and generalized models (with a minimum number of specific assumptions). The first of these, such as the {\tt AGNSED} model \citep{Kubota+Done18} and models taking into account the double (warm and hot) corona 
\citep{Petrucci13}, consider three zones of AGN X-ray formation: an outer standard disc, an inner warm Comptonizing region (to produce the soft X-ray excess) and a hot corona (see Fig.~10 in \cite{Petrucci13}). Generalized models such as the {\tt BMC} \citep{TMK97,tz98,LT99,BRTST99,ST99}, {\tt NTHComp} \citep{Zdziarski96}, {\tt Comptb} \citep{ft11}, {\tt Comptt} \citep{Titarchuk94} are based on the first principles and also consider three X-ray formation regions: an outer standard disc, a hot Comptonizing region (a so called {\it Transition layer}) to reproduce the soft X-ray excess and a thermal Comptoniziation  hump in the source spectrum, and a converging flow region (possibly analogous to a hot corona, to form a hard high-energy tail in the source X-ray spectrum). In this paper, we apply these generalized models to identify the  features of CL-AGN compared to Sy1 and Sy2, without assuming a specific geometry of these sources.

In addition, for better understanding the properties of NGC~1566 during CL events, it is interesting to compare the behavior of this CL-AGN with other AGNs, such as Sy1 and Sy2.  To do this, we used the Sy1 galaxy (1H~0707--495) and the Sy2 galaxy (NGC~7679) to study their differences and similarities during X-ray outbursts in comparison to CL-AGN.

The Narrow-Line Sy 1 galaxy 1H~0707--495 (z = 0.0411, hereafter 1H~0707, see also Table~\ref{tab:parameters_binaries}) is a bright Narrow Line Seyfert 1 (NLS1) galaxy \citep{Leighly99}.  BH mass estimates  in 1H~0707 vary over a wide range from 2$\times 10^6$ to 10$^7$ M$_{\odot}$ \cite{Zhou+Wang05,Kara13,Done+Jin16}. In particular, \cite{Zoghbi11} assumed that most of the radiation is emitted at $\sim 2r_g$ and estimated the inner radius and the  emissivity index using   their  spectral fitting. They interpreting the 30 s lag as the light crossing time and estimated a BH mass of $M_{BH}\sim 2\times 10^6M_{\odot}$, which was consistent with the uncertain mass of this BH quoted in the literature \citep{Zhou+Wang05}.

NGC~7679 is a barred lenticular galaxy seen face on \citep{Yankulova07}. It is located at a distance of about 200 million light years from Earth (see also Table~\ref{tab:parameters_binaries}). It was discovered by Heinrich d'Arrest on September 23, 1864. The nucleus of NGC~7679 turned out to be active and was classified as a Seyfert galaxy. The most common theory for the energy source of Seyfert galaxies is the presence of an accretion disk around a supermassive black hole. NGC~7679 is believed to host a supermassive black hole, with a mass estimated at 5.9$\times 10^6$ M$_{\odot}$ based on velocity dispersion \citep{Alonso-Herrero13}. The X-ray spectrum of NGC~7679 using {\it Beppo}SAX shows no significant absorption above 2 keV, and the K$_{\alpha}$ line of iron was only slightly detected. However, the galaxy shows signs of obstruction to visual light as it lacks broad emission lines. Two possible reasons: the presence of dust or an X-ray emitting accretion disk that is not covered and the broad line region is covered \citep{Risaliti02}. To date, the classification of NGC~7679 as Sy1 or Sy2 remains controversial. On the one hand, this AGN has an optical spectrum without broad emission lines, which allows it to be classified as a Sy2 \citep{Risaliti02}. On the other hand, NGC~7679 has significant variable X-ray emission, typical of Sy1. The main feature of objects like NGC~7679 is not the strength of their starburst, but the apparent optical faintness of the Sy1 nucleus when compared to the X-ray luminosity. 

As a second representative of Sy2 galaxies we used Markarian 3 (hereafter Mrk 3), which  is one of the brightest and best-studied members of the Sy2 class. The host galaxy is classified as an elliptical or {\it S0} galaxy type. 
\cite{Awaki90}, \cite{Awaki91} and \cite{Smith+Done96} revealed an anomalously flat power-law continuum emerging through a tall occultation column ($N_H\sim 6\times 10^{23}$ cm$^{-2}$) from {\it GINGA} observations of Mrk~3. A strong Fe line with high equivalent width ($EW \sim 1.3$ keV) was also detected. Mrk~3 has the hardest spectrum among all 16 Sy2 galaxies studied by \cite{Smith+Done96}, significantly harder than the spectrum of Sy1 galaxies. In {\it ASCA} observations of Mrk 3, the object showed a spectrum with a photon index $\Gamma\sim 1.8$ and a two-component iron line \citep{Iwasawa94}.  

Thus, the dominant component of the K$_{\alpha}$ iron line at 6.4 keV has $EW=0.9$ keV, while the second component at 7 keV has $EW=0.2$ keV. A reanalysis of the spectrum of Mrk~3 using non-simultaneous {\it GiINGA}, {\it ROSAT} and {\it ASCA} observations \citep{Griffiths98} covering a wide spectral band (0.1--30 keV) yielded a typical value for the power law, $\Gamma\sim$~1.7, when either an additional absorption edge at 8 keV (possibly arising from a warm absorber) or reflection was included in the spectral model. Recent observations with {\it Beppo}SAX \citep{Cappi99}, which extend the spectral coverage to 150 keV, indeed confirm a presence of a steep ($\Gamma\sim$~1.8) internal power law. \cite{Turner97} also reanalyzed the {\it ASCA} data and proposed an alternative model in which the internal continuum is resolved through very large absorption column ($N_H > 10^{24}$ cm$^{-2}$), while the reflection component is not obscured. 

\cite{Georgantopoulos99} analyzed the {\it RXTE} data for Mrk~3 and found an agreement with the earlier results of {\it GINGA}. They used a spectral model consisting of a very hard power-law continuum ($\Gamma\sim$~1.1) modified below $\sim$6 keV by a strongly absorbing column ($N_H\sim 6\times10^{23}$ cm$^{-2}$) and an iron line with a high equivalent 
width at 6.4 keV. Their conclusions are consistent with the results by \cite{Turner97} on the complex absorption of a molecular torus.

It is interesting to note that \cite{Boorman18} found an anti-correlation between the equivalence width of the narrow core of the neutral Fe K$\alpha$ fluorescence line, ubiquitously observed in the reflection spectra of obscured AGNs, and the mid-infrared continuum luminosity at 12 $\mu$m. This is consistent with numerous studies of the X-ray Baldwin effect for unobscured and slightly obscured AGNs and challenges the traditional view that the Fe K$\alpha$ line originates from the same region as the underlying reflection continuum, which together make up the reflection spectrum. However, the found anti-correlation  does not apply to Mrk~3, as \cite{Ricci15} find a Compton-thin column density (90\% confidence level) for this source.

\cite{Risaliti02b} discuss another type of X-ray changing look for 
AGNs based on changes in column density (from 20\% to 80\%) during AGN X-ray variability on time scales of months to years. Namely, the AGN switched between thin Compton thin ($N_H < 1.5\times 10^{24}$ cm$^{-2}$) and Compton thick ($N_H > 1.5\times 10^{24}$ cm$^{-2}$) regimes (see also \cite{Matt03}). 

To describe the BH states in AGN during outbursts, it is convenient to use the terminology used to identify the BH states in X-ray binaries (XRBs). Thus, the observed manifestations of BHs in galactic sources are traditionally described in terms of a classification of BH spectral states (see \cite{Klein-Wolt+vanderKlis08,Remillard+McClintock06,Belloni05}, for various definitions of BH states). A general classification of BH states for four main BH states is accepted by the community: the quiescent, the low hard (LHS), the intermediate (IS, sometimes subdivided into Hard intermediate, HIMS, and  Soft intermediate, SIMS) and the high soft (HSS)  states.  

When a BH transient goes into an outburst, it leaves the quiescent state and enters the LHS, a low-luminosity state with an energy spectrum dominated by the thermal component of Comptonization combined with a weak thermal component. The photon spectrum in the LHS is thought to be the result of Comptonization (upward scattering) of soft photons, which originated in the relatively weak inner part of the accretion disk, from electrons in the hot surrounding plasma (see, e.g., \cite{Sunyaev+Titarchuk80}). The HSS photon spectrum is characterized by a pronounced thermal component, which is probably a sign of strong radiation emanating from the geometrically thin accretion disk. The IS is a transition state between the LHS and the HSS. At the same time, the subdivision the IS into SIMS and HIMS states reflects the specifics of a BH source at the entrance and exit from the outburst. As for supermassive BHs in AGNs, despite the different time scale, sizes and sources of accreted matter compared to XRBs, they show a similar pattern of the  spectral changes during their outbursts. Therefore, we use the terminology given above, but note that there is, of course, no direct analogy. In addition, the excess soft radiation of AGNs is another component of the Comptonization effect, since the disk temperature of AGNs is much lower.

X-ray spectroscopy is a very powerful tool for shedding light on the CL-AGN relationship, mainly because X-rays are emitted closer to the primary emission source than optical emission lines, which are reprocessed emission from interstellar gas (e.g., \citet{Terashima09}). In this paper we present the comparative analysis for NGC~1566, 1H~0707--495 and NGC~7679 using the {\it Suzaku}, ASCA, {\it Swift} and {\it Beppo}SAX   observations.    In \S 2 we present the list of observations used in our data analysis while 
in \S 3 we provide the details of X-ray spectral analysis.  We analyze an evolution of X-ray spectral and timing  properties during the state transition in \S 4.   In \S 5 we present  a description of the spectral models used for fitting these data.  In  \S  6 we  discuss  the  main results of the paper. In \S 7 we present our final conclusions.

 \section{Data Selection \label{data}}
NGC~1566 was   observed by {\it Swift} (2007--2023) 
and with {\it Suzaku} (2012). 
1H~0707--495 was   detected by {\it Swift} (2010--2018), 
by $ASCA$ (1995, 1998 and 2005) 
as well as with {\it Suzaku} (2005). 
While NGC~7679 was   observed by {\it Swift} (2017 and 2019), 
by $ASCA$ (1998 and 1999) 
and with {\it Beppo}SAX (1998). 
We extracted these data from the HEASARC archives and found that these data  cover 
a wide range of X-ray luminosities (see Tables \ref{tab:table_Suzaku+ASCA_SAX} and \ref{tab:par_Swift_data}).  We recognized that the well-exposed {\it Beppo}SAX, {\it ASCA} and {\it Suzaku} data are  preferable for the determination of   low-energy photoelectric absorption. 

\subsection{\it Swift data \label{swift data}}

 Using {\it Swift}/XRT data in the 0.3--10 keV energy range, we studied flaring events of NGC~1566, 1H~0707 and NGC~7679 (see the log of observations for all three sources
in Table~\ref{tab:par_Swift_data}).  The data used in this paper are public and available through the GSFC public archive\footnote{https://heasarc.gsfc.nasa.gov}. 

Data were processed using the HEASOFT v6.14, the tool {\tt xrtpipeline} v0.12.84, and the calibration files (CALDB version 4.1). The ancillary response files were created using {\tt xrtmkarf} v0.6.0 and exposure maps were generated by {\tt xrtexpomap} v0.2.7. The source events accumulated in a circular region of optimal radius from 17 to 45{\tt"} centered on the individual source position shown in Table \ref{tab:parameters_binaries}.  The background was estimated in a nearby source-free circular region taking into account the relative areas of the source and background regions. Spectra were rebinned with at least ten 
 counts  in each energy bin using the {\tt grppha} task in order to apply $\chi^2$ statistics. We also used the online XRT data product generator\footnote{http://www.swift.ac.uk/user\_objects/} to obtain the image of the source field of view in order to make a visual inspection and to get rid of possible contamination by nearby sources  \citep{Evans07,Evans09}. 

\subsection{\it BeppoSAX data \label{sax data}}

We used {\it Beppo}SAX data of NGC~7679 carried out on December 6--9, 1998. In Table~\ref{tab:table_Suzaku+ASCA_SAX} (bottom line) we show the log of the {\it Beppo}SAX 
observation analyzed  in this paper. Generally, broad band  energy spectra of the source were obtained combining data from  three {\it Beppo}SAX Narrow Field Instruments (NFIs): the Low Energy Concentrator Spectrometer [LECS; \citet{parmar97}] for 0.3 -- 4 keV, the Medium Energy Concentrator Spectrometer [MECS; \citet{boel97}] for 1.8 -- 10 keV and the Phoswich Detection System [PDS; \citet{fron97}] for 15 -- 60 keV.  The SAXDAS data analysis package is used for processing data. For each of the instruments we performed the spectral analysis in the energy range for which the response matrix is well determined. Both LECS and MECS spectra were accumulated in circular regions of 4{\tt '} radius. The LECS data have been re-normalized based on MECS. Relative normalization of the NFIs were treated as free parameters in  model fitting, except for the MECS normalization that was fixed at a value  of 1. We checked after that  this  fitting procedure  if these normalizations  were in a standard range for each  instruments
\footnote{http://heasarc.nasa.gov/docs/sax/abc/saxabc/saxabc.html}.
In addition,  spectra are rebind  accordingly to  energy resolution of the instruments in order to obtain  significant data points.  We rebin the LECS spectra with a binning factor which is not constant over energy (Sect.3.1.6 of Cookbook for the BeppoSAX NFI spectral analysis) using  re-binnig template files  in GRPPHA of  XSPEC \footnote{http://heasarc.gsfc.nasa.gov/FTP/sax/cal/responses/grouping}. Also we re-binned the PDS spectra with linear binning  factor 2, grouping two bins together (resulting bin width is 1 keV).  Systematic error of 1\% have been applied to these analyzed spectra. 
\subsection{ \it Suzaku data 
\label{suzaku data}}

{\it  Suzaku} observed NGC~1566 and 1H~0707. Table~\ref{tab:table_Suzaku+ASCA_SAX} summarizes the start and  end times, and the MJD interval  for each of these observations,  indicated by {\it green} triangle in top of Figure \ref{ev_1566} for NGC~1566. 
One can see a description of the {\it Suzaku} experiment in \cite{Mitsuda07}. 
For observation obtained by a focal X-ray CCD camera (XIS, X-ray Imaging Spectrometer, \cite{Koyama07}), which is sensitive over  the 0.3--12~keV range, we used  software of  the {\it Suzaku} data processing {\tt pipeline} (ver. 2.2.11.22).  We carried out the data reduction and analysis following the standard procedure using the 
{\tt HEASOFT software package} (version 6.25)  and following  the {\it Suzaku} Data Reduction Guide\footnote{http://heasarc.gsfc.nasa.gov/docs/suzaku/analysis/}. 
The spectra of the source were extracted using spatial regions within the 3.51${\tt '}$-radius circle centered on the source nominal position (Table~\ref{tab:parameters_binaries}),  while a background was extracted from source-free regions 
for each XIS module separately.

The spectrum data were re-binned to provide at least 20 counts per spectral bin to validate the use of the $\chi^2$-statistic. We carried out  spectral fitting  applying XSPEC v12.10.1.  The energy ranges around of 1.75 and 2.23 keV are not  used for spectral fitting because of the known artificial structures in the XIS spectra around the Si and Au edges.  Therefore, for spectral fits we have chosen  the 0.3 -- 10 keV  range  for the XISs (excluding 1.75 and 2.23 keV points).

\subsection{{\it ASCA data} \label{asca data}}
{\it ASCA} observed 1H~0707--495 and NGC~7679 (see Table~\ref{tab:table_Suzaku+ASCA_SAX}, which  summarized the start time, end time, and the MJD interval. One can see a description of the {\it ASCA}  data  by \cite{Tanaka94}.  
The solid imaging spectrometers (SIS)  operated in Faint CCD-2 mode. The {\it ASCA} data were screened using the ftool ascascreen and the standard screening criteria. The
spectrum for the source were extracted using spatial regions with a diameter of 4${\tt '}$ (for SISs) and 6${\tt '}$ (for GISs) centered on the nominal position of the source,
while background was extracted from source-free regions of comparable size away from the source. The spectrum data were re-binned to provide at least 20 counts per spectral bin to validate the use of the $\chi^2$-statistic. The SIS and GIS data were fitted using {\tt XSPEC} in the energy ranges of 0.6 -- 10 keV and 0.8 -- 10 keV, where the spectral responses are well-known.  

\subsection{\it RXTE data \label{rxte data}}

We have also  analyzed the  available data of 1H~0707--495 (1997) obtained with {\it RXTE}~\citep{bradt93}.   
We have  made an analysis of the {\it RXTE} observation  of 1H~0707 during {\it the low/hard state, LHS} (ID=20309-01-01-00). 
We should note that that  the "LHS", "HSS"  notations , are not very often applied to AGNs. Altogether,  we introduce these notations because of the photon index change $\Gamma$.
For example, in similarity with binary systems if the $\Gamma$ index is within the 1.5-1.7 range then we  associate it with the Comptonization photon index  which is typical for the low/hard spectrum in the binaries and vice versa if $\Gamma>2.3$ then we   relate  it with the  high/soft state (HSS)  observed in the binaries. 

Standard tasks of the LHEASOFT/FTOOLS 6.33.2 software package were utilized for data processing using methods recommended by {\it RXTE} Guest Observer Facility according to the  {\it RXTE} Cook Book\footnote{http://heasarc.gsfc.nasa.gov/docs/
xte/recipes/cook\_book.html}. For spectral analysis, we used  data from the Proportional Counter Array (PCA) and High-Energy X-Ray Timing Experiment (HEXTE) detectors. {\it RXTE}/PCA spectra ({\it Standard 2} mode data, 3 -- 50~keV energy range) have been extracted and analyzed using the PCA response calibration (ftool pcarmf v11.1). The relevant deadtime corrections to energy spectra
have been applied. In turn, HEXTE data in the 20--150 keV energy range were used for the spectral analysis in order to exclude the channels with largest uncertainties. We subtracted background corrected in off-source observations. 

We used the data which  are available through the GSFC public archive\footnote{http://heasarc.gsfc.nasa.gov}. In Table~\ref{tab:table_Suzaku+ASCA_SAX} we presented the informations  of the {\it RXTE} observation of 1H~0707 and Mrk~3. Systematic error of 0.5\% have been applied to the analyzed spectrum. 

\subsection{{\it NuSTAR} data\label{nustar data}}
We processed {\it NuSTAR} observations 
using the {\tt NuSTARDAS} (version 2.1.1) and the latest files available in the {\it NuSTAR} Calibration Database
 \footnote{http://heasarc.gsfc.nasa.gov/FTP/caldb/data/nustar/fpm/}. We generated clean event files for each observation of NGC~1566 using the {\tt nupipeline} task and extracted the source and background spectra from circular regions of 60{\tt"} and 90{\tt"} radius, respectively, centered at the source position (Table~\ref{tab:parameters_binaries}). Source spectra 
were extracted using the {\tt nuproduct} task. We binned each spectrum so that there were at least 25 counts per spectral bin.

\subsection{{\it XMM–Newton} data\label{xmm-n data}}
We processed the XMM-Newton data from EPIC-pn (Struder et al. 2001) using SAS software (version 18.0.0)\footnote{https://www.cosmos.esa.int/web/xmm-newton/sas-threads} and the latest calibration files. Following \cite{Jana21}, we corrected for the pile-up by excluding events in the inner 10{\tt"} radius circular region from the clean event lists. We extracted the source spectrum from the annular region (with outer radius 30{\tt"} and inner radius 10{\tt"}\footnote{https://www.cosmos.esa.int/web/xmm-newton/sas-thread-epatplot}) centered at the source position (Table~\ref{tab:parameters_binaries}) and the background spectrum from a 40{\tt"} radius circular region from the source-free region for each observation. The response files ($arf$ and $rmf$ files) were generated using the SAS {\tt arfgen} and {\tt rmfgen} tasks, respectively. We binned each spectrum so that there were at least 25 counts per spectral bin.

\section{Results \label{results}}

\subsection{Images of NGC~1566, 1H~0707--495, NGC~7679  and Mrk~3\label{images}}

The {\it Swift}/XRT (0.3 -- 10 keV) image of the 
NGC~1566, 1H~0707--495, Mrk 3  and  NGC~7679  fields of view (FOVs) are presented in Figs.~\ref{image_1566}$-$\ref{image_7679}, respectively.

{\it Swift} X-ray image of NGC~1566 (Fig.~\ref{image_1566}) is accumulated from 12th  December, 2007 to  1 November 2023 with an exposure time of 230 ks. Yellow contours in this image demonstrate the lack of X-ray jet structure (elongated), as well as minimal contamination by other point sources within 18{\tt" } in the field of view around NGC~1566. The closest next source is 21 {\tt"} (LSXPS J041956.5--545528  is marked with a green circle). 

The field of view for 1H~0707 is shown in Fig.~\ref{image_0707}. $Swift$ X-ray image of 1H~0707--495 (2SXPS J070841.4--493306 -- according to the {\it Swift} catalog), accumulated from April 3, 2010 to April 30, 2018 with a 161 ks exposure time. The gray dashed square with a side of 6.3{\tt'} (160 pixels) in the 1H~0707 image demonstrates the absence of other nearby objects in the 1.3{\tt'} field of view. The next closest source is 2.8{\tt'} away (outside the image).

According to the {\it Swift} catalog),  the field of view for Mrk~3 is shown in Fig.~\ref{image_Mrk3}. $Swift$ X-ray image of Mrk~3 (2SXPS J061536.2+710214 accumulated from May 3, 2006 to April 20, 2015-04-20 with a 91 ks exposure time.  The green contours in this image demonstrate the lack of X-ray jet-like structure around Mrk~3, as well as minimal contamination by other point sources within 18{\tt"} of the source environment. The closest next source is 21{\tt"} (LSXPS J041956.5--545528 is marked with a white circle, 
see Fig.~\ref{image_Mrk3}).

The $Swift$ X-ray image of NGC~7679, presented in Fig.~\ref{image_7679}, is catalog), 
accumulated from July 25, 2015 to October 6, 2017 with a 1.7 ks exposure time. The image segment highlighted with a gray square (6.3{\tt'} to a side) is also shown in more detail in the enlarged panel. Contour levels demonstrate the absence of X-ray jet (elongated) structure and minimal contamination from other point sources and diffuse radiation in the 1.3{\tt'} field of view around NGC~7679. The next closest source, 279{\tt"} is 2LSXPS~J232904+032910.

\subsection{X-ray Light Curves\label{lc}}

All three sources are Seyfert galaxies and the BHs at their centers have approximately the same masses ($\sim$10$^6$ M$_{\odot}$, see Table~\ref{tab:parameters_binaries}). However, they exhibit completely different temporal patterns of activity. To demonstrate this, we compared their long-term behavior in the form of light curves.

\subsubsection{NGC~1566 Light Curve\label{lc_1566}}
\begin{figure}
\centering
\includegraphics[scale=0.9,angle=0]{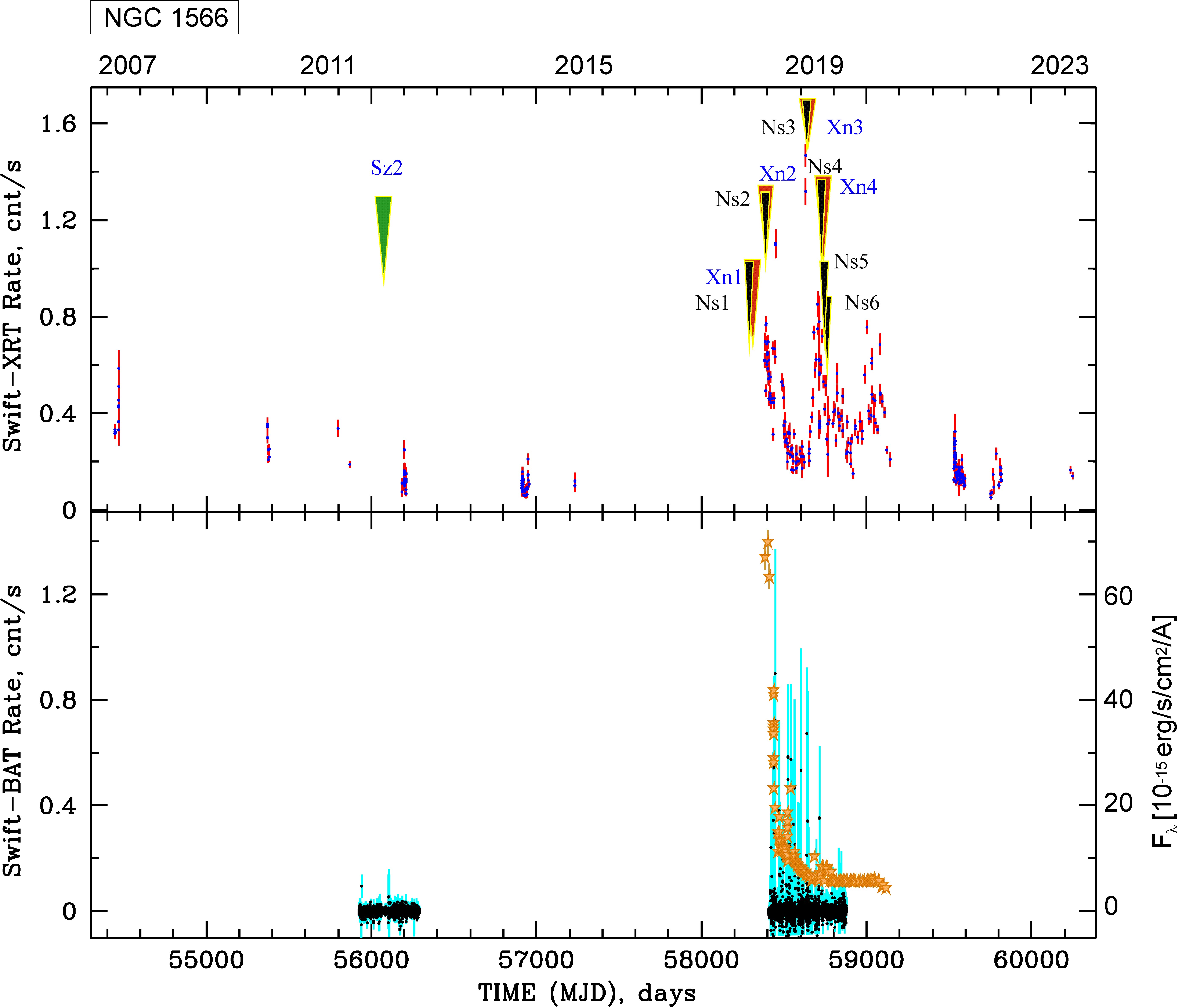}
\caption{Evolution of NGC~1566 during  2007--2023 observations with Swift/XRT (top panel, 0.3--10 keV), Swift/BAT (bottom panel, black points, 15--150 keV) and Swift/UVOT (bottom panel, orange stars, UVW2 band [1600--2300 $\dot A$], right axis). Green, black and red  triangles indicates {\it Suzaku, NuSTAR} and {\it XMM-Newton} observations, respectively, used for in our analysis.
}
\label{ev_1566}
\end{figure}

\begin{figure}
\centering
\includegraphics[scale=0.8,angle=0]{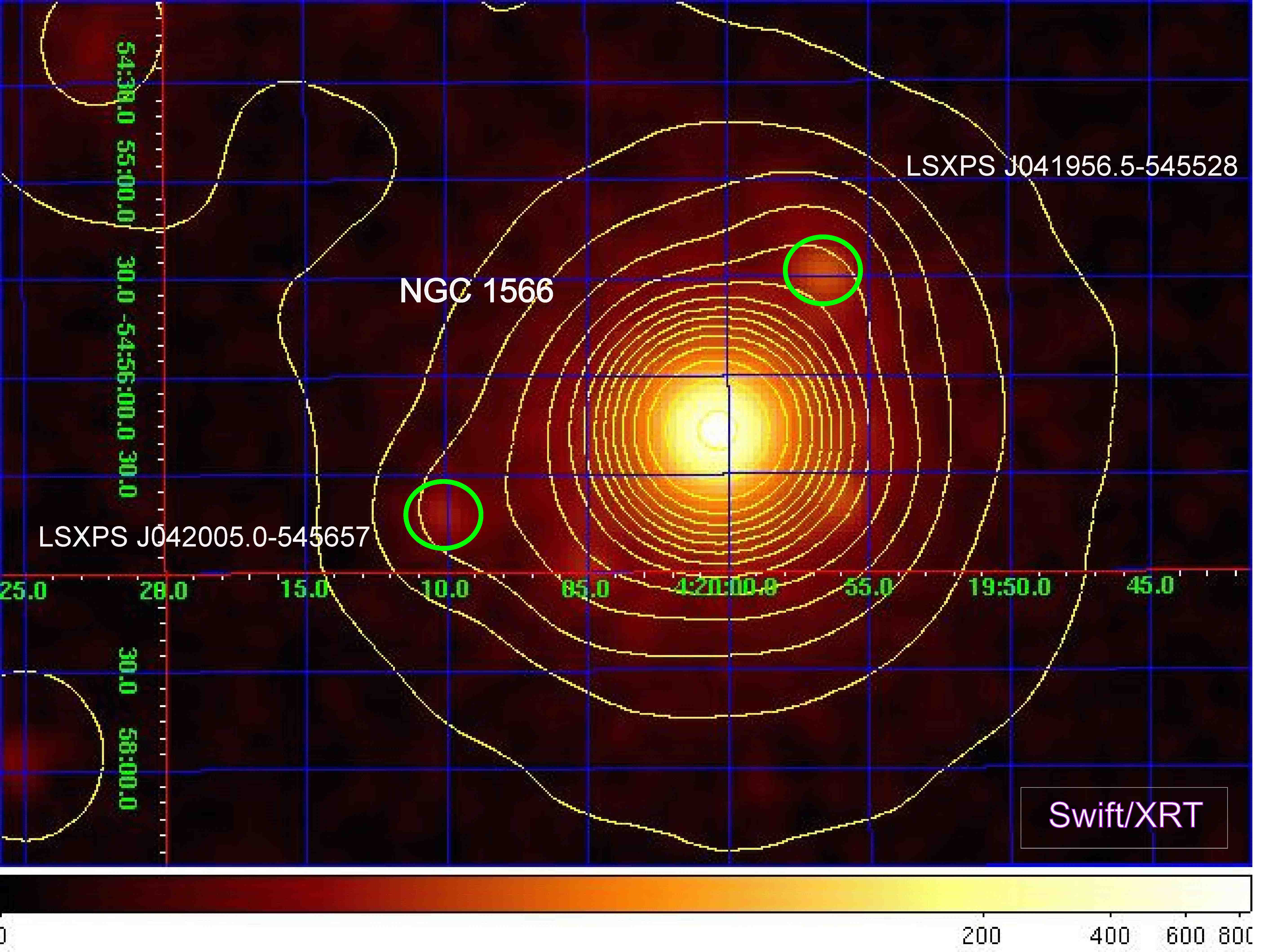}
\caption{
X-ray image of NGC~1566 (in center), 
accumulated by $Swift$/XRT from December 12, 2007 to November 2, 2023 with a 230 ks exposure time. The yellow contours in this image demonstrate the lack of X-ray jet-like structure, 
as well as minimal contamination by other point sources within 18{\tt"}. The closest next source is 21{\tt"} (LSXPS J041956.5--545528 is marked with a green circle). 
}
\label{image_1566}
\end{figure}

\begin{figure}
\centering
\includegraphics[scale=0.8,angle=0]{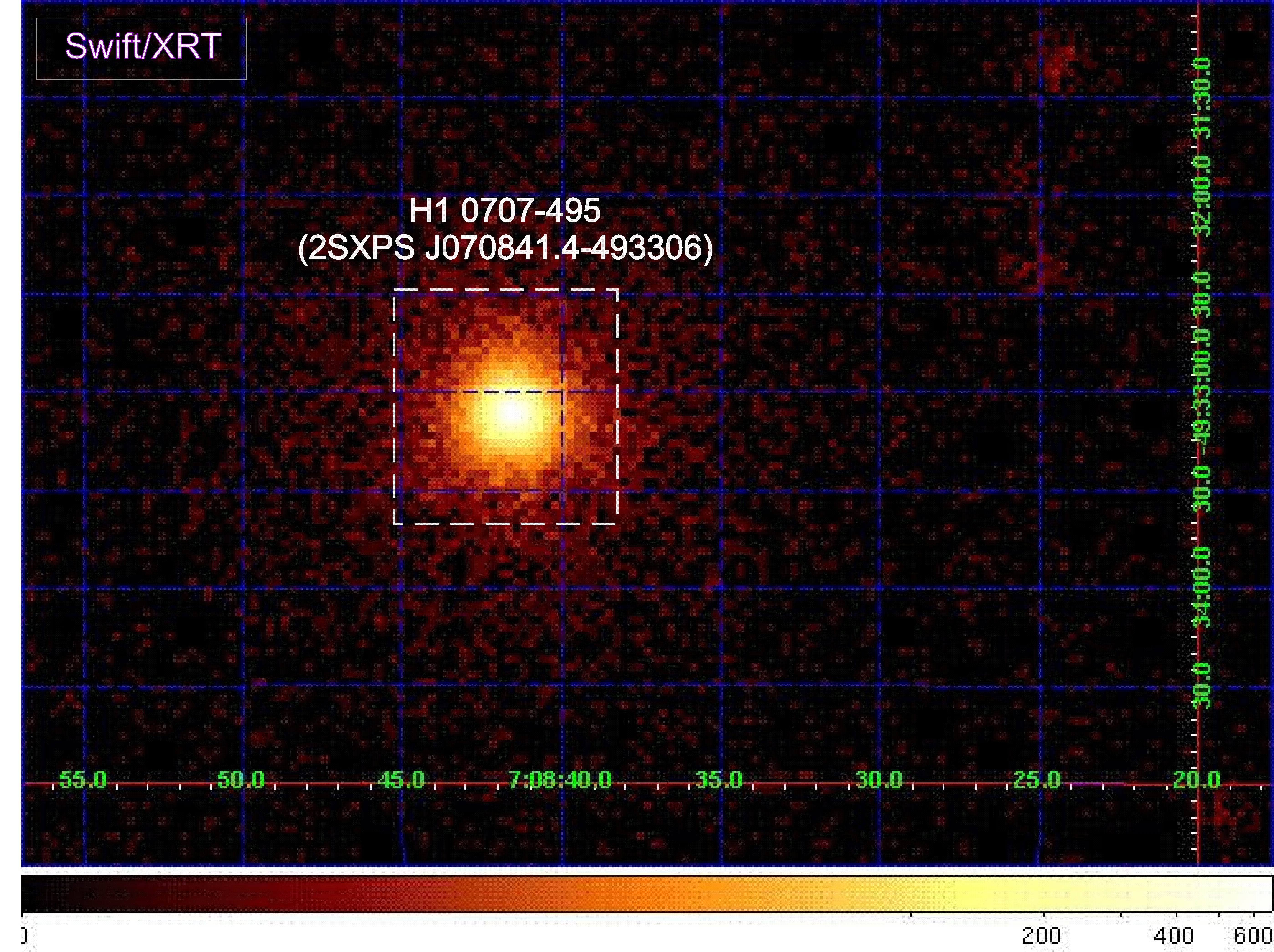}
\caption{
X-ray image of 1H~0707--495 (2SXPS J070841.4-493306 -- according to the {\it Swift} catalog), accumulated from April 3, 2010 to April 30, 2018 with a 160 ks exposure time. The  dashed square with a side of 6.3{\tt'} (160 pixels) in the SDSS~J0752 image demonstrates the absence of other nearby objects in the 1.3{\tt'} field of view. The next closest source is 166{\tt"} away (outside the image).}
\label{image_0707}
\end{figure}

\begin{figure}
\centering
\includegraphics[scale=0.8,angle=0]{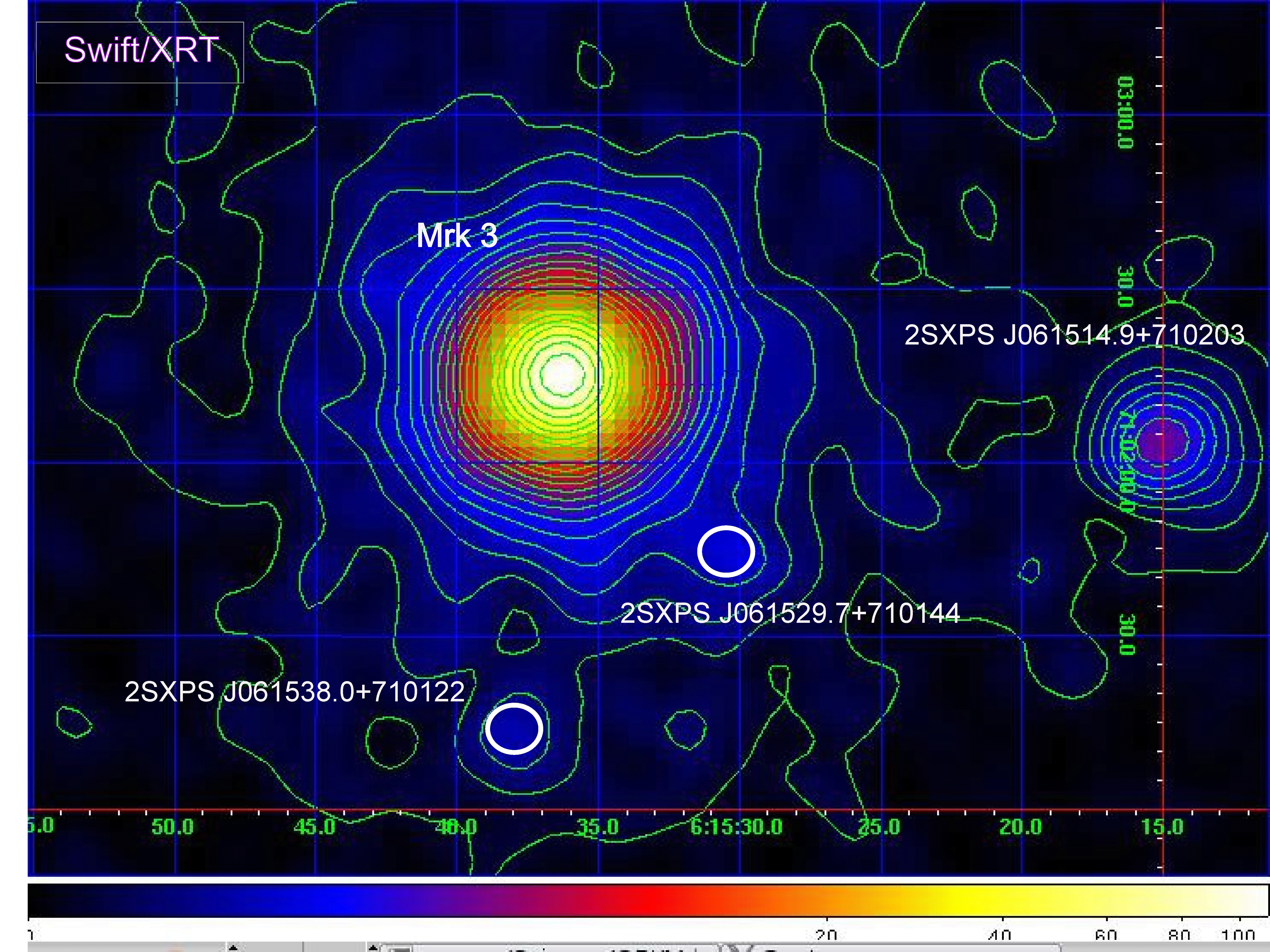}
\caption{
X-ray image of Mrk~3 (in center), accumulated by $Swift$/XRT from March 21, 2006 to April 20, 2015 with a 91 ks exposure time. The green contours in this image demonstrate the lack of X-ray jet-like structure around Mrk~3, as well as minimal contamination by other point sources within 18{\tt"} of the source environment. The closest next source is 21{\tt"} (LSXPS J041956.5--545528 is marked with a white circle). 
}
\label{image_Mrk3}
\end{figure}

\begin{figure*}
\centering
\includegraphics[scale=0.8,angle=0]{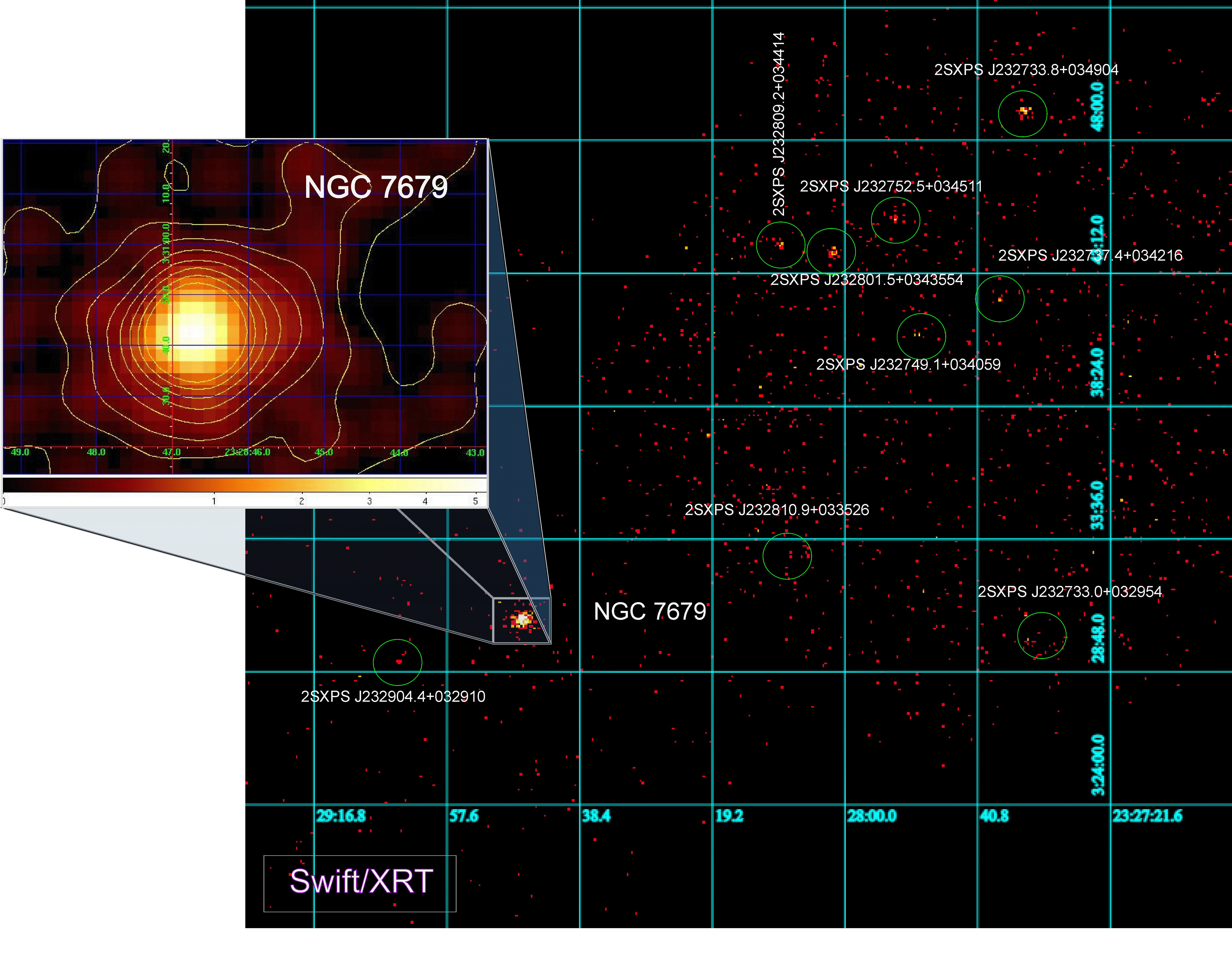}
\caption{
X-ray image of NGC~7679 (LSXPS J232846.7+033042 -- according to the {\it Swift} catalog), accumulated from July 25, 2015 to October 6, 2017 with a 1.7 ks exposure time. The image segment highlighted with a gray square with a side of 6.3{\tt'} (160 pixels) is also shown in more detail in the enlarged panel. Contour levels demonstrate the absence of X-ray jet (elongated) structure and minimal contamination from other point sources and diffuse radiation in the 1.3{\tt'} field of view around NGC~7679. The next closest source is 2LSXPS~J232904+032910, 279{\tt"} away.
}
\label{image_7679}
\end{figure*}

We present a long-term X-ray light curve of   NGC~1566 detected by the XRT on board  of {\it Swift} from 2007 -- 2023 (see Fig. \ref{ev_1566}). In addition, we used {\it Suzaku} data from this AGN and marked the time of its observations with a green arrow in the background of the Swift light curve (in  top panel). In the bottom panel of this figure, we have presented the optical light curve by Swift/UVW2 [1600--2230 $\AA$] (hazel stars) and by Swift/BAT [15--50 keV] (black dots) observations to trace the quiet/active states of NGC~1566 at different wavelengths. Figure~\ref{ev_1566} also 
shows that NGC~1566 was in the 
LHS state from 2007 to 2017. 

Here, along with the countrate light curve in total band [0.3--10 keV], it is also interesting to investigate the hardness count-rate light curves 
without using any model. Thus, we study the light curve of NGC~1566 with a bintime of 1 ks for four energy bands 0.3--10 keV, 0.3--1 keV, 1--2 keV and 3--10 keV. Using these energy-dependent light curves, the soft hardness coefficient ($HR1$) is defined as the ratio of the difference in count rates in the 1--2 keV ($M$) and 0.3--1 keV ($S$) energy bands to their sum, and the hardness coefficient ($HR2$) is defined as the ratio of the difference in count rates in the 2--10 keV ($H$) and 1--2 keV ($M$) energy bands to their sum (see recommendations in 2SXPS\footnote {https://heasarc.gsfc.nasa.gov/W3Browse/swift/swift2sxps.html}): $HR1 = (M-S)/(M+S)$ and 
$HR2 = (H-M)/(H+M)$. 
In this approach, NGC~1566 shows variability in terms of $HR1$ (blue points) and $HR2$ (pink points) during transient events from 2007 to 2023 (see second panel from top in Fig.~\ref{fraq_1566}). 
From 2017 to 2019, the object entered an active state with a powerful outburst, followed by decay, accompanied by a series of repeated re-flares of smaller amplitude. Comparison of the top and bottom panels of Fig.~\ref{fraq_1566} demonstrate that the source light curves in the optical (UVW2) and X-ray (Swift/XRT and Swift/BAT) ranges correlate quite well. Around this time 
a changing-look of the galaxy occurred. It can be seen that during the changing-look events (marked by vertical blue stripes, $F1-F4$) NGC~1566 was characterized by the dominance of the soft component, and only after the outburst peak, a softening of its spectrum occurred during repeated small flares (MJD 58,400--59,000, $F2-F3$) during the decay phase. So this interval is most interesting for subsequent spectral analysis (see Sect.~\ref{spectral analysis}). 

\subsubsection{1H~0707 Light Curve\label{lc_0707}}
The light curve of Sy1 galaxy 1H~0707 is shown in Fig.~\ref{ev_0707} before, during, and after the transient events from 2010 to 2018. It can be seen that in 2011 (MJD 55550--65599) 1H~0707 remained in the stable low/hard state (mean countrate $\sim 0.02$ cnt/s). During the rest of the {\it Swift} observations, the object is highly variable.

Overall, 1H~0707 shows four X-ray outbursts (marked with vertical blue strips) with good coverage of the rise-peak-decay in our sample (see Table~\ref{tab:par_Swift_data}). Furthermore, the object shows variability in terms of $HR1$ (blue dots) and $HR2$ (pink dots) during transient events with a clear dominance of source hard emission from 2010 to 2018 (Fig.~\ref{fraq_0707}).

\subsubsection{NGC~7679 Light Curve\label{lc_7679}}

The Sy2 galaxy NGC~7679 has a modest monitoring history and, according to optical observations (see Fig.~\ref{ev_7679} from Catalina Sky Survey (CSS, V-band, blue dots), the object is weakly variable, at least in the V band. In Fig.~\ref{ev_7679} we also show the time distribution of observations of NGC~7679 using ASCA (grey arrows), {\it Beppo}SAX (green arrow) and {\it Swift}/XRT (bright blue vertical strip with red points) in conjunction with the source optical light curve in V band. Apparently, {\it Swift}/XRT detected an X-ray flare from NGC~7679 around MJD 58033.

\subsubsection{Mrk~3 Light Curve\label{lc_mrk3}}
The light curve of Sy2 galaxy Mrk~3 is shown in Fig.~\ref{fraq_mrk3} before, during, and after the transient events from 2010 to 2018. From this figure we can see that in early 1996 (MJD 50450--50520) Mrk~3 remained in a stable low/hard state (average count rate $\sim 2-3$ cnt/s). Then two outbursts occurred around MJD 50540 and 50550, when the count rate increased to 5 cnt/s. During the rest of the {\it RXTE} observations  the object remained in a stable low/hard state.

Generally, Mrk~3 shows two X-ray outbursts (marked with vertical blue strips) in our sample (see Table~\ref{tab:table_Suzaku+ASCA_SAX}). Furthermore, the object shows variability in X-ray fluxes in 3--10 keV (blue dots) and 10--20 keV (pink dots) bands during transient events with a clear dominance of source hard emission from 1996 to 1997 (Fig.~\ref{fraq_0707}).

\subsection{Spectral Analysis \label{spectral analysis}}

\subsubsection{Model Selection \label{model selection}}
To model AGN spectra, special models such as {\tt AGNSED} \citep{Kubota+Done18}  and dual-coronal \citep{Petrucci13} are created. In fact, they take into account the specific geometry of AGNs with a breakdown into an outer standard disc, an inner warm Comptonization region  and a hot corona \citep{Kubota+Done18}. The last two components are called warm and hot corona (dual-coronal model by \cite{Petrucci13}), which are described by the thermal Comptonization with different plasma temperatures. Petrucci et al. used mainly the warm corona component to describe the soft X-ray excess in the AGN spectra in the low/hard state (see also the discussion in Sect.~\ref{discussion}). 

Titarchuk et al. (1994-2024)  in their papers indicate that the generalized Comptonization model called {\tt XSPEC BMC} [the {\it Bulk Motion Comptonization} is a XSPEC Comptonization by relativistic matter model\footnote{https://heasarc.gsfc.nasa.gov/xanadu/xspec/manual/node142.html}] can be applied to any state and Comptonization type (thermal or dynamic) for the observed X-ray spectra of BHs or NSs \citep{TS24,TSC23,TS23,TS21,TSCO20,TS17,TS16,tsei16b,STS14,tss10,ts09} . Thus,  we apply the {\tt BMC} model as a generalized one for all AGN sources in all spectral states.
The {\tt BMC} model calculates the soft excess and the primordial emission self-consistently. In this model, the total emission is determined by the {\tt BMC} normalization $N_{bmc}$, which is proportional to the mass accretion rate and the spectral index $\alpha$ (or the photon index $\Gamma=\alpha+1$). The disk emission appears as  a color  temperature-corrected blackbody emission at radii $R_{out} > r > R_{TL}$, where $R_{out}$ and $R_{TL}$ are the outer disk radius and the outer radius of the transition layer (TL), respectively. At $r<R_{TL}$, the disk emission appears as the Comptonized emission from the warm and optically thick medium, rather than as  the thermal one. The hot and optically thin TL is located in the inner part of the disk around the BH ($R_{TL}<r<R_{ISCO}$, where $R_{ISCO}$ is the radius of the last stable orbit) and creates a high-energy power-law continuum. The total Comptonized radiation is split between the (hot) TL and the (cold) BH-converging flux (CF, where dynamical Comptonization is effective), and the fraction of (hot) Comptonized radiation ($f$ or $logA$) can be found from a model fit. The seed photon temperature ($kT_s$) and the TL spectral index of the corona
determine the energy of the upward-scattered soft excess radiation. The Compton continuum is approximated as a convolution of the BB blackbody radiation with the Comptonization Green function $G$, which can be described as

\begin{eqnarray}
G(x,x_0) & = & x^{-\alpha} , ~~~\rm {if}~~x>x_0, \\
G(x,x_0) & = &  x^{\alpha+3}, ~~~\rm {if}~~ x<x_0, 
\end{eqnarray}
\noindent where $x = hv/kT_e$, $x_0$ is the breakpoint of $G(x,x_0)$ at the point where the two cases ($x>x_0$ and $x<x_0$) meet each other. Thus, the general model consists of a BBody-like and a Comptonized component (XSPEC models ``BMC", ``COMPTB", ``COMPTT" are the sum of these components: $BB+f\cdot BB*G$). When using the {\tt BMC} model, we included a Gaussian component to account for the Fe emission lines. The model reads in XSPEC as 
{\tt tbabs *
(BMC+Gaussian)}. 
For all sources we used the Comptonization model {\tt BMC} modified (see the description of the model in Fig. \ref{model}) by neutral absorption and the {\it Gaussian} line at $\sim$~6.5 keV.
The parameters of a  {\it Gaussian} component are  a centroid line energy $E_{line}$, the width of the line $\sigma_{line}$  and normalization, $N_{line}$ to fit the data in the 6 -- 8 keV  energy range.  We also use   interstellar absorption with a column density $N_H$  (see Table~\ref{tab:parameters_binaries}).

As for the inclusion of the iron line in the source spectrum fitting model for 1H~0707, some explanations should be given here. The first time the iron fluorescent line in 1H~0707--495  was detected by \cite{Fabian09} using  {\it XMM}-Newton. The previous observations showed only a sharp and deep drop in the spectrum at 7 keV, but did not reveal narrow emission features. This has given rise to two interpretations \citep{Boller02}: either the source is partially shielded by a large layer of iron-rich material (then the sharp drop is due to a photoelectric absorption edge), or it has very strong X-ray reflection \citep{Fabian04} in its innermost regions, where relativistic effects change the observed spectrum (the sharp drop in the spectrum by energy 7 keV is due to the blue wing of the line (see also \cite{sei99}), partly formed by relativistic Doppler shifts). 

Absorption requires the iron abundance to be about 30 times the solar value, while reflection requires this ratio to be between 5 and 10. \cite{Fabian09} showed that the extreme variability in the soft range of the 1H~0707 spectrum does not make sense in the model partial coverage. However, analysis of the spectral variability
of the source showed the spectrum of the source in all states (low and high flux) to be well described by a power-law continuum with a photon index $\Gamma = 3$, with an excess at lower energies ($\le$ 1.1 keV). 
\cite{Fabian09} argued that the H1~0707 spectra are well described by a simple phenomenological model consisting of a power-law continuum, a soft blackbody, two relativistically broad $Laor$ lines \citep{Laor91} and galactic absorption (corresponding to $N_H = 5 \times 10^{20} cm^{-2}$). These lines are characterized by energies of 0.89 and 6.41 keV (in the rest frame), an innermost radius of $1.3R_g$ ($R_g = GM_{BH}/c^2$, wherein  $G$ and $c$ as the common physical constants and $M_{BH}$ as a BH mass), an outer radius of $400R_g$, an emissivity of 4, and an inclination of 55.7 degrees. The rest energies of these lines correspond to ionized iron-L and K, respectively. However, \cite{Fabian09} found the iron abundance to be almost 9 times  of the solar value, which they associated with the possible influence of a dense nuclear star cluster in the vicinity of 1H~0707, which can lead to the formation of massive double white dwarfs that enriched the core with iron-rich supernova  emissions  \citep{Shara+Hurley02}.

In the present paper, we made an assumption that this shape of 
the ionized iron-K line in the 1H~0707 spectrum may be due to the outflowing wind, which, to a first approximation, can be described by an iron line with a {\it Laor} profile. The presence of outflowing gas from the nuclear environment of 1H~0707 is supported by recent near-infrared spectroscopy in the $zJHK$ bands, which revealed the dominance of broadband lines with low ionization. Namely, extensive components in H~I, Fe~II and O~I were found, shifted at a velocity of $\sim$500 km/s. At the same time, most lines have a blue-asymmetric profile, the velocity shift of which is $\sim$826 km/s. This feature is consistent with previous indications of escape gas in 1H~0707, observed in the X-ray and UV lines, and now found in the low ionization lines. \cite{Rodriguez-Ardila24} argue that the wind can propagate far into the region of the narrow line due to the observation of a blue-shifted component in the forbidden [S~III] $\lambda$9531 
$\AA$ line. \cite{Xu21} revealed absorption edges in the fits of the 1H~0707 spectra, which can also be interpreted as an evidence of a clumpy, multi-temperature outflow around 1H~0707. Therefore, to describe the 1H~0707 spectra, we used the {\tt tbabs*(bmc+N$\times$Laor)}. The $Laor$ model parameters are the line energy, $E_L$, the emissivity index, a dimensionless inner disk radius, $r_{in} = R_{in}/R_{g}$, inclination, $i$, and the normalization of the line, $N_L$ (in units of photons cm$^{-2}s^{-1}$).

For the {\it Laor} component we fixed the outer disk radius to the default value of 400 $R_g$ and   vary all other parameters.
We also fixed the emissivity index to 3. The inclination is constrained to a value $i \sim 50^{\circ}$.
As a result, in our spectral  data analysis for all three sources  we use a model which consists a  sum of a  Comptonization   component ({\it BMC}) and {\it Gaussian (Laor)} line component for NGC~1566 and NGC~7679 (1H~0707). 
Let us recall that the {\it BMC} spectral component has the following parameters:   the seed photon  temperature,  $T_s$, the energy index of the Comptonization spectrum $\alpha$ ($=\Gamma-1$),   
a Comptonization  fraction $f$ [$f=A/(1+A)$], which is the relative weight of the Comptonization component  and normalization of the seed photon spectrum, $N_{com}$. When the parameter  $\log(A) =>> 1$ we fix $\log(A) = 2$ because the Comptonized illumination fraction $f = A/(1 + A) \rightarrow1$ and variation of $A$ does not improve the fit quality any more.

\subsubsection{NGC~1566 Spectra \label{spectral analysis_1566}}

Spectral analysis of the NGC~1566 observations by {\it Suzaku} and {\it Swift} 
indicates  that  the source  spectra can be  reproduced by a model with an absorbed  Comptonization component, the XSPEC {\it BMC} model\footnote{https://heasarc.gsfc.nasa.gov/xanadu/xspec/manual/Models.html} with the adding {\it Gaussian} iron line component. 
In Figures \ref{4_spectra_ngc_1566} and \ref{spectrum_ev_7679_all} we present examples of X-ray spectra of NGC~1566 indicating the spectral components in the range 0.3--10 keV and their evolution. 

In the {LHS}, represented by the {\it Swift} observation (ID=00014916001), the spectrum of NGC~1566 is dominated by the hard emission component (Fig. \ref{4_spectra_ngc_1566}, panel $a$), which is well reproduced by the Comptonization model with parameters: $\Gamma$=1.40$\pm$0.02 and $T_s$=110$\pm$12 eV 
(reduced $\chi^2$=0.95 for 563 d.o.f). With better energy resolution in the LHS state, the spectrum of NGC~1566 with the {\it Suzaku} observation  is also well described by the Comptonization component, but with an addition of a noticeable emission feature in the energy range $\sim$6.4--7.0 keV associated with neutral, H-like and He-like K$_{\alpha}$ Fe lines (panel $b$, ID=707002010). The best-fit model parameters are $\Gamma$=1.77$\pm$0.04, $T_s$=130$\pm$1 eV and $E_{line}=6.36\pm$0.04 keV (reduced $\chi^2$=0.95 for 563 d.o.f). In the {\it intermediate} state  of NGC~1566 (ID=00014923002, panel $c$) the best-fit model parameters are $\Gamma$=2.01$\pm$0.04 and $T_s$=118$\pm$5 eV (reduced $\chi^2$=0.95 for 86 d.o.f). In the panel $d$ we present again the IS  spectrum (ID=00035880003) in the HSS, for which the best-fit model parameters  are $\Gamma$=2.1$\pm$0.1, $T_s$=90$\pm$10 eV 
(reduced $\chi^2$=0.99 for 86 d.o.f). The data  are presented by black crosses and the best-fit spectral  model   {\it tbabs*(BMC+Gauss)} by red line.  Bottom: $\Delta \chi$ vs photon energy in keV  (see more details in Table~\ref{tab:fit_table_suz+asca_0707+1655+7679}). In the bottom panels we present $\Delta \chi$ versus photon energy in keV. 

\begin{figure}
\centering
\includegraphics[scale=0.8,angle=0]{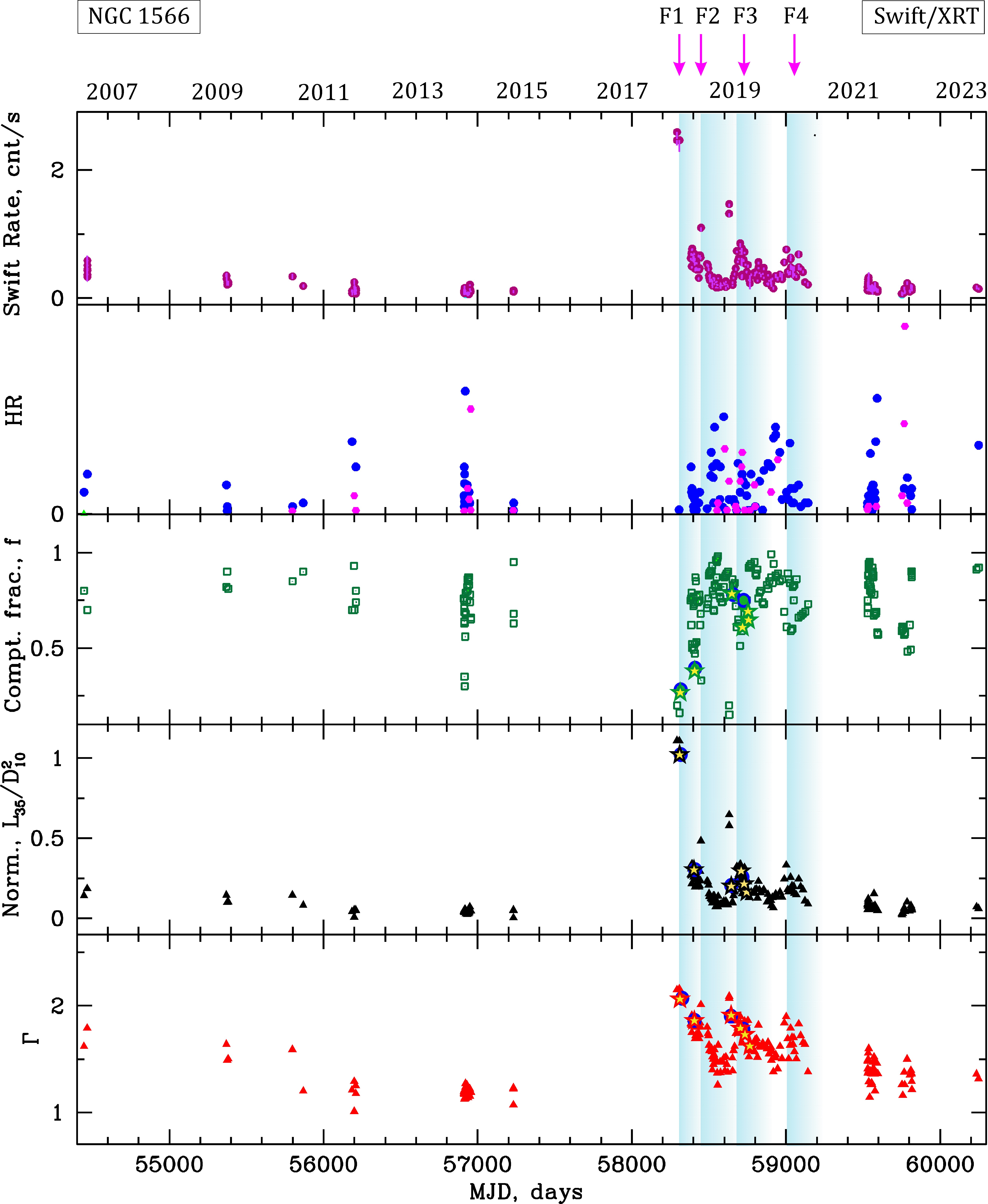}
\caption{
From top to bottom: evolution of the Swift/XRT count rate, hardness ratios HR1 and HR2 (blue and crimson points, respectively), Comptonized fraction $f$, and {\tt BMC} normalization during 2007 and 2023 flare transition of NGC~1566. In the last bottom panel, we present an evolution of the photon index $\Gamma = \alpha +1$. The decay phases of the flares are marked with blue vertical strips. The peak outburst times are indicated by the arrows at the top of the plot. For three bottom panels {\it NuSTAR} and {\it XMM-Newton} observations are indicated with stars and circle points, respectively.
}
\label{fraq_1566}
\end{figure}

\begin{figure}
\centering
\includegraphics[scale=0.9,angle=0]{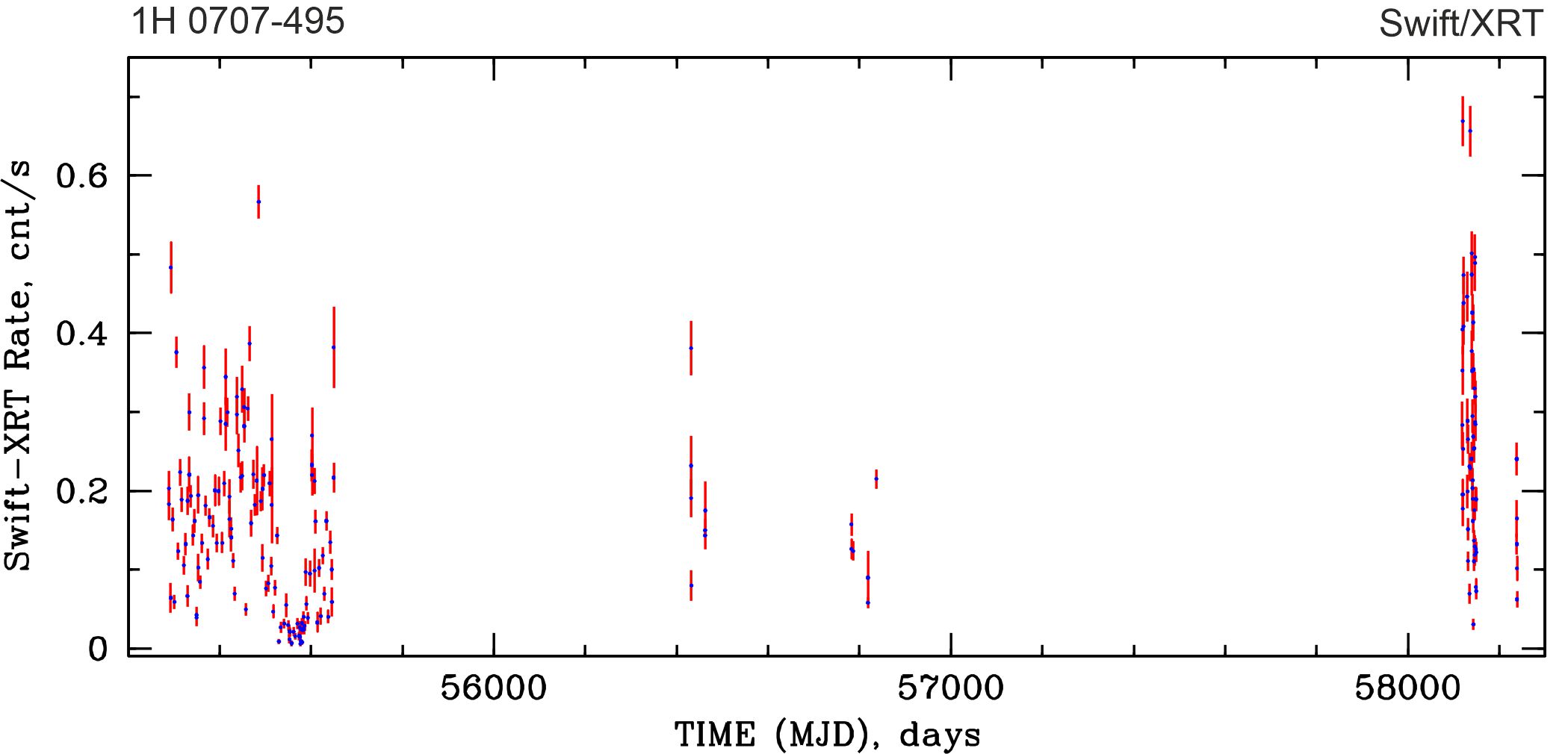}
\caption{Evolution of the Swift/XRT count rate during  1996 -- 2010 observations of 1H~0707--495. }
\label{ev_0707}
\end{figure}

\begin{figure*}
\centering
 \includegraphics[width=15cm]{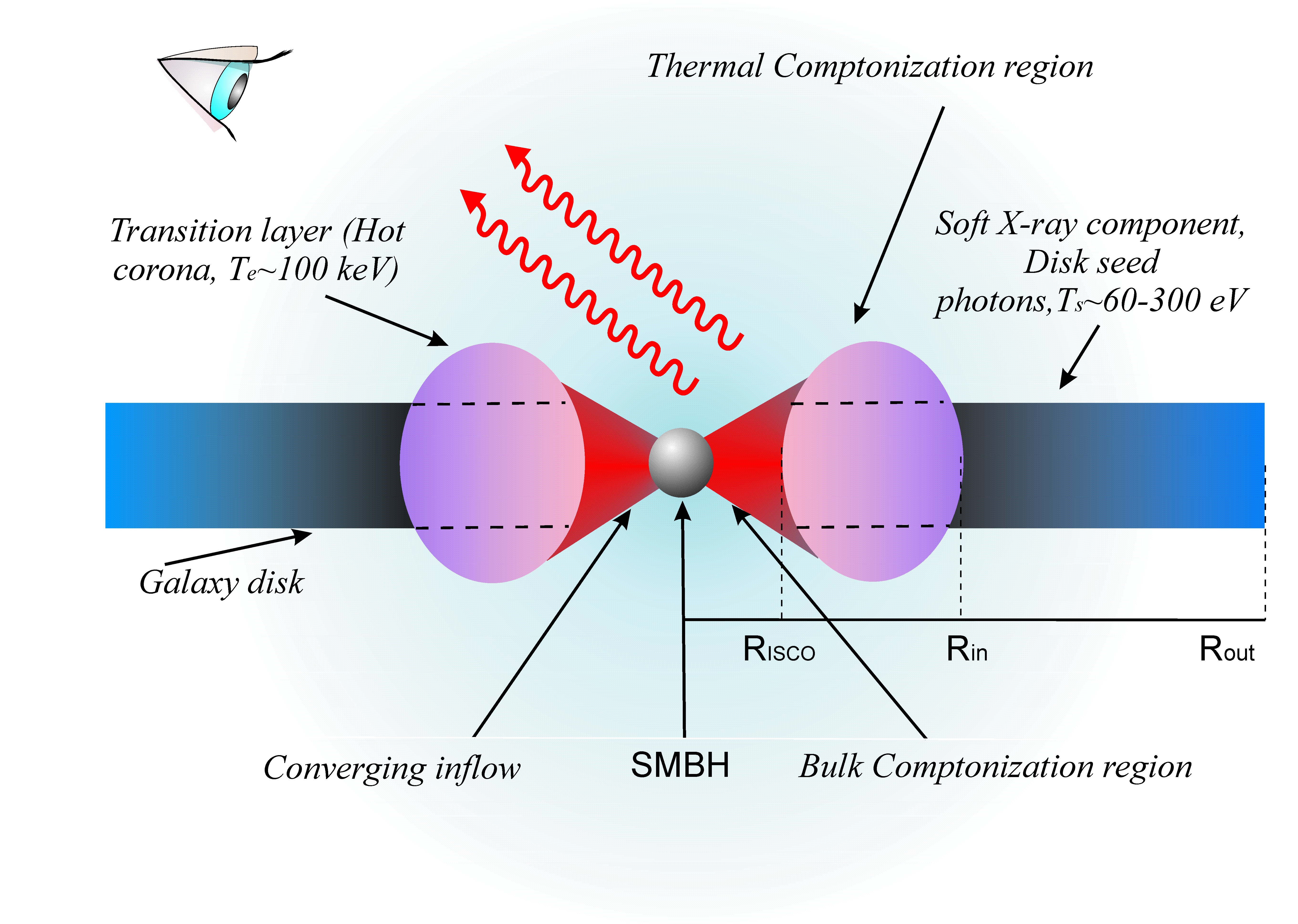}
\caption{A suggested geometry for NGC~1566, 1H~0707, Mrk~3 and NGC~7679 sources. Disk soft photons are upscattered (Comptonized) off relatively hot plasma of the transition layer.
}
\label{model}
\end{figure*}

\begin{figure}
\centering
\includegraphics[scale=0.9,angle=0]{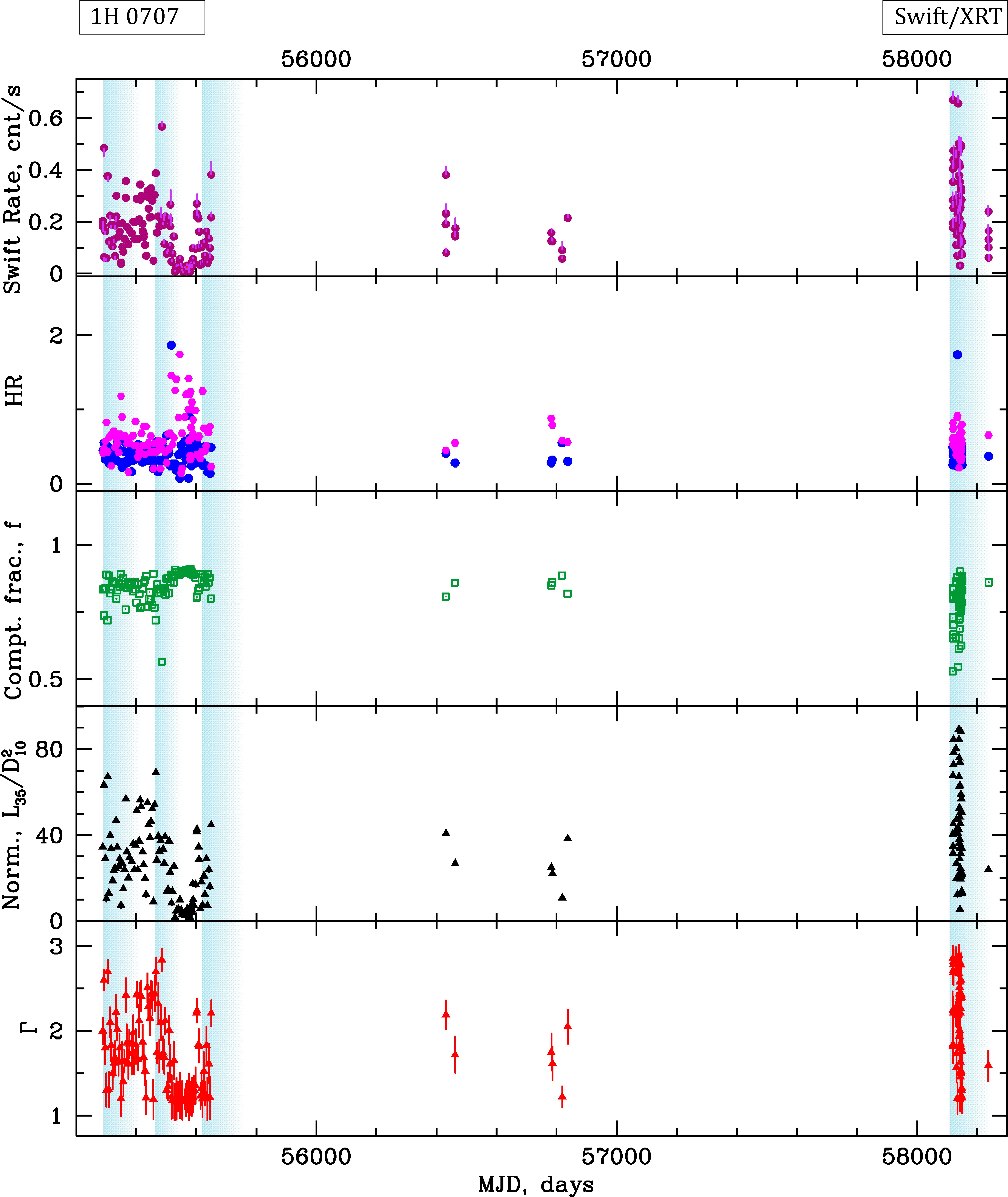}
\caption{ From top to bottom: evolution of the Swift/XRT count rate, hardness ratios HR1 and HR2 (blue and crimson points, respectively), Comptonized fraction $f$, and {\tt BMC} normalization during 2010--2018 flare evatsn of 1H~0707. In the last bottom panel, we present an evolution of the photon index 
$\Gamma = \alpha +1$. The decay phases of the flares are marked with blue vertical strips. 
}
\label{fraq_0707}
\end{figure}

\begin{figure*}
\centering
\includegraphics[scale=0.85,angle=0]{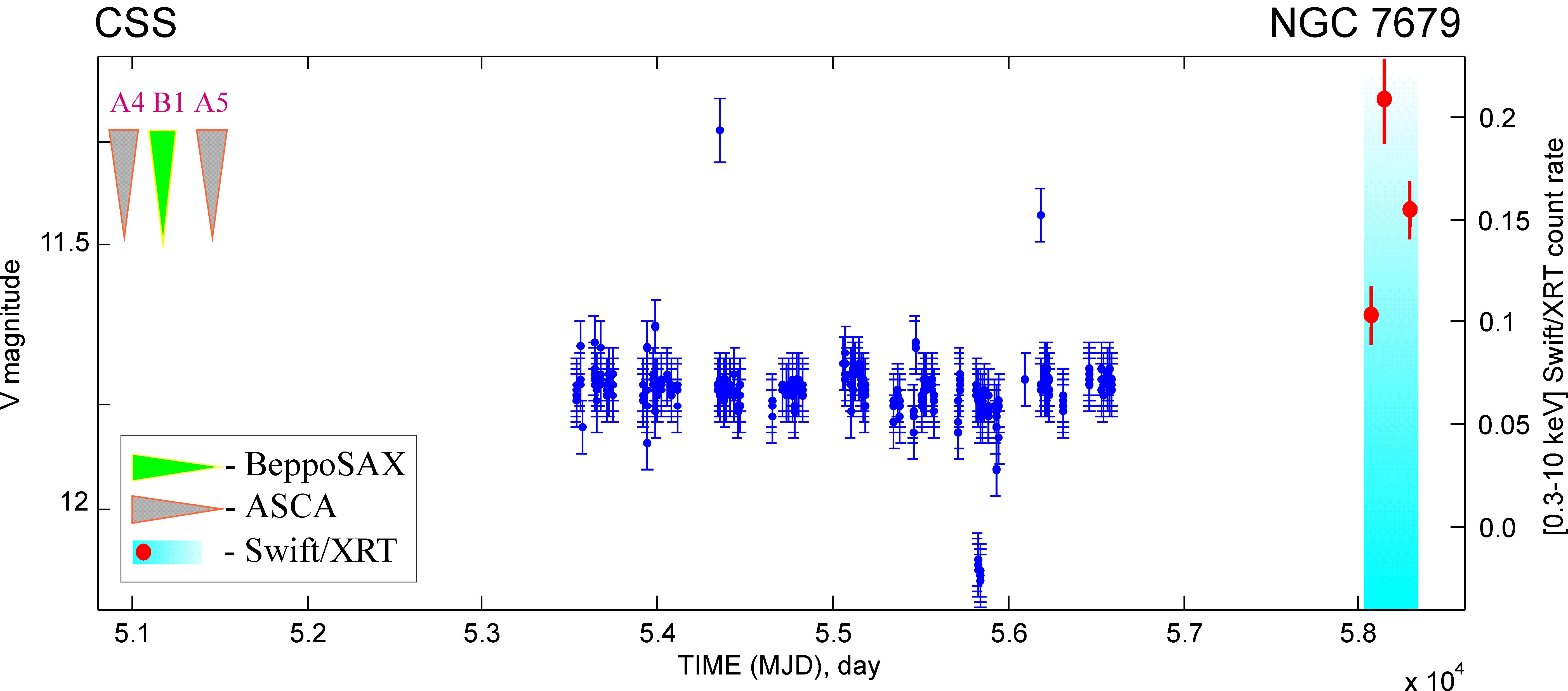}
\caption{The distribution of NGC~7679 observations by {\it Swift}/XRT (bright blue vertical strip with red points), {\it Beppo}SAX (green arrow), and ASCA (grey arrows) is shown along with the CSS V-band light curve (blue points) , 
(see Tables~\ref{tab:table_Suzaku+ASCA_SAX} and \ref{tab:par_Swift_data}).
}
\label{ev_7679}
\end{figure*}

\begin{figure}
\centering
\includegraphics[scale=0.9,angle=0]{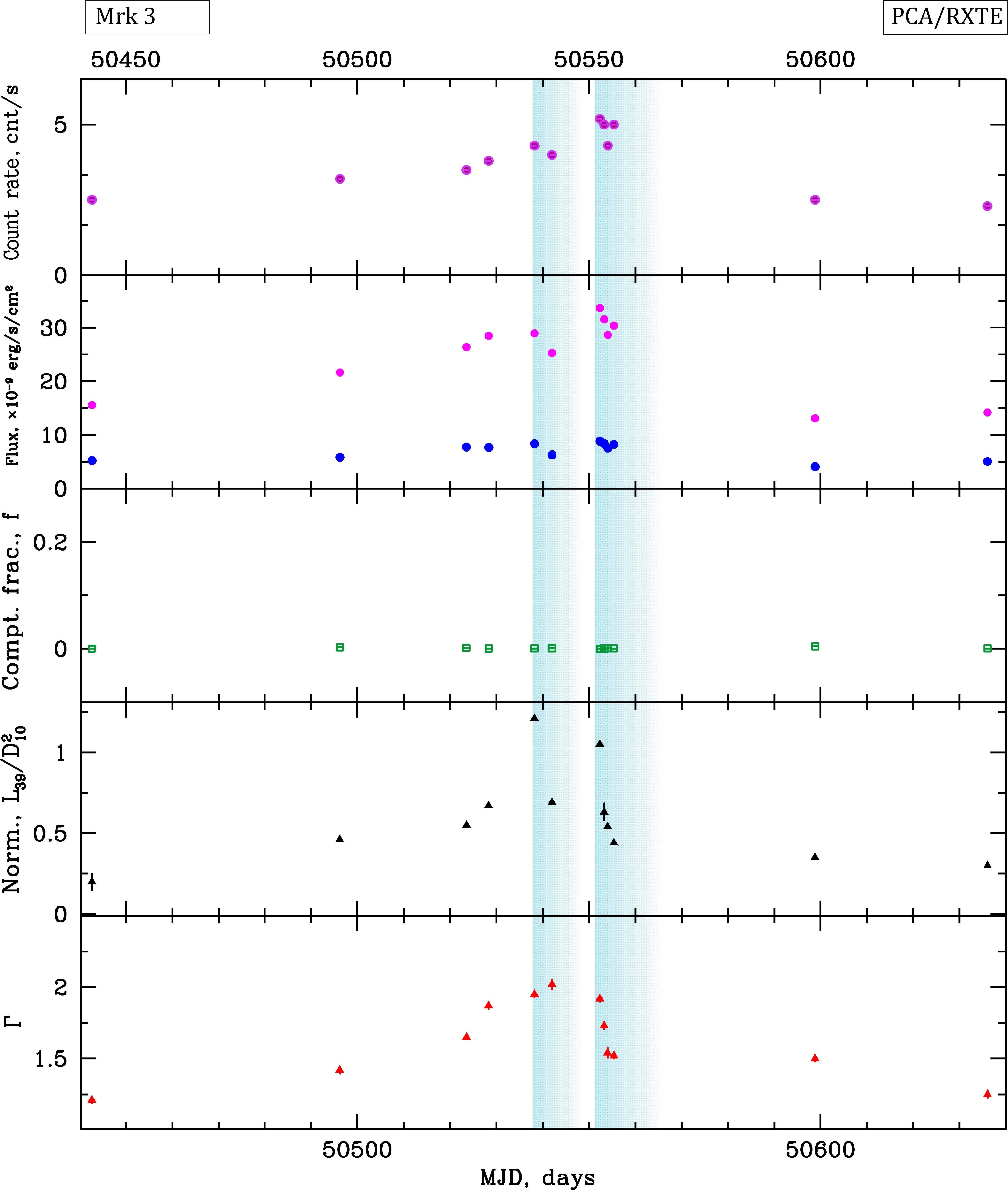}
\caption{ From top to bottom: evolution of the {\it RXTE}/PCA count rate, fluxes in 3--10 keV (blue points) and 10--20 keV (crimson points) bands, Comptonized fraction $f$, and {\tt BMC} normalization during 1997--1997 flare events of Mrk~3. In the last bottom panel, we present an evolution of the photon index $\Gamma = \alpha +1$. The decay phases of the flares are marked with blue vertical strips. 
}
\label{fraq_mrk3}
\end{figure}

\begin{figure*}
\centering
 \includegraphics[width=16cm]{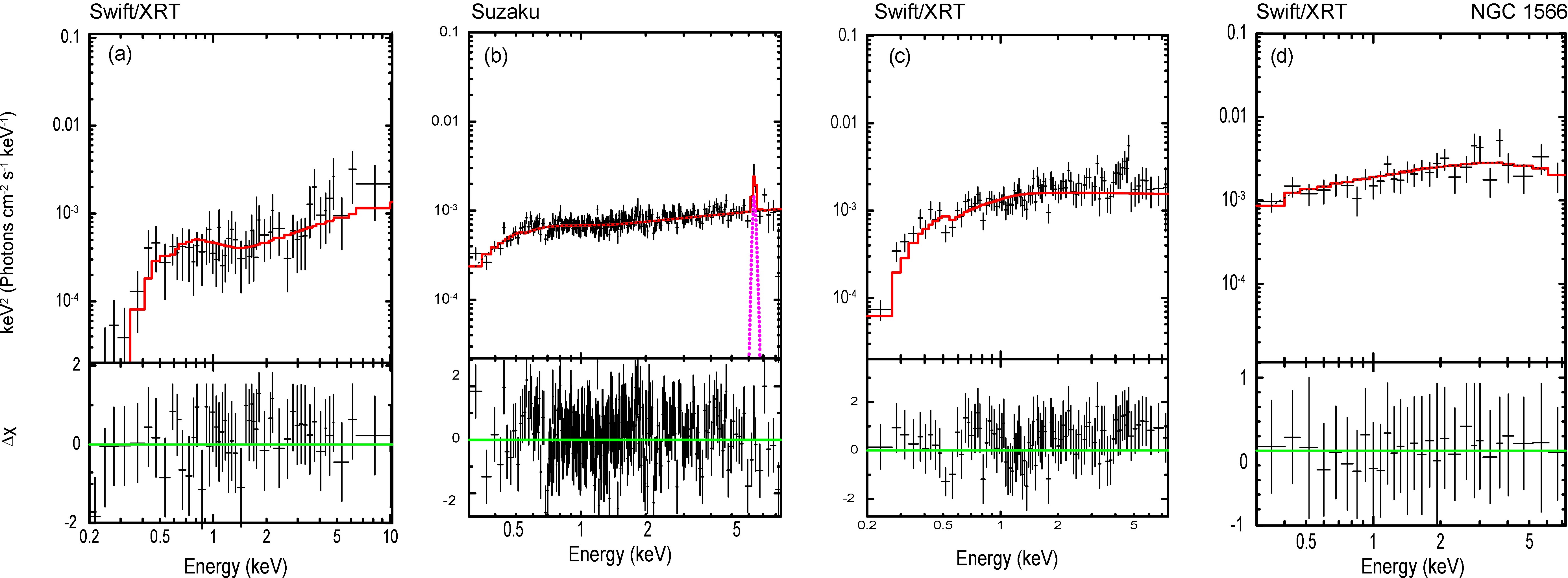}
\caption{Four representative spectra of NGC~1566 from {\it Swift} data in units of $E*F(E)$ with the best-fit modeling for the LHS (ID=00014916001, panel $a$), IS (ID=707002010, 
IS (ID=00014923002, panel $c$) and IS (ID=00035880003, panel $d$) states. The spectrum of NGC~1566 from {\it Suzanne} with the best-fit modeling for the LHS (ID=707002010, panel (b)). The data are denoted by  crosses, while the spectral model is shown by a red histograms for each state.  Bottom: $\Delta \chi$ vs photon energy in keV. 
}
\label{4_spectra_ngc_1566}
\end{figure*}

We clearly see that the change in spectral state from LHS to HSS in NGC~1566 is accompanied by a slight change in the seed photon temperature $T_s$ between 90 and 130~eV and an increase in the $\Gamma$ index from 1.1 to 2.1 (Fig.~\ref{scaling_1566}). 
In details, a number of X-ray spectral transitions of NGC~1566 have been detected by {\it Swift} during 2007--2023. We have searched for common spectral and timing features that can be revealed during these spectral transition events. The X-ray light curve of NGC~1566 shows complex behavior in a wide range of timescales: from hours to years (e.g., \cite{Oknyansky20,Oknyansky19}). Here we discuss the source variability on the timescales of hours. In Figure \ref{fraq_1566} we demonstrate the model parameter evolution for all analyzed outburst spectral transitions. As one can see from the $HR$ panel (second from the top), all outbursts of NGC~1566 are characterized by a significant increase of the soft component ($HR1$, blue points). Some events also demonstrate an increase of hard component ($HR2$, pink points) in the decay outburst phase (see particular events at MJD 57,000 and 58,400--59,000).

We paid special attention to the flaring events of 2018 -- 2019 (see Fig.~\ref{fraq_1566}). Following the terminology of the astronomical community, which previously discussed the changing-look in NGC~1566, we have identified four events $F1$, $F2$, $F3$ and $F4$ that occurred after the main flare of 2018 (it was not observed using {\it Swift}), indicated in Fig.~\ref{fraq_1566} with using the pink arrows on top of the figure. Event $F1$ occurred in December 2018 \citep{Grupe18}, event $F2$ occurred at the end of May 2019 \citep{Grupe19}, event $F3$ occurred in August 2019 \citep{Oknyansky20} and event $F4$ in May 2020 \citep{Jana21}.

The 2018 outburst itself began in March 2018 (58,200 MJD) according to the V-band ASSAS-SN and {\it Master} data\citep{Oknyansky19} data. {\it Swift}/XRT detected this outburst just a few months later, in June 24, 2018 (MJD 58,293.7), when the X-ray intensity increased to 30 times its quiescent state (see Figure~\ref{fraq_1566}). 
For this outburst with a good peak-decay coverage, the enhancement of the soft component (pink points in secondpanel from the top) during the decay phase (Fig.~\ref{fraq_1566}, MJD 58,452 -- 58,749) correlates with the disappearance of broad lines and [Fe~X] $\lambda$ 6374 $\AA$ to 58,561 in optical spectrum of NGC~1566 \citep{Oknyansky20}, which may be due to an increase in the illumination of the accretion disk by soft X-rays. We can also suggest that the  NGC~1566 spectral state evolution can be traced by the illumination fraction $f$. In fact, the parameter $f$ increased greatly during  these days ($0.7<f<1$). The photon index $\Gamma$ is well traced by soft X-ray flux (compare the bottom and two upper panels of Figure \ref{fraq_1566}). In addition, this may be related to the moderate mass accretion rate regime. In fact, the BMC Normalization parameter ($N_{com}$) and photon index ($\Gamma$) for these dates are some lower (MJD 58,452--58,749) than for events at the peak of the 2018 outburst (MJD 58,293.

\subsubsection{1H~0707--495 Spectra\label{spectral analysis_0707}}

\begin{figure*}
\centering
\includegraphics[scale=0.95,angle=0]{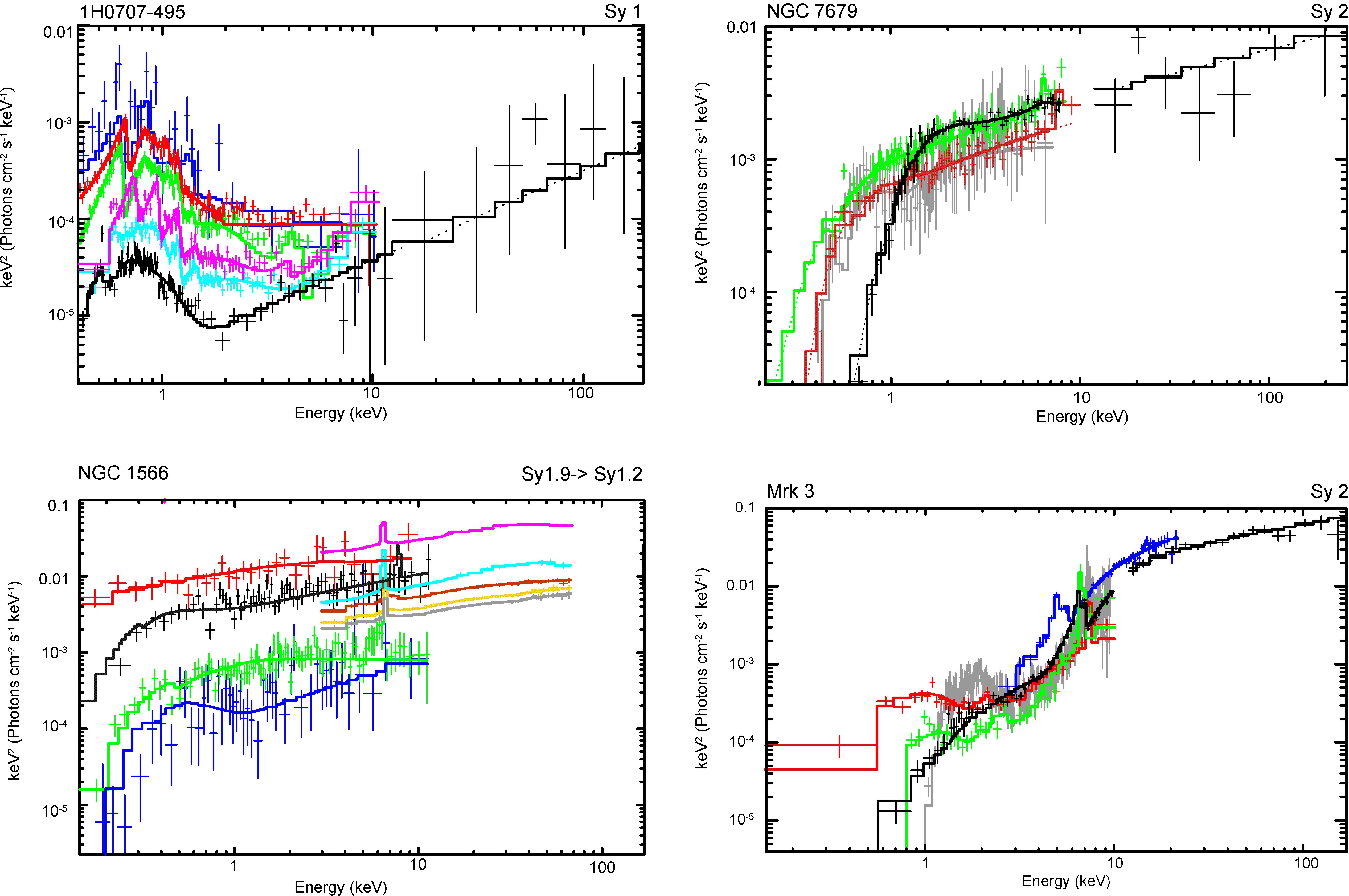}
\caption{Representative $E*F(E)$ spectral diagrams that are related to different spectral states for 1H~0707--495 (top left) using Suzaku observation 00091623 (black, LHS) combined with {\it RXTE}/HEXTE observation 20309010100 (black, LHS), {\it ASCA} observations 73043000 (brigh blue, IS), 
73043000 (pink, IS), 763100 (green, IS), Swift observation 0090393 (red, IS) and Swift obs 0091623 (blue, HSS);  NGC~1566 (bottom left) using {\it Swift} observations 0001496 (blue, LHS), 00014923 (green, LHS), 00031742 (black, IS), 00035880 (red, IS), and {\it NuSTAR} observations 80301601002 (pink, IS), 80401601002 (bright blue, IS), 80502606002 (brown, IS), 60501031004 (yellow, IS) and 60501031006 (grey, IS);  NGC~7679 (top right) using 40631001 (black, LHS, from {\it BeppoSAX}), 00088108002 (grey, LHS, from {\it Swift}), 66019010 (red, HIMS, from {\it ASCA}), and 66019000 (green, IS, from {\it ASCA}) and Mrk~3 (bottom right) using {\it ASCA} observation 70002000 (LHS, red), 70002000 (LHS, red), Suzaku observation 709022010 (LHS, green),  {\it RXTE} observation 20330-01-09-00 (blue, IS), BeppoSAX observation 50132002 (IS, black).
}
\label{spectrum_ev_7679_all}
\end{figure*}

Spectral analysis of 
1H~0707 observations by {\it Suzaku}, ASCA, {\it Swift} and  {\it RXTE}   
indicates  that source  spectra can be  reproduced by a model with a  Comptonization component, the XSPEC {\it BMC} model. 
In Figures \ref{spectrum_ev_7679_all}  and \ref{Swift_spectra_0707} we present examples of X-ray spectra of 1H~0707 indicating the spectral components in the range 0.3--10 keV and their evolution. The BMC model provides a description of the X-ray emission from the source, which is processed by the accretion disk and/or the surrounding wind environment, with the disk producing the observed K$_{\alpha}$ iron fluorescence lines and hump in the continuum.  It is clearly seen that the energy range $\sim$6.4--7.0 keV is also modified by other lines associated with neutral, H-like and He-like K$_{\alpha}$ Fe lines. We found a number of positive excesses in the spectrum and added a number of additional lines to better describe the spectrum. Using this approach, we found lines at energies of 6.4, 6.8, 2.9, 1.02, 0.85, 0.65 and 0.5 keV, which are easily associated with Fe~I--XXII, Fe~ XXV/Fe~XXVI K$_{\alpha}$, lines S~XVI, Ne~X, Fe~XVII, O~III and N~XVII, respectively.

On panel ($a$) of Figure~\ref{Swift_spectra_0707} we demonstrate  the best-fit {\it Suzaku} spectrum of 1H~0707 which is a typical for  the LHS using  our  model for observation ID=00091623 carried out on 3--6 December, 2005.  The best-fit model parameters are $\Gamma$=1.79$\pm$0.13, $T_s$=121$\pm$4 eV and $E_{line}$=0.83$\pm$0.02 keV (reduced $\chi^2$=1.00 for 802 d.o.f). In panels ($b$) and ($c$) of this figure, we have presented the source spectra for the IS 
using  the Swift observation (ID=00091623002 and observation ASCA ID=73043000). The best-fit model parameters for ID=00091623002 are $\Gamma$=2.00$\pm$0.10, $T_s$=120$\pm$5 eV and $E_{line}$=0.85$\pm$0.08 keV (reduced $\chi^2$=1.06 for 189 d.o.f). The best-fit model parameters for ID=73043000 are $\Gamma$=2.03$\pm$0.09, $T_s$=107$\pm$4 eV and $E_{line}$=0.83$\pm$0.09 keV (reduced $\chi^2$=1.01 for 86 d.o.f). 

Finally, in the panel ($d$) we showed the spectrum of 1H~0707 in the HSS using observation Swift ID=00080720048, for which the best-fit model parameters  are $\Gamma$=2.9$\pm$0.1, $T_s$=230$\pm$10 eV and $E_{line}$=0.86$\pm$0.09 keV (reduced $\chi^2$=0.97 for 158 d.o.f). The data  are presented by black crosses and the best-fit spectral  model   {\it tbabs*(BMC+N*Laor)} by red line.  Bottom: $\Delta \chi$ vs photon energy in keV  (see more details in Table~\ref{tab:fit_table_suz+asca_0707+1655+7679}). 
Thus, we clearly see that the change in spectral state from the LHS to the HSS in 1H~0707 is accompanied by a slight change of the seed photon temperature $T_s$ between 100 and 230~eV and an increase in the $\Gamma$ index from 1.1 to 2.9 (Fig.~\ref{scaling_0707}). For the LHS, we combine the  Suzaku and {\it  RXTE} observations (see Table \ref{tab:table_Suzaku+ASCA_SAX}) to demonstrate that the source spectrum varies over a wide energy range from 0.3 to 200 keV (Fig.~\ref{spectrum_ev_7679_all}). In this case, the dominance of the hard (10--100 keV) emission of 1H~0707 with a slight excess of the soft component at 0.3--2 keV is obvious.

\subsubsection{NGC~7679 Spectra\label{spectral analysis_7679}}
In Figure~\ref{3_spectra_ngc_7679} we show three representative spectra of NGC~7679 for the LHS, HIMS and IS. We again simulate the spectra using the plasma Comptonization process (BMC) with an addition of a {\it Gaussian} iron line component. The best-fit modeling of the NGC~7679 spectrum in the LHS is presented on the left panel
using from {\it Beppo}SAX data (ID=40631001) in units of 
$E*F( E)$. The broadband spectrum [0.3--200 keV] demonstrates a hard tail and has the best-fit parameters: $\Gamma$=1.60$\pm$0.02, $T_s$=350$\pm$60 eV and $E_{line}$=6.4$\pm$0.5 keV (reduced $\chi^2$=1.04 for 82 d.o.f). In central panel of this Figure, we presented the source spectra which are typical  for  {\it hard intermediate state} (HIMS) using ASCA observation  (ID=66019000). The best-fit parameters are  $\Gamma$=1.62$\pm$0.02, $T_s$=150$\pm$70 eV and $E_{line}$=6.5$\pm$0.1 keV (reduced $\chi^2$=1.09 for 145 d.o.f). Finally, on  the right panel we present the IS spectrum (ID=00088108002) of NGC~7679 observed by $Swift$/XRT. 
The best-fit model parameters are $\Gamma$=1.93$\pm$0.07 and $T_s$=226$\pm$9 eV 
(reduced $\chi^2$=0.97 for 194 d.o.f).  The data are denoted by  crosses, while {\it red} and {\it pink} lines stand for the {\it BMC} and {\it Gaussian} components, respectively. Bottom panels demonstrate $\Delta \chi$ versus photon energy in keV. 
\subsubsection{Mrk~3 Spectra\label{spectral analysis_mrk3}}

In Figure~\ref{4_spectra_ark_3}
 we show four representative spectra of Mrk~3 
for the LHS and IS states. We simulate these spectra using the BMC model with an addition of a {\it Gaussian} iron line component. The best-fit model of the Mrk~3 spectrum of the LHS is presented in  panel ($a$)  
 using from {\it ASCA} data (ID=70002000) in units of $E*F( E)$. The best-fit parameters are  $\Gamma$=1.102$\pm$0.02, $T_s$=180$\pm$10 eV and $E_{line}$=6.41$\pm$0.09 keV (reduced $\chi^2$=1.05 for 52 d.o.f). In panels ($b-d$) of the Figure, we present the source spectra which are typical  for  the IS  using {\it Suzaku}   (ID=100040010, panel $b$), {\it RXTE} (ID=20330-01-09-00, panel $c$) and {\it Beppo}SAX (ID=50132002, panel $d$) observations. The best-fit parameters for this {\it Suzaku} observation  are   $\Gamma$=1.10$\pm$0.02, $T_s$=100$\pm$10 eV and $E_{line}$=6.41$\pm$0.05 keV (reduced $\chi^2$=0.94 for 629 d.o.f). The best-fit parameters for the {\it RXTE} observation  are   $\Gamma$=1.11$\pm$0.01, $T_s$=300$\pm$20 eV and $E_{line}$=6.38$\pm$0.08 keV (reduced $\chi^2$=0.86 for 47 d.o.f). The best-fit model parameters for {\it Beppo}SAX spectrum are $\Gamma$=1.55$\pm$0.04, $T_s$=230$\pm$30 eV and $E_{line}$=6.40$\pm$0.08 keV (reduced $\chi^2$=0.90 for 84 d.o.f).   The data are denoted by  crosses, while 
{\it red} and {\it pink} lines stand for the {\it BMC} and {\it Gaussian} components, respectively.  Bottom panels demonstrate $\Delta \chi$ versus photon energy in keV. 

\subsection{Comparative analysis of X-ray patterns of CL-, Sy1- and Sy2-AGNs during transient events 
\label{evolution_all}}

Spectral analysis of the X-ray spectra for CL-, Sy1- and Sy2-AGN showed that the emission of all these AGNs can be reproduced by the absorption model and  the Comptonization  one applying in addition to linear characteristics. However, the Comptonization hump parameters and emission features are very different for Sy1 and Sy2. CL-AGN is a combination of  Sy1 and Sy2 properties depending on the Eddington-normed X-ray luminosity level of CL-AGN. The Comptonization hump, characterized  primarily by the $\Gamma$ index, shows different shapes and energy positions in the spectrum. To make this more convincing, we have added the results found in the literature [e.g. \cite{Weng20}, \cite{Hernandez15}]. As a result, for Sy1 $\Gamma$ varies in moderate limits (from 1.1 to 2.9), while for Sy2 $\Gamma$ varies in a wider range: from 1.6 to 2.9. For CL $\Gamma$ varies in rather narrow limits: from 1.3 to 1.9. Taking into account the different masses of the objects (see Fig.~\ref{gam_LEdd}), we compare them and their X-ray luminosities normalized to the Eddington luminosity $L_x/L_{Edd}$ for 1H~0707, NGC~1566, NGC~7679 and Mrk~3 (all indicated by black arrows in Fig.~\ref{gam_LEdd}).
 In this Figure 
 we can see the different positions of Sy1 (blue squares), Sy2 (red stars) and CL-AGN (marked by black background circles) 
 on the $L_x/L_{Edd}$ diagram for NGC~1566 (in the purple box), 1H~0707 (in the pink box) and NGC~7679 and Mrk3 (in the green box). 
 There  the grey arrow shows the critical value of $3.5\times10^{-4}L_x/L_{Edd}$ which separates Sy1 from Sy2. In addition to this obvious difference in the relative X-ray luminosities of Sy1 and Sy2, their difference in photon index is immediately visible. Sy2 shows a wide $\Gamma$ range, while Sy1 is somewhat smaller (tending to high $\Gamma$) and CL-AGN is very narrow. At the same time, an interesting property of CL-AGN is revealed in contrast. Namely, with a small change in $\Gamma$ they reach the relative X-ray luminosities of both Sy1 and Sy2. Apparently, CL-AGN is capable of making such a jump in $L_x/L_{Edd}$ that at low relative luminosities it looks like Sy2 (in the green box region), and at relatively high luminosities (in the pink box region). 

It is  worth noting  that Sy1 and Sy2  AGNs are not differentiated by a BH mass ($2\times10^5-5\times10^8$) in our AGN sample (Fig.~\ref{gam_mass}), but show different tracks on the $\Gamma-M_{BH}$ diagram: Sy1 (blue squares) tend to $\Gamma\sim1.5-2$, while Sy2 (red stars) have a larger spread in $\Gamma$ (1.6--2.8).

It is worth noting  that the emission from NGC~1566 and NGC~7679 is strongly subject to reprocessing  in the inner parts of the disk, the Compton cloud, for which $0.3<f<1$ and  thus, only some fraction of disk emission component ($1-f$) is directly seen by the Earth observer. While a degree of irradiation of 1H~0707 remains low all the time ($0.05<f<0.08$). It is interesting that the temperature of the disk seed photons, $kT_s$ is almost the same for all three objects of different subclasses and varies slightly from 100 to 300 eV (see Table \ref{tab:fit_table_suz+asca_0707+1655+7679}).

We also found that during the transition from Sy2 to Sy1, the NGC~1566 spectrum shows an increase in the soft excess (Fig.~\ref{spectrum_ev_7679_all}), accompanied by a decrease in the illumination degree ($0.1<f<0.5$), which can be associated with the entry into the Sy1 state during the outburst (Fig.~\ref{fraq_1566}), so thereby blurring the line between the two subclasses. Our results strongly suggest that the large diversity in  the behavior observed among CL, Sy1 and Sy2 AGNs with different Eddington-normed X-ray luminosities can be explained by changes in a single variable parameter, namely the mass accretion rate, without any need for additional differences in Sy AGN parameters or its inclination.

\subsection{A BH mass estimate}
\label{mass_estimate}

We used a scaling technique to estimate a BH mass $M_{BH}$, previously developed specifically for BH weighing~\cite{TS24,TS23,TSC23,STU18,SCT18,STV17}. 
We  used the $\Gamma-N$ correlation to estimate the mass of BHs (for details see \cite{st09}, ST09). This method ultimately ({\it i}) identifies a pair of BHs for which $\Gamma$ correlates with an {increasing} normalization of $N$ (which is proportional to a mass accretion rate $\dot M$ and a BH mass $M$, see ST09, Eq.~(7)) and for which the saturation levels $\Gamma_{sat},$ are the same and ({\it ii}) calculates the scaling factor $s_{N}$, which allows us to determine a black hole mass of the target object. It should also be emphasized that to estimate a BH mass  using the following equation for the scale factor, the ratio of the distances to the {\it target} and {\it reference} sources is necessary:
\begin{equation}
 s_N=\frac{N_r}{N_t} =  \frac{m_r}{m_t} \frac{d_t^2}{d_r^2}{f_G},
\label{mass}
\end{equation}
where $N_r$ and $N_t$ are normalizations of the spectra, $m_t=M_t/M_{\odot}$ and  $m_r=M_r/M_{\odot}$ are the dimensionless  BH masses with respect to a solar mass, and $d_t$ and $d_r$  are distances to  the {\it target} and {\it reference} sources, correspondingly. 
A   geometrical factor, $f_G=\cos i_r/\cos i_t$, where $ i_r$ and $ i_t$ are the disk inclinations for   the {\it reference}  and {\it target} sources, respectively (see ST09, Eq.~(7)).

\subsubsection{A BH mass estimate in NGC~1566}
\label{mass_estimate_1566}

For appropriate scaling, we need to select X-ray sources (reference sources), which also show the effect of  the index saturation, namely at the same $\Gamma$ level as in NGC~15660707--495 (target source). For reference sources, a BH mass, inclination, and distance must be well known. We found that NGC~4051, GX~339--4, GRO~J1655--40, Cyg~X--1 and 4U~1543--47 can be used as the reference sources because these sources met all aforementioned  requirements to estimate a BH mass of the target sources NGC~1566 and NGC~7679 (see  items (i) and (ii) above). 

\begin{figure}
\centering
\includegraphics[scale=0.9,angle=0]{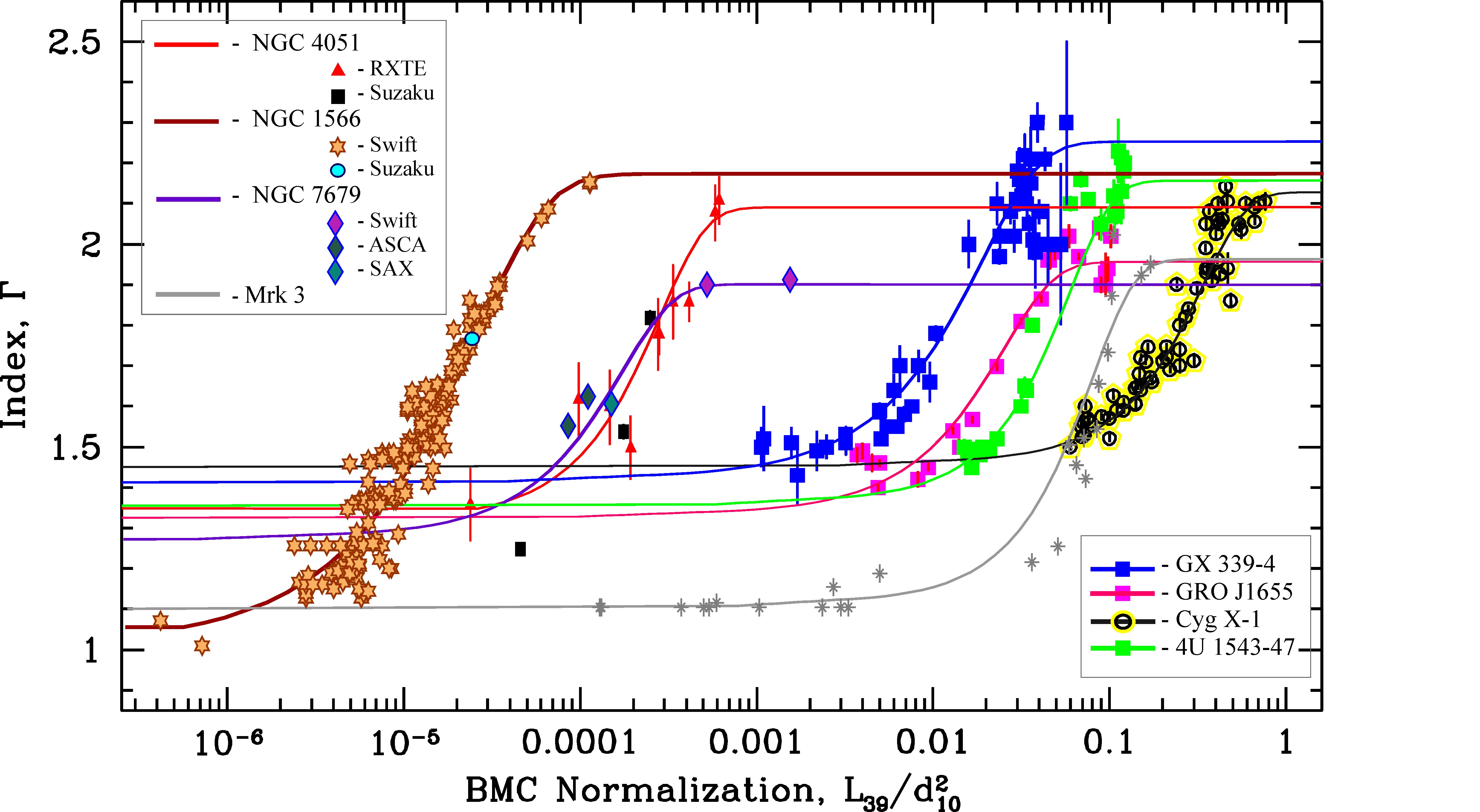}
\caption{Scaling of photon index $\Gamma$ for NGC~1566 (with brown line for target source) and NGC~7679 (with violet line for target source) with NGC~4051, GX~339--4, GRO~J1655--40, Cyg~X--1 and 4U~1543--47 
as reference sources).
}
\label{scaling_1566}
\end{figure}

\begin{figure*}
\centering
 \includegraphics[width=16cm]{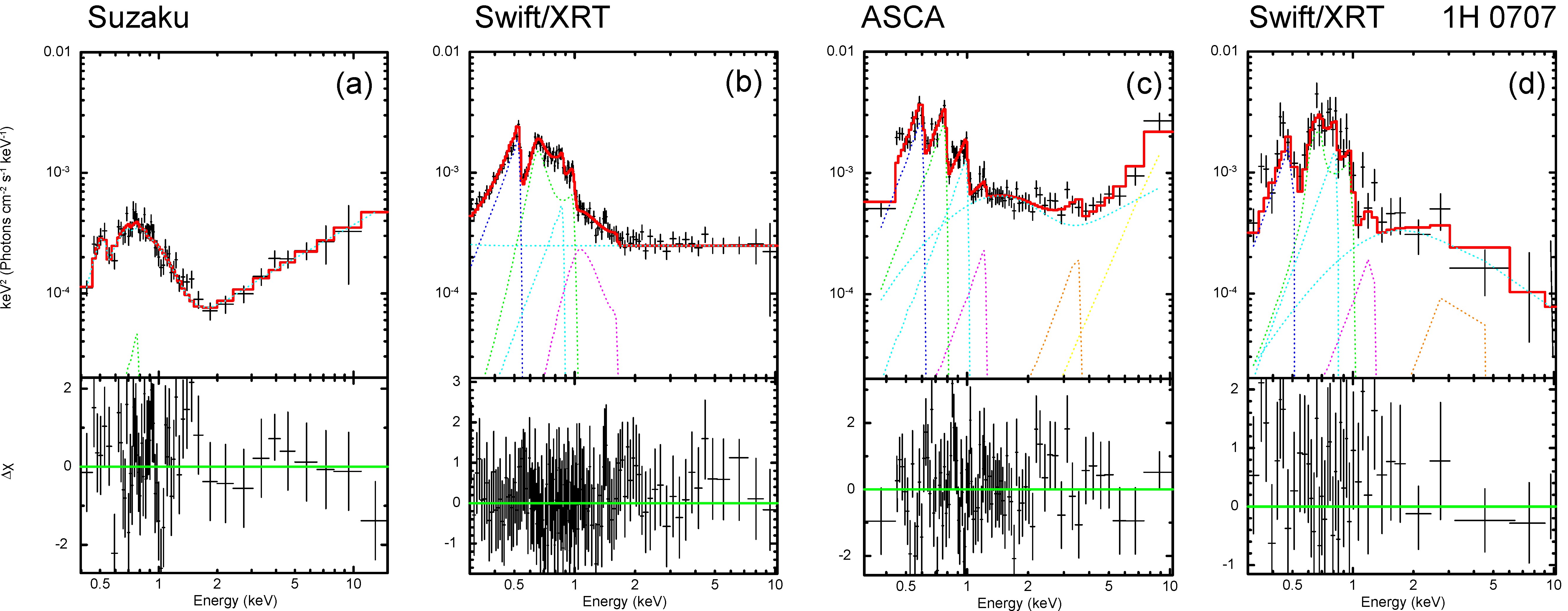}

\caption{The best-fit spectrum of 1H~0707--495 in $E*F(E)$ units during: (a) LHS using {\it Suzaku} observation 00091623; (b)  (IS)  using Swift observation 00091623002; (c)  IS using ASCA observation 73043000 and (d)  HSS   using Swift observation 00080720048.
  The data  are presented by crosses and the best-fit spectral  model   {\it tbabs*(BMC+N*Laor)} by red line.  The Comptonization hump component is shown by the dotted sea-green line. To model the $Laor$ line components, we used the N~XVII (blue), O~III (green), Fe~XVII (bright blue), Ne~X (pink), S~XVI (orange), Fe~I--XXII  and Fe~ XXV/Fe~XXVI K$_{\alpha}$ (yellow) lines with energies of 0.5, 0.65, 0.85, 1.02, 2.9, 6.4, and 6.8 keV, respectively.
 Bottom: $\Delta \chi$ vs photon energy in keV. 
}
\label{Swift_spectra_0707}
\end{figure*}

In Figure \ref{scaling_1566} we demonstrate how the photon index $\Gamma$ evolves with normalization $N$ (proportional to the mass accretion rate $\dot M$) in NGC~1566 ({\it target} source) and NGC~4051, GX~339--4, GRO~J1655--40, Cyg~X--1 and 4U~1543--47 ({\it reference} sources),  where  $N$  is presented in units of $L_{39}/d^2_{10}$ ($L_{39}$ is the source luminosity in units of $10^{39}$ erg/s and $d_{10}$ is the distance to the source in units of 10 kpc). As we can see from this Figure 
that these sources have almost the same index saturation level $\Gamma$. 
We estimated a BH mass for NGC~1655 using the scaling approach  (see e.g.,  ST09). In Figure~\ref{scaling_1566} we illustrate  how the scaling method works shifting one correlation versus another.     From these correlations we could estimate $N_t$, $N_r$ for NGC~1566 and   for the reference sources (see Table~\ref{tab:par_scal_1566+7679}). A value of $N_t=1.04\times10^{-4}$,  $N_r$ in units of $L_{39}/d^2_{10}$  is determined in the beginning of the $\Gamma$-saturation  part (see Fig. \ref{tab:par_scal_1566+7679}, ST07,  ST09, \cite{STS14,TS16,tsei16b,ts09}).

A value of   $f_G=\cos {i_r}/\cos{i_t} $  for the {\it target} and {\it reference} sources  can be obtained  using trial inclination for NGC~1566 $i_t=60^{\circ}$  and for $i_r$   (see Table \ref{tab:par_scal_1566+7679}). As  a result of the estimated target  mass  $m_t$   (in NGC~1566),we find  that
\begin{equation}
m_t= f_G\frac{m_r}{s_N} \frac{d_t^2}{d_r^2}, 
\label{mass_target1}
\end{equation}
where we used  values of $d_t=21.3$ Mpc  (see Table \ref{tab:parameters_binaries}).

 Applying   Eq.(\ref{mass_target1}), we can estimate  $m_t$ (see Table \ref{tab:par_scal_1566+7679})  and we  find  that the secondary BH mass  in NGC~1566 is about $1.9\times(1\pm 0.20)\times10^5$ M$_{\odot}$. To obtain this estimate  with appropriate error bars, we need to consider  error bars for $m_r$ and $d_r$ assuming, in the first approximation, errors for  $m_r$ and $d_r$ only.
 We rewrote Eq. (\ref{mass_target1}) as
 \begin{equation}
m_t(1+\Delta m_t/m_t)= f_G\frac{m_r}{s_N} \frac{d_t^2}{d_r^2}(1+\Delta m_r/m_r)(1+ 2 \Delta d_r/ d_r).
\label{mass_target_expand}
\end{equation}
Thus, we obtained errors for the $m_t$ determination (see Table \ref{tab:par_scal_1566+7679}, second column for the {\it target} source), such that 
\begin{equation}
\Delta m_t/m_t \sim \Delta m_r/m_r + 2 \Delta d_r/ d_r.
\label{mass_target_errors}
\end{equation} 
\noindent As a result, we find that M$_{1566} \sim 1.9 \times 10^5$ M$_{\odot}$ ($M_{1566} = M_t$) assuming $d_{1566}= 21.3$ Mpc for NGC~1566. 
We present all these results in Table~\ref{tab:par_scal_1566+7679}. In order to calculate  the  dispersion ${\mathcal D}$ of the arithmetic  mean  $\bar{ m_t}$ for a BH  mass estimate using different reference sources ${\mathcal D}$  (see   Table \ref{tab:par_scal_1566+7679}), one should keep in mind  that 
\begin{equation}
{\mathcal D} (\bar{m_t})= D/n,
\label{dispersion_mean}
\end{equation} 
where $D$ is the dispersion of $m_r$  using  each of the reference sources and $n=5$ is a number of  the reference sources. As a result we determined that  the mean deviation  of the arithmetic mean  
\begin{equation}
\sigma  (\bar{m_t})= \sigma/\sqrt{n}\sim 0.20
\label{sigma_arithm_mean}
\end{equation}
 and finally  we came to the following conclusion (see also Table \ref{tab:par_scal_1566+7679}):
 \begin{equation}
\bar{m_t} \sim1.9\times(1\pm 0.20)\times10^5 ~~~\rm M_{\odot}. 
\label{arithm_mean}
\end{equation}

\subsubsection{A BH mass estimate in NGC~7679}
\label{mass_estimate_7679}
Figure~\ref{scaling_1566} we illustrate  how the scaling method works shifting one correlation versus another.     From these correlations we could estimate $N_t$, $N_r$ for NGC~7679 and   for the reference sources (see Table~\ref{tab:par_scal_1566+7679}). A value of $N_t=5\times10^{-4}$,  $N_r$ in units of $L_{39}/d^2_{10}$  is determined in the beginning of the $\Gamma$-saturation  part (see Fig. \ref{scaling_1566}). 

In a similar way and using the same reference sources as for NGC~1566, we determined a BH mass in NGC~7679 (see also Table \ref{tab:par_scal_1566+7679}):
 \begin{equation}
\bar{m_t} \sim8.4\times(1\pm 0.20)\times10^6 ~~~\rm M_{\odot}, 
\label{arithm_mean}
\end{equation}
\noindent where we used  values of $d_t=57.28$ Mpc and source inclination $i_t=30^{\circ}$  (see Table \ref{tab:parameters_binaries}).

\subsubsection{A BH mass estimate in Mrk~3}
\label{mass_estimate_Mrk3}
In Figure~\ref{scaling_1566} we  illustrate how the scaling method works for Mrk~3. In this case, we used five reference sources NGC~4051, GX~339--4, GRO J1655--40, Cyg~X--1 and 4U~1543--47, as well as for NGC~1566 and NGC~7679. From these correlations we were able to estimate $N_t$, $N_r$ for Mrk~3 and for the reference sources (see Table~\ref{tab:par_scal_1566+7679}). The value $N_t=0.1$, $N_r$ in units of $L_{39}/d^2_{10}$ is determined at the beginning of the $\Gamma$-saturation part (see Fig. \ref{scaling_1566}).

In a similar manner to Sects. \ref{mass_estimate_1566} and \ref{mass_estimate_7679} as well as using the same reference sources as for NGC~1566 and 7679, we determined the mass of the BH in Mrk~3 (see also Table \ref{tab:par_scal_1566+7679}): 
\begin{equation}
\bar{m_t} \sim2.2\times(1\pm 0.20)\times10^8 ~~~\rm M_{\odot}, 
\label{arithm_mean}
\end{equation}
\noindent where we used  values of $d_t=63.2$ Mpc and source inclination $i_t=50^{\circ}$  (see Table \ref{tab:parameters_binaries}).

\subsubsection{A BH mass estimate in 1H~0707--495}
\label{mass_estimate_0707}

We found that SDSS~J0752, OJ~287, M101 ULX--1 and ESO~243 HLX--1 can be used as the reference sources because these sources met all aforementioned  requirements to estimate a BH mass of the target source 1H~0707. 

\begin{figure*}
\centering
\includegraphics[scale=0.8,angle=0]{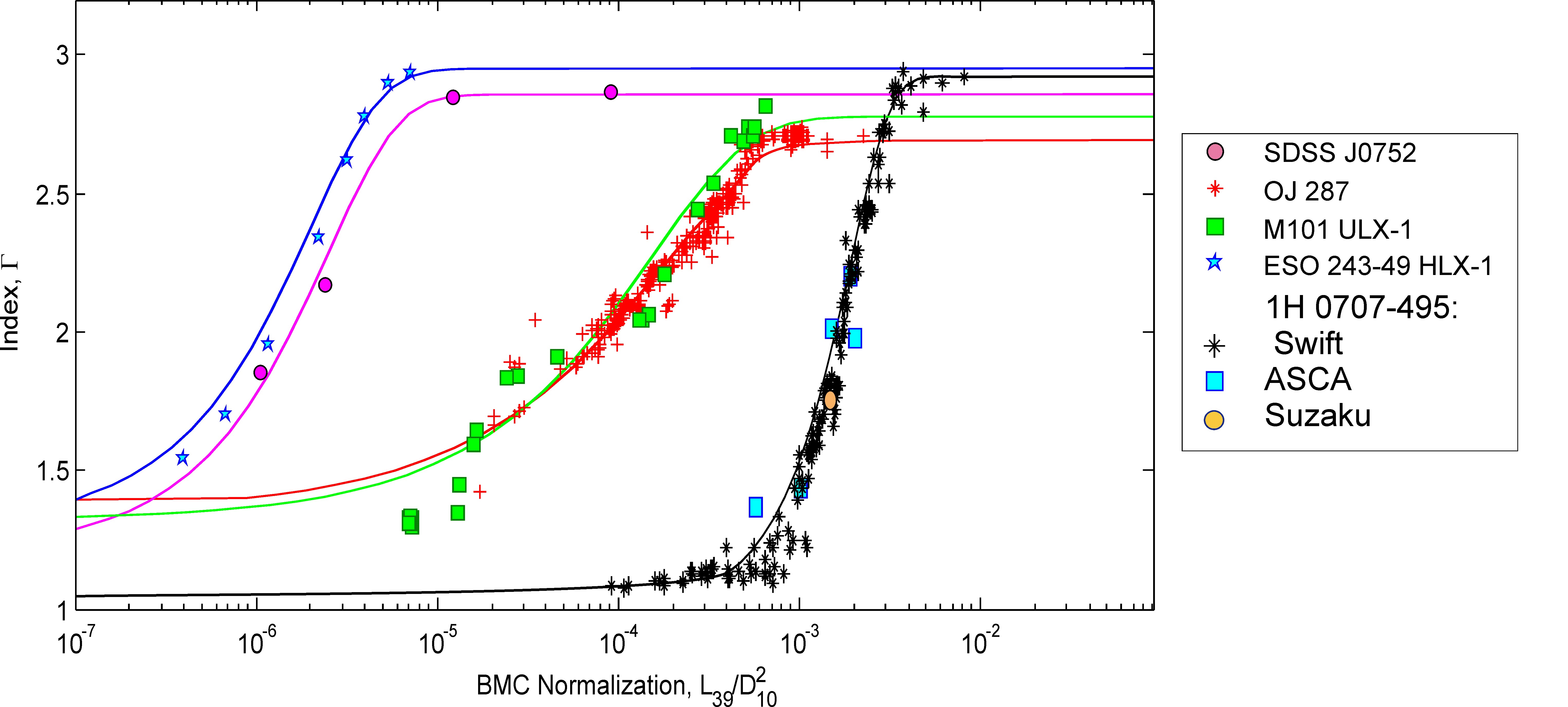}
\caption{Scaling of photon index $\Gamma$ for 1H~0707--495 (with black line for target source) and SDSS~J0752, OJ~287 and M101~ULX--1 (as reference sources).
}
\label{scaling_0707}
\end{figure*}

\begin{figure}
\centering
 \includegraphics[width=16cm]{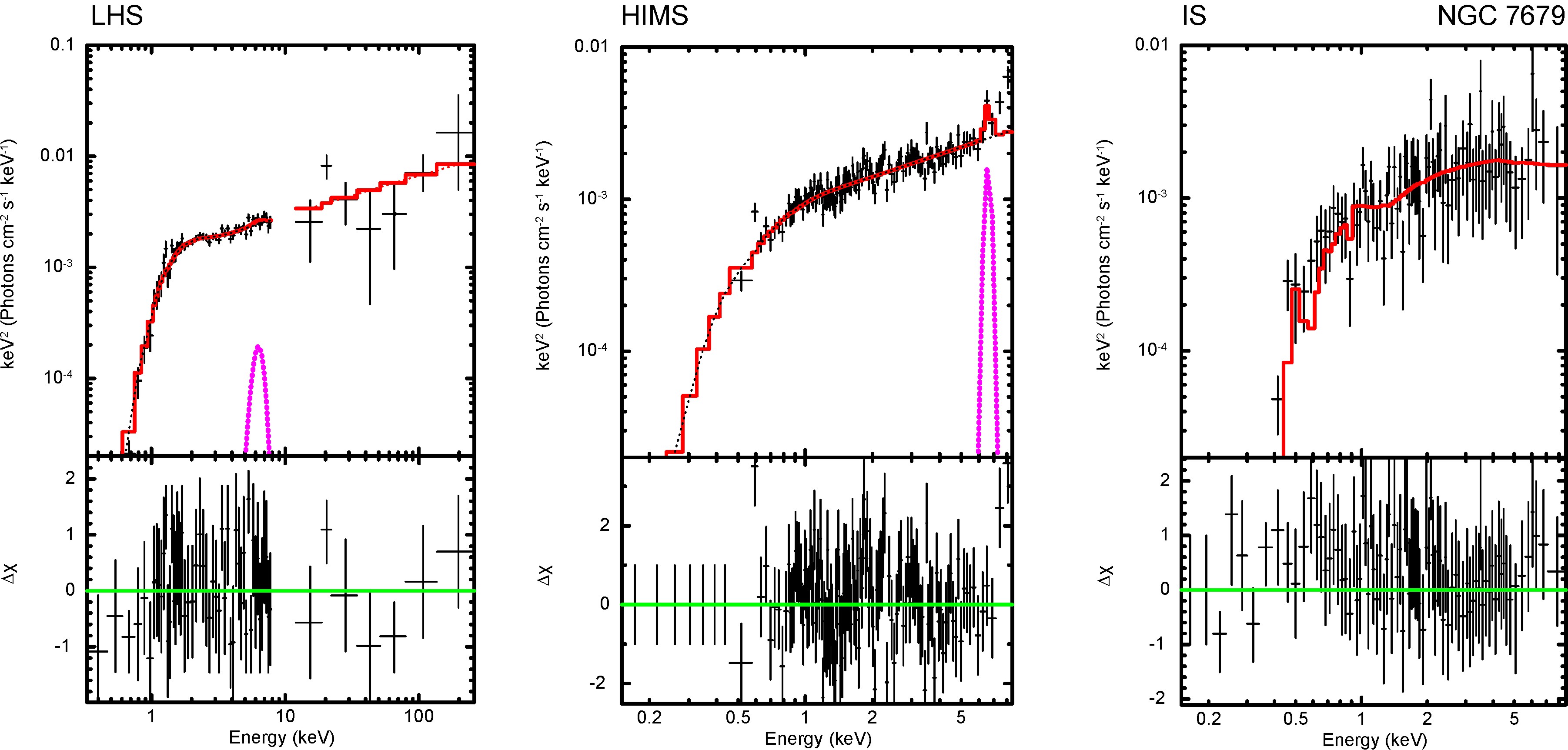}
\caption{Three representative spectra of NGC~7679 with the best-fit modeling for the LHS (ID=40631001, left) using the {\it Beppo}SAX data,  HIMS (ID=66019000, center) from ASCA and for the IS (ID=00088108002, right) from $Swift$ in units of $E*F(E)$ using 
the  {\tt tbabs*(bmc+gauss)} model, respectively.  The data are denoted by  crosses, while the spectral model is shown by a red histograms for each state.
o model the line component, we used Fe~ XXV/Fe~XXVI K$_{\alpha}$ (pink) lines with energies of 6.7/6.9 keV.
Bottom: $\Delta \chi$ vs photon energy in keV. 
}
\label{3_spectra_ngc_7679}
\end{figure}

\begin{figure*}
\centering
\includegraphics[width=16cm]{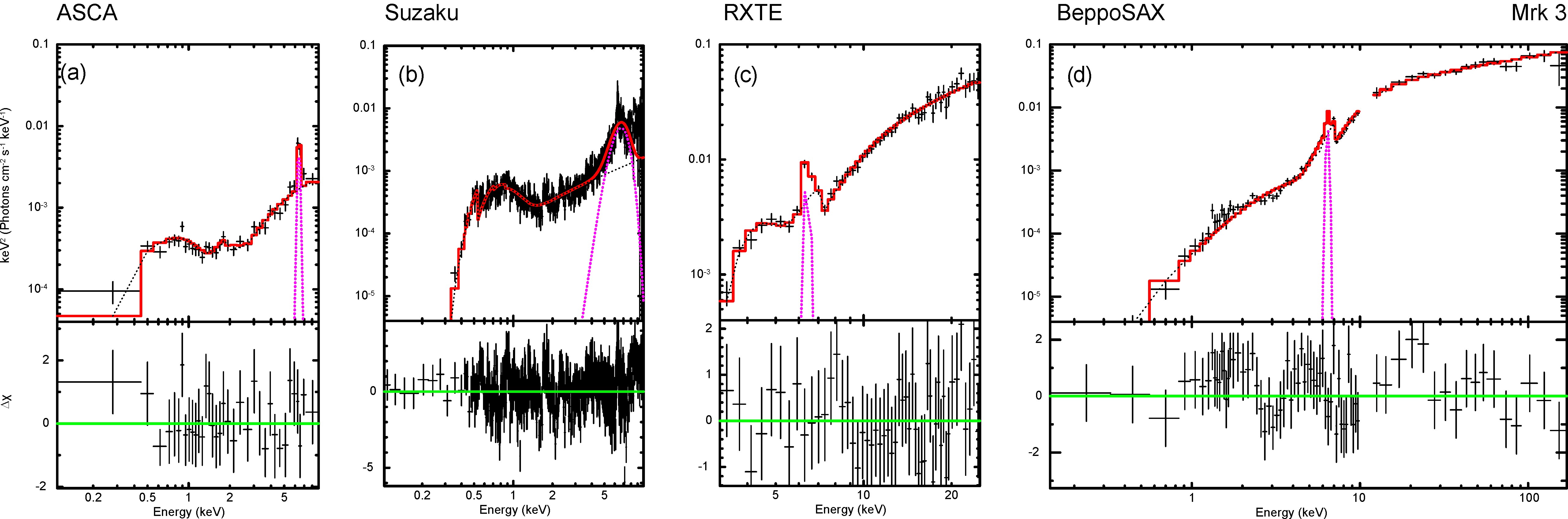}

\caption{Four representative spectra of Mrk~3 
in units of $E*F(E)$ with the best-fit modeling for the LHS ({\it ASCA} observation with ID=70002000, panel $a$), IS ({\it Suzaku} observation with ID=100040010, panel $b$),  IS ({\it RXTE} observation with ID=20330-01-09-00, panel $c$) and IS ({\it Beppo}SAX observation with ID=50132002, panel $d$) states. 
The data are denoted by  crosses, while the spectral model is shown by a red histograms for each state.
 Bottom: $\Delta \chi$ vs photon energy in keV. 
}
\label{4_spectra_ark_3}
\end{figure*}

\begin{figure}
\centering
\includegraphics[scale=0.8,angle=0]{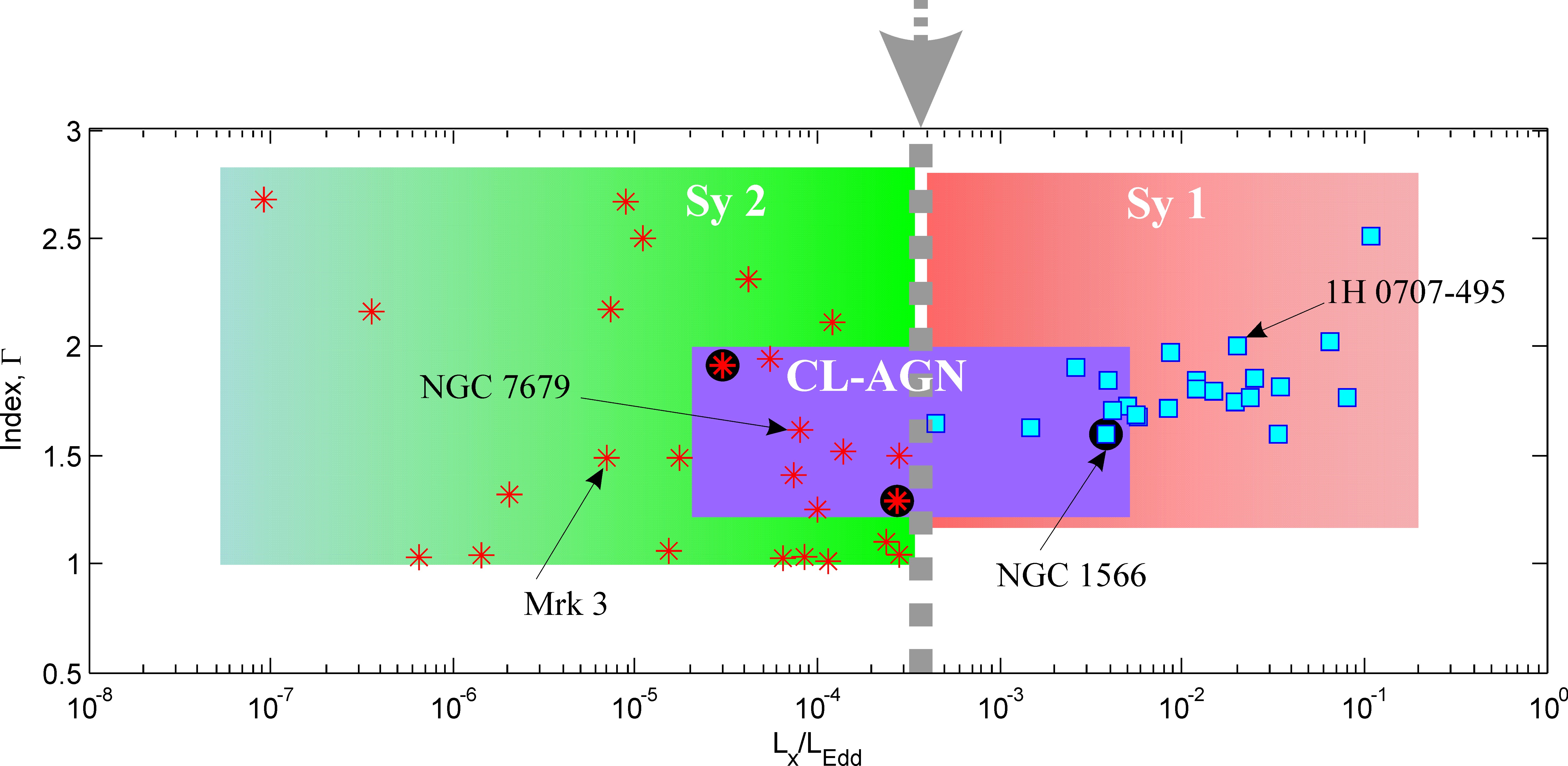}
\caption{Photon index, $\Gamma$ plotted versus $L_x/L_{Edd}$ for Sy 1 (red box), Sy 2 (green box) and CL-AGN (violet box). The grey dotted arrow indicates the critical value of $L_x/L_{Edd}$, separating Sy1 and Sy2. It is evident that the CL-AGN box covers both the Sy1 and Sy2 AGN regions.
}
\label{gam_LEdd}
\end{figure}

\begin{figure}
\centering
\includegraphics[scale=0.8,angle=0]{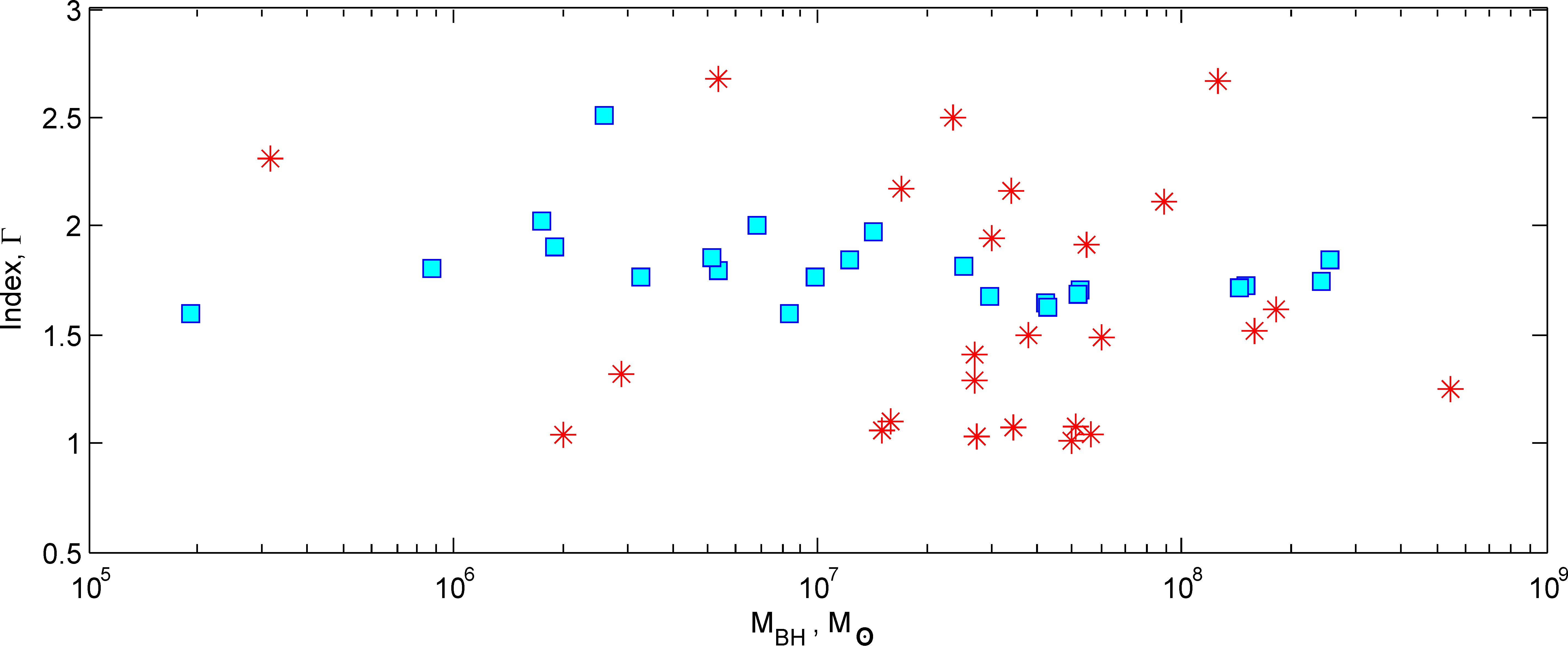}
\caption{Photon index, $\Gamma$ plotted versus a BH mass for Sy 1 (blue squares) and Sy 2 (red stars) AGNs.
}
\label{gam_mass}
\end{figure}

In Figure \ref{scaling_0707} we demonstrate how the photon index $\Gamma$ evolves with normalization $N$ in 1H~0707 ({\it target} source) and SDSS~J0752, OJ~287, M101 ULX--1 and ESO~243 HLX--1 ({\it reference} sources),  where  $N$  is presented in units of $L_{39}/d^2_{10}$. As we can see from this Figure 
that these sources have almost the same index saturation level $\Gamma$ about 2.8--2.9. We estimated a BH mass for 1H~0707--495 using the scaling approach  (see e.g., ST09). In Figure~\ref{scaling_0707} we illustrate  how the scaling method works shifting one correlation versus another.     From these correlations we can estimate $N_t$, $N_r$ for 1H~0707 using   the reference sources (see Table~\ref{tab:par_scal_0707}). A value of $N_t=3.5\times10^{-3}$,  $N_r$ in units of $L_{39}/d^2_{10}$  is determined in the beginning of the $\Gamma$-saturation  part (see Fig. \ref{tab:par_scal_0707}). 

To determine the distance to 1H~0707-495 we used the formula (for $z < 1$) 
\begin{equation}
d_{0707} = z_{0707}c/H_0\simeq 160~~Mpc, 
\end{equation}
\noindent where the redshift $z_{0707} = 0.004$ for 1H~0707, $H_0 = 70.8 \pm 1.6  {\rm km s^{-1} Mpc^{-1}}$ is the Hubble constant and $c=3\times 10^5$ km/c is the speed of light. This distance $d_{0707}$ agrees with the luminosity distance estimate using Ned Wright's Javascript Cosmology Calculator\footnote{https://www.astro.ucla.edu/~wright/CosmoCalc.html} $d^{NW}_{1H0707}\sim 170$ Mpc \cite{Wright06}. 

A value of   $f_G$ for the {\it target} and {\it reference} sources  can be obtained  using the trial inclination for 1H~0707 $i_t=55^{\circ}$  and for $i_r$   (see Table \ref{tab:par_scal_0707}). 

 We estimated the  target  BH mass, $m_t$ (1H~0707)  (see also Table \ref{tab:par_scal_0707}):
 \begin{equation}
\bar{m_t} \sim 6.8\times(1\pm 0.27)\times10^7 ~\rm M_{\odot},  
\label{arithm_mean}
\end{equation}
\noindent where we used  values of $d_t=160$ Mpc. 
Here we also calculated  the  dispersion ${\mathcal D}$ of the arithmetic  mean  $\bar{ m_t}$ for a BH  mass  using different reference sources ${\mathcal D}$  (see   Table \ref{tab:par_scal_0707})
\begin{equation}
\sigma  (\bar{m_t})= \sigma/\sqrt{n}\sim 0.27, 
\label{sigma_arithm_mean}
\end{equation}
\noindent wherein $n=4$.

\section{Discussion \label{discussion}} 
We argued that the emission of all  considered AGNs (NGC~1566, 1H~0707, NGC~7679  and Mrk~3) shows variability and has a flaring nature, presumably due to a change in the rate of accretion of matter in the central regions of each AGN. We found that the X-ray spectra for these CL, Sy1 and Sy2 AGNs can contribute to our understanding of the differences between these AGN subclasses. Their differences immediately follow from different patterns of variability and different luminosity levels of the sources. 
However, we found differences and similarities between them. Emission from NGC~1566 during changing-look can be associated with different combinations of hard and soft components of the spectra depending on the level of relative luminosity ($L_x/L_{Edd}$) of NGC~1566. This source uniquely combines the properties of Sy1 and the properties of Sy2: at high relative luminosities NGC~1566 demonstrates behavior typical for Sy1, and at low relative luminosities we observed a transition of NGC~1566 from Sy1 behavior to Sy2 one.

Furthermore, as a result of the comparative analysis, we found spectral distinctive features between Sy1 and Sy2, which are easy to detect. In fact, the evaluation of two observable quantities (the photon index $\Gamma$ and relative X-ray luminosity $L_x/L_{Edd}$ of AGNs shows a strong difference for Sy1 and Sy2. AGNs that fall in both regions of Sy1 and Sy2 in terms of $L_x/L_{Edd}$ are candidates for CL-AGNs (see Fig.~\ref{gam_LEdd}).

An intriguing question is what would be the mechanism for such abrupt switching on or off of CL AGN power generation. It was studied by \cite{Katebi19,Lyuty84,Penston+Perez84,Runnoe16,Lightman+Eardley74,Liu20,MacLeod19,Parker19,Ruan19,Sniegowska21,Oknyansky20}. Several different possibilities have been considered: variable dimming, disk accretion flares, tidal disruption events, and a supernova event (see, for example, Okniansky20). We found that a BH mass  in CL-AGN NGC~1566, determined by scanning X-ray properties and the detected saturation of the index in NGC~1566 ($\alpha_{1566}\sim1.7$), turned out to be one or two orders of magnitude lower than a BH mass, determined from optical observations. This may indicate a possible duality of the SMBH at the level of the center of NGC~1566. This fact may be one of the reasons for a rapid change in the accretion rate, which leads to a change observed in the NGC~1566 galaxy.

During the outbursts, we found the index saturation in all three sources, which is consistent with the spectral signature of the presence of a BH in these sources (TC09). Based on the saturation effect, we estimated a BH mass in each source. It should be noted that a BH mass in NGC~1566 indicates a possible nature of the variability of NGC~1566 associated with the presence of a second, more massive BH in the center of NGC~1566. Interestingly, ta BH mass of the BH in 1H~0707 is slightly larger than those  based on reverberation delays \cite{Zhou+Wang05,Kara13,Done+Jin16}, which may indicate a presence of an additional soft X-ray source, for example, associated with enhanced star formation \cite{Sani10,Zoghbi10}. A BH mass in NGC~7679 is consistent with the estimates by other authors available in the literature \cite{Woo+Urry02,Elagali19,Fabian09}.

In addition, we compared our results with those of other authors. For example,  \cite{Jana21}  performed an analysis of the X-ray emission from NGC~1566 during its June 2018 outburst observed by {\it Swift}. They analyzed quasi-simultaneous observations of this source obtained with {\it NuSTAR} and {\it XMM-Newton} using models ({\tt power-law} and {\tt NthCOMP}) different from ours ({\tt BMC+Gauss}). 

Moreover, our results are consistent with their conclusions, according to which a strong excess of soft X-ray emission was detected in the spectra during the outburst. We also found a soft excess, clearly visible in the central panel of Fig.~\ref{spectrum_ev_7679_all} (red spectrum) and panel $d$ in Fig.~\ref{4_spectra_ngc_1566}. However, our model makes it possible to compensate  it increasing a fraction of disk soft photons, caused by an enlargement of  the irradiation parameter $f$. Indeed, soft excess emission occurs in the thermal Comptonization region of the inner accretion disk (Compton cloud, see \cite{TS23,TSC23,TS21,TSCO20,TS17,TS16,tsei16b,STS14,tss10}. 
\cite{Jana21} found that an increase of the mass accretion rate is responsible for a sudden growth of  source luminosity. They based on the {\it q}-shaped shape of the hardness-intensity diagram that is commonly found in flaring BH X-ray binaries. We also came to the conclusion that the main driver of a sharp flare is due to a sharp increase of the mass accretion rate. We explain  this behavior using  normalization parameter $N_{com}$ of the Comptonization  radiation and its typical growth with a flare up to the  peak flare saturation   \citep{ts09}.

\begin{figure*}
\centering
\includegraphics[scale=0.8,angle=0]{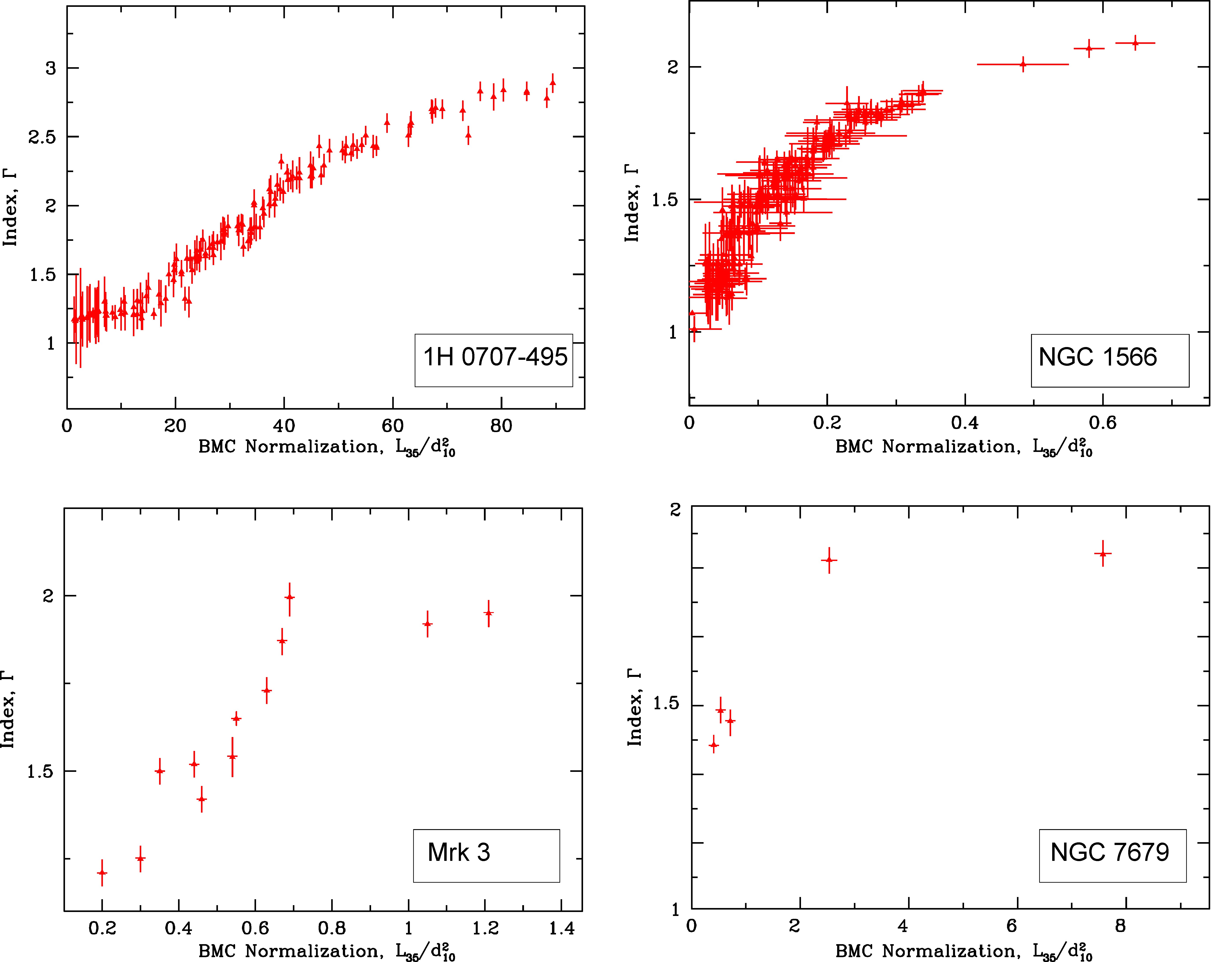}
\caption{Photon index, $\Gamma$ plotted versus BMC normalization (which is proportional to $\dot M$) for 1H~0707 (top left panel), NGC~1566 (top right panel), Mrk~3 (bottom left panel) and NGC~7679 (bottom right panel).
}
\label{saturation_all}
\end{figure*}

It was worthwhile to emphasize  that \cite{Jana21} found a constancy (or weak variability) of the index during the {\it NuSTAR} and {\it XMM-Newton} observations of NGC~1566 at the level of $\Gamma = 1.7-1.9$. This differs from our results for the joint {\it Swift}, {\it NuSTAR} and {\it XMM-Newton} data analysis, when the $\Gamma$ index varies widely from 1 to 2. It is possible that such a difference in the  $\Gamma$  behavior, discovered by us, is related with the use of different spectral models. However, this difference is due to the specific distribution of the observational time for NGC~1566.
These  {\it NuSTAR} observations were rare and fell precisely on the peaks of the primary outburst and the peaks of the secondary outbursts or close to them (see Fig.~\ref{fraq_1566}).  
They fall exactly on the saturation region of the $\Gamma-N_{BMC}$ plot (top left panel in Fig.~\ref{saturation_all}), when the index is maintained at  $\Gamma=2$ for high source luminosities (Figs. \ref{scaling_1566}, \ref{scaling_0707} and \ref{saturation_all}).

The use of different models for modeling AGN spectra sheds light on different features of AGNs. In this case, the use of special models such as {\tt agnsed} \citep{Kubota+Done18} and dual-coronal \citep{Petrucci13}, which take into account the geometry of AGNs with different zones (outer standard disk, inner warm Comptonizing region and hot corona),
determine well the thermal component of Comptonization to reveal the soft X-ray excess in the spectra of AGNs in the 
LHS state. \cite{Petrucci13} used mainly the thermal Comptonization component to describe the soft X-ray excess in the AGN spectra in the LHS.  Their Figure 10 correctly represents the geometry of the accretion flow in the inner region of the galaxy Mrk~509. However, Petrucci et al. (2013) do not find any spectral evolution, for example, in the source Mrk~509. In contrast, the photon index in our {\tt BMC} or {\tt COMPTB} models has an upper limit of 3 with all the signs of photon index saturation. This is clearly confirmed by our spectral numerical and analytical calculations of the convergent flow spectra (see \cite{tz98,LT99}).

Interestingly, \cite{Jana21} discusses a scenario in which the central nucleus of NGC~1566 could be a merging SMBH. We also come to a similar conclusion, but we based on an estimate a BH mass at the center of NGC~1566 applying   X-ray data.
Since our ``X-ray BH mass estimate'' turned out to be two orders of magnitude lower than the optical one, we can assume a presence of a binary system of SMBHs of different masses in the center of NGC~1566. Similar situations have been observed before \citep{S23,TSC23,TS24}. In these cases, a lower BH mass  (secondary BH), being more mobile, orbiting around a more massive BH (primary BH), causes X-ray outbursts  as it passes through the disk around the primary BH. The X-ray method for a BH mass estimate determines a BH mass only of the secondary BH. It is clear that the primary BH does not appear in the X-ray range, although it does appear in the optical range and here its mass dominates the contribution of the secondary BH, especially when one estimates a BH mass using stellar kinematics and dynamics methods \citep{Woo+Urry02,Elagali19}.

\section{Conclusions \label{summary}} 
We present our analysis  of the spectral properties observed in X-rays from 
the  NGC~1566, 1H~0707--495, NGC~7679 and Mrk~3 during long-term transitions
between the faint phase and the bright phase. 

We analyze transition episodes for these sources  observed by {\it Swift},{\it Beppo}SAX, Suzaku, ASCA, {\it NuSTAR}, {\it XMM-Newton}  and {\it RXTE}. We show that the X-ray spectra for CL, Sy1 and Sy2 during all spectral states can be well fitted by a  composition of  the  Comptonization and   {\it Gaussian/Laor} components. We argued that these changes in the X-ray properties of NGC~1566 are characterized by a different combination of Sy1 and Sy2 properties. At high luminosities NGC~1566 exhibits  a behavior which is  typical for Sy1, and at low luminosities we observed a transition of NGC~1566 from the Sy1 behavior to the Sy2 one.

 We discovered the saturation of the  photon  index for these four AGNs during outbursts ($\Gamma_{1566}\sim2.1$, 
 $\Gamma_{0707}\sim3$, $\Gamma_{7679}=1.9$  and $\Gamma_{mrk3}=1.9$) and estimated the masses 
 of the  SMBHs in the centers of these AGNs: M$_{0707}\sim (6.8\pm0.27)\times 10^7$ M$_ {\odot}$, M$_{7679}\sim(8.4\pm0.2)\times 10^6$ M$_{\odot}$,  M$_{mrk3}\sim(2.2\pm0.2)\times 10^8$ M$_{\odot}$ and M$_{1566} \sim(1.9\pm0.2)\times 10^5 $ M$_{\odot}$ by the scaling method. Notably, the X-ray scaled  BH mass  in CL-AGN NGC~1566 is one to two orders of magnitude lower than the BH mass determined from optical observations, indicating a possible duality of the SMBH at the center of NGC~1566.  This may be one of the reasons for a rapid change in the accretion rate, which leads to a changing look observed in the NGC~1566 galaxy. 

We also find that when CL-AGN transits from Sy2 to Sy1, the spectrum of NGC~1566 shows an increase of the soft excess accompanied by a decrease in the luminosity fraction ($0.1<f<0.5$), which is consistent with a typical behavior of BH sources during X-ray outburst decay.

Our results strongly suggest that the large diversity in the  behavior observed among CL, Sy1, and Sy2 AGNs with different X-ray luminosities can be explained by changes of  a single variable parameter, such as the ratio of the AGN's X-ray luminosity to its Eddington mass accretion rate, without any need for additional differences in Sy AGN parameters, such as its inclination, thereby blurring the distinction between the Sy1, Sy2 and CL-AGN subclasses.

\section*{Acknowledgements}
This research has made using the  data and/or software provided by the High Energy Astrophysics Science Archive Research Center (HEASARC), which is a service of the Astrophysics Science Division at NASA/GSFC and the High Energy Astrophysics Division of the Smithsonian Astrophysical Observatory.  We appreciate valuable remarks  by Chris Schrader on the paper  and thank Juliana Seifina for her assistance in preparing the figures.  
We acknowledge the interesting remarks and points of the referee. 
The {\it Swift} data is available to download through the UK Swift Data Science website \url{https://www.swift.ac.uk/archive}. 


\appendix
\section{Tables}

%
%

\begin{footnotesize}
\begin{table*}
 \caption{Basic information on 1H~0707--495, NGC~1566 and NGC~7679}
 \label{tab:parameters_binaries}
 \centering 
\setlength{\tabcolsep}{1pt}
\resizebox{1.\textwidth}{!}{%
 \begin{tabular}{lllll}
 \hline\hline                        
Source parameter               &     1H~0707--495     &   NGC~1566      &  Mrk~3 & NGC~7679\\
      \hline
 \\

Class of AGN         & Sy1, NLS1$^{~(a)}$     & CL-AGN, SAB(s)bc$^{~(b)}$   & Sy2, S0, HBLR$^{(c)}$& Sy2, SB0, SB+AGN\\

Mass of BH, M$_{\odot}$ & 2 $\times$ 10$^6$--10$^{7~(d)}$ &8.3 $\times$ 10$^{6}-1.3\times10^{7~(e,f)}$ & 5.5$\times$ 10$^{8~(g)}$ & 5.9$\times$ 10$^{6~(h)}$\\
Inclination,  $i$,  deg           & 55-65$^{~(i)}$                           & 60 $\pm$ 5$^{~(j)}$, & 50$^{~(c)}$  & 30$^{~(k)}$\\
Distance, $D$,  Mpc       & 160$^{(l)}$               &  21.3$^{(m)}$ & 63.2$^{(c)}$& 58.633$\pm$ 1.910$^{(k)}$\\
Redshift, $z$ & 0.041$^{(d)}$                          &0.005$^{(j,m)}$ &0.013$^{(c)}$& 0.017$^{(d)}$\\
RA (J2000), $\alpha$       & 07$^h$ 08$^m$ 41.48$^{s~(l)}$               & 04$^h$ 20$^m$ 00.52$^{s~(l)}$ &06$^h$ 15$^m$ 36.3$^{s~(l)}$&23$^h$ 28$^m$ 46.73$^{s~(l)}$\\
Dec (J2000), $\delta$       & --49$^{\circ}$ 33{\tt '} 06.4{\tt ''}$^{~(l)}$               & --54$^{\circ}$ 56{\tt '} 17.1{\tt ''}$^{~(l)}$ & +71$^{\circ}$ 2{\tt '} 15{\tt ''}$^{~(l)}$& +03$^{\circ}$ 30{\tt '} 42.1{\tt ''}$^{~(l)}$\\
Mean count rate, ct/s & 0.1975$\pm$0.0013$^{(l)}$                          &0.2984$\pm$0.0013$^{(l)}$ & 0.192$\pm$0.012$^{(l)}$& 0.192$\pm$0.012$^{(l)}$\\

Mean flux, & 4.533$\pm$0.030$^{(l)}$  &11.74$\pm$0.05$^{(l)}$ & 9.7$\pm$0.7$^{(l)}$& 7.9$\pm$0.5$^{(l)}$\\
$\times 10^{-12}$ erg/s/cm$^2$ &   & & & \\
{\it Swift} name, 2SXPS...  &     J070841.4--493306                    & J0420.0--5457 & J061536.2+710214& J232846.7+033042\\
Alternative names &  1H~0659--494,  & ESO 157--20, & UGC 03426&
 Arp~216, Mrk~534, \\
                        & 4U 0708--49  & IRAS 04189--5503, & &
 PGC~71554, VV 329a,\\
                        &   &  PGC~14897 & &
 UGC~12618\\
$N_{H,Gal}$, cm$^{-2}$   &8.0$\times$10$^{20~(k)}$    & 1.1$\times$10$^{20~(k)}$  & 1.8$\times$10$^{20~(l)}$& 4.5$\times$10$^{20~(m)}$\\
 \hline                                             
 \end{tabular}
}
{\\NLS1 -- Narrow-line Seyfert 1  galaxy; SAB(s)bc -- barred spiral galaxy; 
CL-AGN -- changing-look AGN; SB0 -- early-type barred galaxy; S0 --  lenticular galaxy; HBLR -- hidden broad-line-region; Mean count rate and flux of the source is measured in 0.3--10 keV range. }
{ $^{(a)}$ \cite{Zoghbi11} ; 
$^{(b)}$ \cite{Vaucouleurs73}; 
\cite{Shobbrook66}, \cite{Vaucouleurs91}; 
$^{(c)}$ \cite{Veron06,Ricci14}; 
$^{(d)}$ \cite{Zhou+Wang05,Kara13,Done+Jin16,Zoghbi11}; 
$^{(e)}$ \cite{Woo+Urry02}; 
$^{(f)}$ \cite{Elagali19}; 
$^{(g)}$ \cite{Fabian09}; 
$^{(h)}$ \cite{Hancock23};
$^{(i}$ \cite{Aguero04}; 
$^{(j)}$ \cite{Kewley01,Springob05}; 
$^{(k)}$ Galactic neutral hydrogen column density along the line of sight from the Leiden/Dwingeloo Survey \cite{Hartman+Burton97}; 
$^{(l)}$ https://www.swift.ac.uk/2SXPS/;
and $^{(m)}$\cite{DellaCeca01}. 
}
\end{table*}
\end{footnotesize}

%
%
\begin{table*}
 \centering 
 \caption{Details of {\it Suzaku}, ASCA, {\it Beppo}SAX, {\it RXTE}, {\it NuSTAR} and {\it  XMM-Newton}  observations of AGNs. 
}
   \label{tab:table_Suzaku+ASCA_SAX}
\setlength{\tabcolsep}{1pt}
\resizebox{1.\textwidth}{!}{%
 \begin{tabular}{llllllr}
 \hline\hline  
Source & Epoch & Mission & Obs. ID& Start time (UT)  & End time (UT) & MJD interval \\

      \hline
                     & A1 & ASCA           & 73043000  & 1995 Mar 15 23:01:26 & 1995 Mar 17 03:00:19 &49792.0--49793.1$^1$ \\
               & A2 & ASCA           & 76031000  & 1998 May 16 19:24:15 & 1998 May 20 07:10:38  &50949.8--50953.3\\ 
1H~0707       & A3 & ASCA           & 87043000  & 1999 Mar 13 00:23:59 & 1999 Mar 14 14:48:30 &51250.0--51251.6\\

                     &Sz1 & {\it Suzaku} & 700008010& 2005 Dec 3 04:13:19 & 2005 Dec 6 02:20:20  &50873.0--50873.5$^1$ \\
                    & R1 & RXTE     & 20309010100  & 1997 Mar 15 06:03:28 & 1997 Mar 15 13:01:36 &50522.2--50522.6\\ 
 \hline
     &Sz2 & {\it Suzaku}  & 707002010  & 2012 May 19 01:36:44 & 2012 May 20 06:34:19 &56066.3--56067.2$^2$ \\
     &Ns1 & {\it NuSTAR}  & 80301601002  & 2018 June 26 20:06:09 & 2018 June 28 02:21:07 &58295.8--58297.0$^2$ \\
     &Xn1 & {\it XMM-Newton}  & 0800840201  & 2018 June 27 00:13:32& 2018 June 28 02:00:33&58296.0--58297.0$^2$ \\
     &Ns2 & {\it NuSTAR}  & 80401601002  & 2018 Oct 4 12:16:09 & 2018 Oct 6 06:24:57 &58395.5--58397.2$^2$ \\
     &Xn2 & {\it XMM-Newton}  & 0820530401 & 2018  Oct 4 12:52:23& 2018 Oct 5 18:29:26&58395.5--58396.7$^2$ \\
     &Ns3 & {\it NuSTAR}  & 80502606002  & 2019 June 5 07:46:09 & 2019 June 6 16:09:57 &58639.3--58640.6$^2$ \\
NGC~1566     &Xn3 & {\it XMM-Newton}  & 0840800401  & 2019  June 5 08:11:23& 2019 June 6 09:55:03&58639.3--58640.4$^2$ \\
     &Ns4 & {\it NuSTAR}  & 60501031002  & 2019 Aug 8 11:26:09 & 2019 Aug 9 16:29:57 &58703.4--58704.6$^2$ \\
     &Xn4 & {\it XMM-Newton}  & 0851980101  & 2019  Aug 11 16:45:53& 2019 Aug 11 16:53:37&58706.6--58706.7$^2$ \\
     &Ns5 & {\it NuSTAR}  & 60501031004  & 2019 Aug 18 05:01:09 & 2019 Aug 19  18:19:57 &58713.2--58714.7$^2$ \\
     &Ns6 & {\it NuSTAR}  & 60501031006  & 2019 Aug 21 10:21:09 & 2019 Aug 23 03:41:11 &58716.4--58718.1$^2$ \\

\hline
               & A4 & ASCA   & 66019000  & 1998 July 1 15:52:42 & 1998 July 3 10:30:42&50995.6--50997.4$^3$ \\ 
NGC~7679         & A5 & ASCA   & 66019010  & 1999 June 13 23:33:07 & 1999 June 14 13:38:03 &51342.9--51343.6$^3$ \\ 
          & B1 & {\it Beppo}SAX    & 40631001  & 1998 Dec 6 15:27:33 & 1998 Dec 9  03:39:11&51153.6--51156.2$^3$ \\ 
\hline
            & A6  & ASCA    & 70002000  & 1993 Apr 21 06:27:06 & 1993 Apr 22 03:00:58 &49098.2--49099.1$^4$ \\
          & A7  & ASCA    & 74041000  & 1996 Oct 26 14:02:48 & 1996 Oct 27 03:50:10 &50382.5--50383.1$^4$ \\
          & R2  & RXTE    & 20330010100  &1996 Dec 25 16:17:04 & 1996 Dec 25 17:39:44  &50442.6--50442.7$^4$ \\ 
          & R3  & RXTE   & 20330011000  & 1997 Feb 17 05:17:29 & 1997 Feb 17 07:10:40 &50496.2--50496.3$^4$ \\ 
          & R4  & RXTE   & 20330010900  & 1997 Mar 16 13:03:44 & 1997 Mar 16 15:14:40 &50523.5--50523.6$^4$ \\ 
          & R5  & RXTE   & 20330010800  & 1997 Mar 21 08:06:40 & 1997 Mar 21 10:07:44 &50528.3--50528.4$^4$ \\ 
          & R6  & RXTE   & 20330010700  & 1997 Mar 31 05:30:08 & 1997 Mar 31 07:20:48 &50538.2--50538.3$^4$ \\ 
          &  R7 & RXTE   & 20330010600  &  1997 Apr 4 00:22:40 & 1997 Apr 4 11:13:04 &50542.0--50542.4$^4$ \\ 
          &  R8 & RXTE    & 20330010300  & 1997 Apr 15 06:53:36 & 1997 Apr 15 08:44:32 &50553.2--50553.3$^4$ \\ 
Mrk 3  & R9  & RXTE    & 20330010200  & 1997 Apr 14 08:28:32 & 1997 Apr 14 10:07:44 &50552.3--50552.4$^4$ \\ 
          &  R10 & RXTE   & 20330010400  &  1997 Apr 16 01:17:20 & 1997 Apr 16 02:24:32 &50554.0--50554.1$^4$ \\ 
          &  R11 & RXTE   & 20330010500  &  1997 Apr 17 09:09:20 & 1997 Apr 17 11:13:04 &50555.3--50555.4$^4$ \\ 
          &  R12 & RXTE  & 20330011100  & 1997 May 30 19:40:32 & 1997 May 30 21:31:44 &50598.8--50598.9$^4$ \\ 
          &  R13 & RXTE  & 20330011200  & 1997 July 7 01:01:20 & 1997 July 7 02:57:36 &50636.0--50636.1$^4$ \\ 
          & B2 & {\it Beppo}SAX & 50132002  & 1997 Apr 16 14:46:38 & 1997 Apr 18 10:21:50&50554.6--50556.4$^4$ \\ 
          &Sz3 & {\it Suzaku} & 100040010& 2005 Oct 22 02:02:09 & 2005 Oct 24 06:26:14  &53665.1--53667.2$^1$ \\
          &Sz4 & {\it Suzaku} & 709022010& 2014 Oct 1 22:08:01 & 2014 Oct 2 10:02:09  &56931.9--56932.4$^4$ \\
          &Sz5 & {\it Suzaku} & 709022020& 2014 Oct 7 19:20:38 & 2014 Oct 8 01:48:17  &56937.8--56938.1$^4$ \\
          &Sz6 & {\it Suzaku} & 709022030& 2014 Oct 23 03:37:31 & 2014 Oct 23 16:09:23  &56953.2--56953.7$^4$ \\
          &Sz7 & {\it Suzaku} & 709022040& 2015 Mar 23 05:48:56 & 2015 Mar 23 15:02:23  &57104.2--57104.6\\
          &Sz8 & {\it Suzaku} & 709022050& 2015 Apr 4 11:21:15 & 2015 Apr 4 23:42:19  &57116.4--57116.9\\
      \hline
      \end{tabular}
}
{\\Columns 1 -- 7 denote Source name, Epoch, Observation ID, Mission Name, Start time (UT), End time (UT), MJD interval, respectively. See the text for details.}
{
(1) \citet{Zoghbi11};  
(2) \cite{Jana21,Oknyansky19,Oknyansky20,Tripathi+Dewangan22a}; 
(3) \citet{DellaCeca01}; 
(4) \citet{Griffiths98,Georgantopoulos99,Matt00,Ricci14}.
}
\end{table*}

%
%

\begin{table*}
  \centering 
 \caption{Details of {\it Swift} observations of 1H~0707--495, NGC~1566, NGC~7679  and Mrk~3. 
}
\setlength{\tabcolsep}{1pt}
\resizebox{1.\textwidth}{!}{%
 \begin{tabular}{lllllcc}
  \hline\hline
Source&Observation ID  & Start time (UT) & End time (UT)&  MJD interval \\
 \hline
&00090393(001-023,025-063, & Apr 3,   2010 & Apr 30, 2018  & 55289 -- 58238\\
&065-068,070-075,077-085, &  &   & \\
&087-091,093-105)&  &   & \\
1H~0707--495&0004031700(1,3)  &  Dec 22, 2010 18:43:44   & Jan 27, 2011 12:08:54          & 55552.78 -- 55588.50 \\
&0009162300(1,2)  & May 19,  2013 & June 19, 2013  & 56431.34 --  56462.05 \\
&000807200(01-13,15-16,19-59) & May 5, 2014 & Jan 31, 2018   & 56782 --  58150 \\ 
\hline 
&00035880(002-004,006,046-051,    & Dec 12,   2007 & July 31, 2019  & 54446 -- 58695$^{1,2}$\\
&053-072,084-089, 092,094-117,&  &   & \\
&119,137-144)&  &   & \\
&000317420(01,03-13,15) & June 23,   2010 & Oct 21, 2020  & 55370 -- 59143$^{1,2}$\\ 
&000456040(01,02,04-08,10-17,& Aug 25,   2011 & Nov 2, 2023  & 55798 -- 60250\\ 
&20-28,31-33,35-51)&  &   & \\ 
&000334110(01-05,07-09,11-30)   & Sep 11,   2014 & July 30, 2015  & 56911 -- 57233$^{1,2}$\\ 
NGC~1566&0008891000(1-3)& Aug 8,   2019$^{1,3}$ & Aug 19, 2021  & 58703 -- 58716$^1$\\ 
&000149160(01,03-19,21-29)& Nov 11,   2021 & Jan 17, 2022  & 59529 -- 59596\\ 
&0311166600(1-9)& June 21,   2022 & Aug 19, 2022  &59751 -- 59810\\ 
&0001492300(1,2)& Nov 19,   2021 00:02:52 & Nov 19,   2021 00:04:54  & 59537.001 -- 59537.003\\ 
&0001530200(1-4)& Aug 19,   2022 & Aug 22, 2022  & 59814 -- 59817\\ 
      \hline
NGC~7679&00088108002& Oct 6,   2017 10:00:34 & Oct 6,   2017 10:280:35  & 58032.417 -- 58032.436\\
              &03105479004& Jan 16,   2019 13:51:34 & Jan 16,   2019 10:16:33  & 58499.577 -- 58499.580\\
      \hline
&0003546000(1-5,7-8)  &  Mar 21, 2006  & May 6, 2012  & 53815.3 -- 56053.1 \\
Mrk~3&00037226001  & Jan 22, 2008 00:40:07 & Jan 22, 2008 02:37:05        & 54487.0 --  54487.1\\
&0008036800(1- 6,8-10)& Sep 7, 2014 & Apr 20, 2015 & 56907.7 -- 57132.7 \\
\hline
 \end{tabular}
}
    \label{tab:par_Swift_data}
{Columns 1 -- 5 denote Source name, Observation ID,  Start time (UT), End time (UT), MJD interval, respectively. See the text for details.}
{
(1) \citet{Jana21}  and (2) \citet{Okn22}. 
}
\end{table*}

%
%
\begin{table*}
  \centering 
 \caption{Best-fit parameters of spectral analysis of ASCA, {\it Suzaku} (shaded in gray), {\it Beppo}SAX (shaded in blue), {\it NuSTAR} (shaded in red), RXTE (shaded in green)  and {\it  XMM-Newton} (shaded in yellow) observations of AGNs
$^{\dagger}$.
}
\setlength{\tabcolsep}{1pt}
\resizebox{0.83\textwidth}{!}{%
 \begin{tabular}{lccccccccccccc}
      \hline\hline
Obs. ID&MJD,& $\alpha=$&$T_s$,& $\log(A)$ & N$_{com}^{\dagger\dagger}$ & E$_{line}$,&$N_{line}^{\dagger\dagger}$ & $\chi^2_{red}$ \\
           & day &$\Gamma-1$& keV&               &                                              & keV           &                                           &            (dof) \\
\hline
                &           &     &  1H~0707--495 &  &           &                        &             &                 \\
73043000   & 49792.0  & 1.03$\pm$0.09 & 0.107$\pm$0.004 & -1.14$\pm$0.07 &   2.31$\pm$0.9 &  0.83$\pm$0.09& 11.02$\pm$0.04&  1.01(86)\\
76031000 & 50949.8   &0.43$\pm$0.10  & 0.111$\pm$0.005 &-1.30$\pm$0.09 &   0.75$\pm$0.3 &   0.84$\pm$0.05& 2.12$\pm$0.06&    1.19(80)\\
87043000  &  51250.0 &  1.20$\pm$0.10  & 0.065$\pm$0.004& -1.06$\pm$0.03&    8.39$\pm$0.6 &   0.82$\pm$0.08& 5.37$\pm$0.02&   1.04(69)\\
\rowcolor[gray]{.9}700008010 &  53707.355 & 0.79$\pm$0.13 & 0.121$\pm$0.004 &  -1.14$\pm$0.05  & 7.6$\pm$0.2  &    0.83$\pm$0.02& 1.04$\pm$0.02 &1.00(802)\\
\hline
               &           &     &  NGC~1566 &  &           &                        &             &                 \\
\rowcolor[gray]{.9}707002010  &  56066.3 & 0.77$\pm$0.04 &  0.13$\pm$0.01 &  0.42$\pm$0.04 &  0.22$\pm$0.03 & 6.36$\pm$0.04 &  0.096$\pm$0.003 & 0.95(563) \\
\rowcolor[rgb]{1,0,0} 
80301601002  &  58295.8 & 1.01$\pm$0.03 &  0.14$\pm$0.02 &  0.12$\pm$0.02 &  1.07$\pm$0.04 & 6.36$\pm$0.05 &  0.098$\pm$0.007 & 0.99(963) \\
\rowcolor[cmyk]{0,0,1,0}
0800840201   &  58296.0 & 1.05$\pm$0.04 &  0.13$\pm$0.01 &  0.11$\pm$0.05 &  1.11$\pm$0.05 & 6.37$\pm$0.06 &  0.095$\pm$0.009 & 0.99(430) \\
\rowcolor[rgb]{1,0,0}
80401601002  &  58395.5 & 0.95$\pm$0.06 &  0.14$\pm$0.03 &  0.18$\pm$0.09 &  0.26$\pm$0.07 & 6.5$\pm$0.1 &  0.08$\pm$0.01 & 1.04(968) \\ 
\rowcolor[cmyk]{0,0,1,0}
0820530401   &  58395.5 & 0.92$\pm$0.04 &  0.13$\pm$0.01 &  0.18$\pm$0.05 &  0.26$\pm$0.05 & 6.37$\pm$0.06 &  0.095$\pm$0.009 & 0.99(430) \\
\rowcolor[rgb]{1,0,0}
80502606002  &  58639.3 & 1.15$\pm$0.06 &  0.14$\pm$0.03 &  0.42$\pm$0.09 &  0.34$\pm$0.07 & 6.5$\pm$0.1 &  0.08$\pm$0.01 & 1.04(968) \\ 
\rowcolor[cmyk]{0,0,1,0}
0840800401  &  58639.3 & 1.12$\pm$0.05 &  0.15$\pm$0.02 &  0.20$\pm$0.04 &  0.57$\pm$0.06 & 6.4$\pm$0.07 &  0.09$\pm$0.02 & 1.02(435) \\ 
\rowcolor[rgb]{1,0,0}
60501031002& 58703.4 & 0.89$\pm$0.05 &  0.14$\pm$0.03 &  0.61$\pm$0.05 &  0.37$\pm$0.05 & 6.46$\pm$0.03 &  0.096$\pm$0.003 & 0.89(831) \\   
\rowcolor[cmyk]{0,0,1,0}
0851980101  & 58706.6 & 0.85$\pm$0.06 &  0.13$\pm$0.02 &  0.64$\pm$0.04 &  0.29$\pm$0.04 & 6.42$\pm$0.09 &  0.096$\pm$0.003 & 0.97(210) \\  
\rowcolor[rgb]{1,0,0}
60501031004& 58713.2 & 0.75$\pm$0.09 &  0.13$\pm$0.01 &  0.62$\pm$0.03 &  0.35$\pm$0.03 & 6.43$\pm$0.05 &  0.096$\pm$0.003 & 0.93(723) \\    
\rowcolor[rgb]{1,0,0}
60501031006& 58716.4 & 0.65$\pm$0.07 &  0.14$\pm$0.03 &  0.72$\pm$0.05 &  0.21$\pm$0.06 & 6.40$\pm$0.06 &  0.096$\pm$0.003 & 0.87(722) \\    
\hline
              &           &     &  NGC~7679 &  &           &                        &             &                 \\
66019000  &  50995.6 & 0.62$\pm$0.02& 0.15$\pm$0.07&  2.00$^{\dagger\dagger\dagger\dagger}$ &  0.30$\pm$0.01  &  6.5$\pm$0.1 & 0.21$\pm$0.01 &  1.09(145)\\66019010   & 51432.9 & 0.55$\pm$0.04 & 0.13$\pm$0.04 &  0.08$\pm$0.01 &  0.37$\pm$0.02 &  6.41$\pm$0.09& 0.41$\pm$0.02 & 0.97(66)\\\rowcolor[cmyk]{1,0,0,0}
40631001  &  51153.6 & 0.60$\pm$0.02 &0.35$\pm$0.06 &  0.26$\pm$0.03&  0.71$\pm$0.03 & 6.4$\pm$0.5 & 0.075$\pm$0.02 &  1.04(82)\\   
\hline
               &           &     &  Mrk 3 &  &           &                        &             &                 \\
70002000& 49098.2& 0.10$\pm$0.02& 0.18$\pm$0.01&  -1.43$\pm$0.02&   0.001$\pm$0.001&6.41$\pm$0.09  &0.54$\pm$0.06& 1.05(52)\\
74041000& 50382.5& 0.10$\pm$0.02& 0.21$\pm$0.01&  -1.59$\pm$0.03&   0.002$\pm$0.001& 6.39$\pm$0.07& 0.63$\pm$0.02& 1.09(41)\\
\rowcolor[rgb]{0,1,0}
20330010100& 50442.6& 0.21$\pm$0.02&  0.28$\pm$0.01&  -3.47$\pm$0.04&   0.46$\pm$0.89&  6.41$\pm$0.08&   1.99$\pm$0.09&  0.83(47)\\
\rowcolor[rgb]{0,1,0}20330010200& 50552.3& 0.54$\pm$0.02&0.29$\pm$0.02&  -3.40$\pm$0.05&   1.21$\pm$0.18&  6.42$\pm$0.05&   1.9$\pm$0.2&  0.92(47)\\
\rowcolor[rgb]{0,1,0}20330010300& 50553.2& 0.31$\pm$0.02&  0.28$\pm$0.02&  -3.25$\pm$0.06&   0.63$\pm$0.09&  6.41$\pm$0.06&   2.4$\pm$0.6&   0.93(47)\\
\rowcolor[rgb]{0,1,0}20330010400& 50554.0& 0.82$\pm$0.03&  0.31$\pm$0.05&  -2.91$\pm$0.09&   0.54$\pm$0.06&  6.41$\pm$0.07&   2.8$\pm$0.3& 1.04(47)\\
\rowcolor[rgb]{0,1,0}20330010500& 50555.3& 0.17$\pm$0.02&  0.28$\pm$0.02&  -3.06$\pm$0.08&   0.44$\pm$0.07&  6.35$\pm$0.09&   1.73$\pm$0.09&  0.89(47)\\
\rowcolor[rgb]{0,1,0}20330010600& 50542.0& 1.02$\pm$0.03&  0.23$\pm$0.02&  -2.82$\pm$0.08&   0.59$\pm$0.07&  6.37$\pm$0.08&   3.83$\pm$0.08&   1.10(47)\\
\rowcolor[rgb]{0,1,0}20330010700& 50538.2& 0.87$\pm$0.02&  0.31$\pm$0.03&  -3.15$\pm$0.07&   1.05$\pm$0.19&  6.42$\pm$0.04&   3.93$\pm$0.09& 0.96(47) \\
\rowcolor[rgb]{0,1,0}20330010800& 50528.3& 0.21$\pm$0.02&  0.29$\pm$0.03&  -3.28$\pm$0.09&   0.67$\pm$0.17&  6.39$\pm$0.05&   2.99$\pm$0.08&  0.94(47) \\
\rowcolor[rgb]{0,1,0}20330010900& 50523.5& 0.11$\pm$0.01&  0.30$\pm$0.02&  -2.79$\pm$0.11&   0.55$\pm$0.07&  6.38$\pm$0.08&   2.20$\pm$0.09&  0.86(47)\\
\rowcolor[rgb]{0,1,0}20330011000& 50496.2& 0.95$\pm$0.02&  0.28$\pm$0.01&  -2.56$\pm$0.11&   0.20$\pm$0.07&  6.39$\pm$0.09&   4.33$\pm$0.08& 0.94(47)\\
\rowcolor[rgb]{0,1,0}20330011100& 50598.8& 0.50$\pm$0.02&  0.24$\pm$0.01&  -2.33$\pm$0.11&   0.03$\pm$0.07&  6.40$\pm$0.06&   1.84$\pm$0.05&  0.89(47)\\
\rowcolor[rgb]{0,1,0}20330011200& 50636.0& 0.51$\pm$0.01&  0.31$\pm$0.03&  -3.19$\pm$0.16&   0.30$\pm$0.04&  6.40$\pm$0.07&   2.55$\pm$0.09&  0.98(47)\\
\rowcolor[cmyk]{1,0,0,0}          
50132002& 50554.6& 0.55$\pm$0.04&  0.23$\pm$0.03&  -2.83$\pm$0.16&   0.04$\pm$0.01&  6.40$\pm$0.08&   0.68$\pm$0.04& 0.90(84)\\
\rowcolor[gray]{.9}
100040010& 53665.1& 0.10$\pm$0.02&  0.10$\pm$0.01&  -0.47$\pm$0.02& 0.0001$\pm$0.0001& 6.41$\pm$0.05&  0.54$\pm$0.06&  0.94(629)\\
\rowcolor[gray]{.9}
709022010& 56931.9& 0.11$\pm$0.01&  0.10$\pm$ 0.01&  -0.90$\pm$0.02& 0.0011$\pm$0.0001& 6.41$\pm$0.02&  1.03$\pm$0.01& 0.98(238)\\
\rowcolor[gray]{.9}
709022020& 56937.8& 0.09$\pm$0.02&  0.13$\pm$ 0.01&  0.27$\pm$0.02& 0.00015$\pm$0.0001& 6.50$\pm$0.08&  5.06$\pm$0.01& 1.05(258)\\
\rowcolor[gray]{.9}
709022030& 56953.2& 0.10$\pm$0.02&  0.14$\pm$ 0.01&  0.53$\pm$0.02& 0.00014$\pm$0.0001& 6.57$\pm$0.09&  5.31$\pm$0.01&  1.02(300)\\
\rowcolor[gray]{.9}
709022040& 57104.2& 0.11$\pm$0.03&  0.16$\pm$0.01&  0.02$\pm$0.01& 0.00003$\pm$0.0001& 6.48$\pm$0.07&  3.62$\pm$0.01&  1.54(293)\\
\rowcolor[gray]{.9}
709022050& 57116.4& 0.10$\pm$0.01&  0.15$\pm$0.02& -0.03$\pm$0.01& 0.00003$\pm$0.0001& 6.49$\pm$0.04&  3.15$\pm$0.01& 1.47(154)\\
\hline
 \end{tabular}
}
    \label{tab:fit_table_suz+asca_0707+1655+7679}
{\\
\scriptsize Parameter errors correspond to 1$\sigma$ confidence level. 
$^\dagger$ The spectral model is  {\it tbabs*(BMC + Gaussian/Laor)}. 
$^{\dagger\dagger}$ Normalization parameters of {\it BMC} component is in units of $L_{35}^{\rm soft}/d^2_{10}$, $erg/s/kpc^2$, 
where $L_{35}^{\rm soft}$ is  
the soft photon  luminosity in units of 10$^{35}$;  
$d_{10}$ is the distance to the source in units of 10 kpc. 
$^{\dagger\dagger\dagger}$ {\it Gaussian/Laor} component is in units of $10^{-4}\times$ total photons cm$^{-2}$s$^{-1}$. $^{\dagger\dagger\dagger\dagger}$ When parameter log(A)$>>$1, it is fixed to a value 2.0 (see comments in the text).  
$N_H$ was fixed at the value 2.5$\times 10^{20}$ cm$^{-3}$ (for NGC~1566), 
3$\times 10^{21}$ cm$^{-3}$ (for 1H~0707) and 
4$\times 10^{21}$ cm$^{-3}$ (for NGC~7679).
For 1H~0707 we presented the strongest {\it Laor}-line component  among all line features in the source spectrum, although we used the full composition of 
N~XVII, O~III, Fe~XVII, Ne~X, S~XVI,  Fe~I--XXII and Fe~ XXV/Fe~XXVI K$_{\alpha}$ lines  with energies of  0.5, 0.65,  0.85, 1.02, 2.9, 6.4 and 6.8 keV, respectively.
}
\end{table*}

%
%
\begin{table*}
 \caption{BH mass scaling for 1H~0707--495.}
 \label{tab:par_scal_0707}
 \centering 
\setlength{\tabcolsep}{1pt}
\resizebox{1.\textwidth}{!}{%
 \begin{tabular}{lllrrll}
  \hline\hline                        
Reference sources   & $m_r$,  M$_{\odot}$       & $i_r^{(a)}$, deg  & $N_r$,               $L_{39}/d^2_{10}$      & $d_r^{(b)}$, kpc  \\

      \hline
HLX--1~ESO 243--49$^{(1)}$ & $(7.2\pm 0.7)\times 10^4$ & 75 & $4.2\times 10^{-6}$ & $ 95\pm 10$\\
M101 ULX--1$^{(2)}$ & $(3.7 \pm 0.6)\times 10^4$ & $18$ & $3\times 10^{-4}$ & $6.9\pm 0.7$\\
OJ 287$^{(3)}$ & $(1.25 \pm 0.5)\times 10^8$ & $50$ & $2.4\times 10^{-4}$ & $1037\pm 10$\\
SDSS~J0752$^{(4)}$ & $\sim9\times(1\pm 0.31)\times10^7$ & $70$ & $1.1\times 10^{-2}$ & $\sim$500\\

 \hline\hline                        
Target source   & $m_{t}$ (M$_{\odot}$) & $i_t^{(a)}$ (deg) & $d_t^{(b)}$ (Mpc) &    \\

                      & M$_{\odot}$ &  deg             & Mpc &    \\
      \hline
1H~0707  & $\sim6.8\times(1\pm 0.45)\times10^7$ &  55 &  160    &    that  using  ESO 243-49  as a ref. source\\
1H~0707  & $\sim6.8\times(1\pm 0.45)\times10^7$ &  55 &  160    &    that  using  M101 ULX-1  as a ref. source\\
1H~0707  & $\sim6.8\times(1\pm 0.45)\times10^7$ &  55 &  160    &    that  using  OJ 287  as a ref. source\\
1H~0707  & $\sim6.8\times(1\pm 0.45)\times10^7$ &  55 &  160    &    that  using  SDSS~J0752  as a ref. source\\
1H~0707  & Final  estimate                                      &  55 &  160   &   as a standard deviation for a mean: \\
           &  $\sim6.8\times(1\pm 0.27)\times10^7$ &      &      &   $0.45/4^{1/2}=0.27$\\
           \hline

 \end{tabular}
}
\\
(a)  System inclination in the literature and  
(b) source distance found in the literature. 
(1) \cite{tsei16b}; 
(2)  \cite{TS16}; and 
(3) \cite{TSC23}.
 \end{table*}

%
%
\begin{table*}
 \caption{BH mass scaling for NGC~1566 and NGC~7679.}
 \label{tab:par_scal_1566+7679}
 \centering 
\setlength{\tabcolsep}{1pt}
\resizebox{1.\textwidth}{!}{%
 \begin{tabular}{lllrrll}
 \hline\hline                        
Reference sources   & $m_r$,  M$_{\odot}$       & $i_r^{(a)}$, deg  & $N_r$,               $L_{39}/d^2_{10}$      & $d_r^{(b)}$, kpc  \\
 \hline
NGC~4051$^{(1)}$ & 6$\times 10^5$  & 50 & $9\times 10^{-4}$ & 9800\\
GX~339--4$^{(2)}$ & $\sim$6  & ... & $5.1\times 10^{-2}$ & $7.5\pm 1.6$\\
GRO~J1655--40$^{(3)}$ & 6.3$\pm 0.3$ & 70$\pm$ 1 & $5.2\times 10^{-2}$ & 3.2$\pm$0.2\\
Cyg~X--1 $^{(4)}$ & 6.8--13.3 & 35$\pm$5 & 0.9 & 2.5$\pm$ 0.3 \\
4U~1543--47 $^{(5)}$ & $\sim$11 & 20.7$\pm$1.5 & $0.12$ & 7.5$\pm$1\\
 \hline\hline                       
Target source   & $m_{t}$,  M$_{\odot}$  & $i_t^{(a)}$, deg  & $d_t^{(b)}$, Mpc&    \\
 \hline
NGC~1566  & $\sim1.9\times(1\pm 0.45)\times10^5$ &  37.5 &  21.3    &    that  using  NGC~4051  as a ref. source\\
NGC~1566 & $\sim1.9\times(1\pm 0.45)\times10^5$ &  37.5 &  21.3    &    that  using  GX~339--4 as a ref. source\\
NGC~1566  & $\sim1.9\times(1\pm 0.45)\times10^5$ & 37.5 &  21.3    &    that  using  GRO~J1655--40  as a ref. source\\
NGC~1566  & $\sim1.9\times(1\pm 0.45)\times10^5$ &  37.5 &21.3    &    that  using  Cyg~X--1  as a ref. source\\
NGC~1566  & $\sim1.9\times(1\pm 0.45)\times10^5$ & 37.5 & 21.3   &    that  using  4U~1543--47   as a reference source\\
NGC~1566  & Final  estimate                                      &  37.52& 21.3   &   as a standard deviation for a mean: \\
           &  $\sim1.9\times(1\pm 0.20)\times10^5$ &      &      &   $0.45/5^{1/2}=0.20$\\
\hline
NGC~7679  & $\sim8.4\times(1\pm 0.45)\times10^6$ &  30 &  57.28    &    that  using  NGC~4051  as a ref. source\\
NGC~7679 & $\sim8.4\times(1\pm 0.45)\times10^6$ &  30 &  57.28    &    that  using  GX~339--4 as a ref. source\\
NGC~7679  & $\sim8.4\times(1\pm 0.45)\times10^6$ &  30 & 57.28    &    that  using  GRO~J1655--40  as a ref. source\\
NGC~7679  & $\sim8.4\times(1\pm 0.45)\times10^6$ &  30 & 57.28    &    that  using  Cyg~X--1  as a ref. source\\
NGC~7679  & $\sim8.4\times(1\pm 0.45)\times10^6$ & 30 & 57.28    &    that  using  4U~1543--47   as a reference source\\
NGC~7679  & Final  estimate                                      &  30 &57.28   &   as a standard deviation for a mean: \\
           &  $\sim8.4\times(1\pm 0.20)\times10^6$ &      &      &   $0.45/5^{1/2}=0.20$\\
\hline
Mrk~3  & $\sim2.2\times(1\pm 0.45)\times10^8$ &  50 &  63.2    &    that  using  NGC~4051  as a ref. source\\
Mrk~3 & $\sim2.2\times(1\pm 0.45)\times10^8$ &  50 &  63.2    &    that  using  GX~339--4 as a ref. source\\
Mrk~3  & $\sim2.2\times(1\pm 0.45)\times10^8$ &  50 & 63.2    &    that  using  GRO~J1655--40  as a ref. source\\
Mrk~3  & $\sim2.2\times(1\pm 0.45)\times10^8$ &  50 & 63.2    &    that  using  Cyg~X--1  as a ref. source\\
Mrk~3  & $\sim2.2\times(1\pm 0.45)\times10^8$ & 50 & 63.2    &    that  using  4U~1543--47   as a reference source\\
Mrk~3  & Final  estimate                                      &  50 &63.2   &   as a standard deviation for a mean: \\
           &  $\sim2.2\times(1\pm 0.20)\times10^8$ &      &      &   $0.45/5^{1/2}=0.20$\\
\hline
 \end{tabular}
}
\\
(a)  System inclination in the literature and  
(b) source distance found in the literature. 
(1) \cite{Christopoulou97,McHardy04,Haba08,Pounds+King13,Lobban11,Terashima09,Denney09,SCT18}; ;
(2)  \cite{Munoz-Darias08,Hynes04}
(3) \cite{Green01,Hjellming+Rupen95};
(4) \cite{Herrero95,Ninkov87}; and 
(5) \cite{Orosz03,Park04}.
 \end{table*}

%
%

\begin{table*}
  \centering 
 \caption{General properties of the Sy1 (left part), Sy2 (right part) and CL-AGN (indicated by yellow color)  sample galaxies. }
\setlength{\tabcolsep}{1pt}
\resizebox{1.\textwidth}{!}{%
 \begin{tabular}{lllll|llllrllcc}
      \hline\hline
 Sy1, CL-AGN& D$^\dagger$,    & M$_{BH}$,  & L$_{x,44}^{\dagger\dagger}$     & $\bar\Gamma$ &     Sy2 AGN& D$^\dagger$,    & M$_{BH}$,  & L$_{x,40}^{\dagger\dagger}$     & $\bar\Gamma$ \\
source         & Mpc & $\times$10$^6$M$_{\odot}$& &                      &         source         & Mpc & $\times$10$^6$M$_{\odot}$& &                          \\

\hline
NGC~7314$^{(1)}$    & 15.4 &  0.87 &   0.11&  1.80$\pm$0.15 &     NGC~7679 & 58.6 &  8.4  &      3200&   1.60$\pm$0.06 \\
NGC~3227$^{(2,3)}$    & 20.4 &  42.2 &   0.19&  1.65$\pm$0.12 & Mrk~348$^{(13)}$  & 63.90 & 38.01 &   575.4 &  1.50$\pm$1.56\\
NGC~3516$^{(2,3)}$    & 37.5 &  42.7 &  0.62 &  1.63$\pm$0.12 &   NGC~424$^{(13)}$  & 47.60 & 60.5 &   54.95 & 1.49$\pm$0.90  \\
NGC~4593$^{(2,3)}$   & 42.0 &  5.36 &     0.8 &  1.79$\pm$0.14 &  Mrk~573$^{(13)}$  & 71.30 & 23.44 &   44.6 &  2.50$\pm$2.78\\
NGC~3783$^{(2,3)}$   & 44.7 &   29.8&     1.7 &  1.68$\pm$0.08 &  NGC~788$^{(13)}$    & 56.10 & 26.91 &   128.8 & 1.41$\pm$1.67\\
NGC~7469$^{(2,3,4)}$     & 62.7&   12.2&    1.47 &  1.84$\pm$0.10  & ESO~417--G06$^{(13)}$& 65.60 & 27.54 & 288.4 &1.03$\pm$1.25
 \\
IC~4329A$^{(2,3)}$     & 70.5 &   9.9&     1.06&   1.76$\pm$0.03 & Mrk~1066$^{(13)}$ & 51.70 & 16.98 &   25.11 &  2.17$\pm$2.40  \\
NGC~5548$^{(5)}$    & 74.5 &  52.2 &   2.9 &  1.69$\pm$0.12 &    3C~98.0$^{(13)}$     & 124.90 & 56.23 &   1348.9 &1.04$\pm$0.81   \\
Mrk~79$^{(2,6)}$        & 94.3 &  52.4&    2.2 &  1.71$\pm$0.21 &  Mrk~3$^{(13)}$     & 63.20 & 549.54 &    173.7  &  1.25$\pm$1.33 \\
Ark~120$^{(2,6)}$      &  138 &   150 &     7.6 &  1.73$\pm$0.17 & Mrk~1210$^{(13)}$& 53.60 & 50.12 &    204.4  &  1.01$\pm$1.40   \\
Mrk~509$^{(2,3)}$       & 141 &   143 &   2.41 &  1.72$\pm$0.08 & IC~2560$^{(13)}$     & 34.80 & 2.88 &   3.7  &  1.32$\pm$1.50     \\
Fairall~9$^{(2,3)}$      & 198  &  255  &    10.1 &  1.84$\pm$0.18 & NGC~3393$^{(13)}$  & 48.70 & 125.89 &   43.6   &  2.67$\pm$3.04  \\
MR~2251-178$^{(7)}$& 271 &   240 &    46.6 &  1.75$\pm$0.11&    NGC~4507$^{(13)}$  & 46.00 & 181.97 &   109.6 &  1.62$\pm$1.70  \\

NGC~4051$^{(2,3,8)}$   & 12.7 &  1.91&     0.05 &  1.90$\pm$0.10 & NGC~4698$^{(13)}$  & 23.40 & 33.88 &   1.38 &  2.16$\pm$1.98   \\
NGC~5506$^{(3,9)}$   & 29.1 &  5.1  &    1.29&   1.85$\pm$0.06 & NGC~5194$^{(13)}$  & 7.85 & 5.37 &   0.34  &    2.68$\pm$3.01   \\
MCG~63015$^{(10)}$ & 35.8 &  3.26&     0.78 &  1.76$\pm$0.09 & Mrk~268$^{(13)}$  & 161.50 & 89.12 &   21.8  & 2.11$\pm$3.07  \\
Mrk~766$^{(11)}$     & 57.6 &  1.76&     1.15 &  2.02$\pm$0.20 &  Circinus$^{(13)}$    & 4.21 & 51.28 &   0.63  &    0.98$\pm$0.75  \\
Ark~564$^{(12)}$      & 98.5 &  2.6  &     2.8 &  2.51$\pm$0.09 &   NGC~5643$^{(13)}$  & 16.90 & 1.99 &   2.75  &  0.99$\pm$1.42  \\
Mrk~335$^{(2,6)}$     & 103  &  14.2&     1.24 &  1.97$\pm$0.51 & Mrk~477$^{(13)}$  & 156.70 & 15.84 &   398.1  & 1.10$\pm$1.48\\
Mrk~110$^{(2,6)}$     & 151  &  25.1&     8.71 &  1.81$\pm$0.16 & IC~4518A$^{(13)}$   & 65.20 & 30.19 &   114.8  &  1.94$\pm$1.66 \\
1H~0707-495 & 160 & 6.8 &     1.38&   2.00$\pm$1.06 &  ESO~138--G01$^{(13)}$& 36.00 & 0.31 & 169.8 &  2.31$\pm$2.40    \\
\fcolorbox{red}{yellow}{NGC 1566} & 21.3 &  0.19 &      0.06&   1.60$\pm$0.06  & NGC~6300$^{(13)}$  & 14.43 & 15.13 &   20.89  & 1.00$\pm$0.99  \\
\fcolorbox{red}{yellow}{Mrk~273$^{(1)}$}  & 156.70 & 54.95 &   0.002  & 1.91$\pm$0.07  & NGC~7172$^{(13)}$ & 33.90 & 158.48 &    316.2  & 1.52$\pm$1.58    \\
\fcolorbox{red}{yellow}{NGC~7319$^{(1)}$} & 77.25 & 726.91 &    0.098 &  1.29$\pm$1.20  & NGC~7212$^{(13)}$ & 111.80 & 34.67 &   64.56  & 1.07$\pm$0.47   \\
\hline
      \end{tabular}
}
    \label{tab:compar-all-AGN}
{
$^\dagger$ Luminosity distance ($D$) are taken from the NED data base.
$^{\dagger\dagger}$ 
$L_{x,44}$ and $L_{x,40}$ are 
the soft photon  luminosities in 2--10 keV energy range in units of 10$^{44}$ and 10$^{40}$ erg/s, respectively. 
{
$^{(1)}$ \cite{Gu06}, \cite{Elvis94}; 
$^{(2)}$ \cite{Peterson04}; 
$^{(3)}$ \cite{Ricci13}; 
$^{(4)}$ \cite{STU18}; 
$^{(5)}$ \cite{Bentz+Katz15}; 
$^{(6)}$ \cite{Woo+Urry02};
$^{(7)}$ \cite{Dunn14}; 
$^{(8)}$ \cite{TSCO20};  
$^{(9)}$ \cite{Nikolajuk09}; 
$^{(10)}$ \cite{Hu16}; 
$^{(11)}$ \cite{Bentz09}; 
$^{(12)}$ \cite{Botte04}; and 
$^{(13)}$\cite{Hernandez15}. 
}
}
\end{table*}

\newpage
%
%

%
%

%
%

%
%
%
%

%
%

%
%

%
%

%
%

%
%

%
%

%
%

%
%

%
%

%
%

%
%

%
%

%
%

%
%

%
%

%
%

\end{document}